\documentclass[pdflatex,sn-mathphys-num]{sn-jnl}%
\makeatletter
\def\breakurldefns{}
\makeatother
\usepackage{graphicx}%
\usepackage{multirow}%
\usepackage{indentfirst}%
\usepackage{amsmath,amssymb,amsfonts}%
\usepackage{amsthm}%
\usepackage{mathrsfs}%
\usepackage[title]{appendix}%
\usepackage{xcolor}%
\usepackage{textcomp}%
\usepackage{manyfoot}%
\usepackage{booktabs}%
\usepackage{algorithm}%
\usepackage{algorithmicx}%
\usepackage{algpseudocode}%
\usepackage{listings}%
\usepackage{caption}%
\usepackage{enumitem}%
\usepackage{float}%
\usepackage{longtable}%
\usepackage{array}%
\usepackage{makecell}%
\usepackage{lscape}%
\renewcommand{\figurename}{Figure}
\renewcommand{\tablename}{Table}
\theoremstyle{thmstyleone}%
\theoremstyle{thmstyletwo}%
\theoremstyle{thmstylethree}%
\begin{document}

\title[AI Subfields: A Bibliometric View]{Horizontal and Longitudinal Comparisons Among AI Subfields: A Bibliometric Perspective}


\author[1]{\fnm{Zeyu} \sur{Li}}\email{lzyfun@cuc.edu.cn}

\author[2]{\fnm{Yalan} \sur{Jin}}\email{yalanjin@cuc.edu.cn}

\author[3]{\fnm{Shuyu} \sur{Chen}}\email{csy@mails.cuc.edu.cn}

\author[4]{\fnm{Tingxin} \sur{Jiang}}\email{aciyao@mails.cuc.edu.cn}

\author[5]{\fnm{Xinyi} \sur{Chang}}\email{xian@cuc.edu.cn}

\author*[2]{\fnm{Lu} \sur{Yuan}}\email{yuanlucuc@cuc.edu.cn}

\affil[1]{\orgdiv{State Key Laboratory of Media Convergence and Communication}, \orgname{Communication University of China}, \orgaddress{\street{No. 1, Dingfuzhuang East Street}, \city{Chaoyang District}, \postcode{100024}, \state{Beijing}, \country{China}}}

\affil*[2]{\orgdiv{School of Journalism}, \orgname{Communication University of China}, \orgaddress{\street{No. 1, Dingfuzhuang East Street}, \city{Chaoyang District}, \postcode{100024}, \state{Beijing}, \country{China}}}

\abstract{Recent artificial intelligence has developed rapidly with significant interdisciplinary expansion, yet existing studies often treat it as a whole, lacking systematic long-term subfield comparisons and structural analyses, thereby limiting understanding of internal differences and evolutionary mechanisms. To address this gap, we employ bibliometric methods, using expert interviews and indicator screening to construct an analytical framework. Twelve bibliometric indicators are selected across three dimensions: Impact and Dissemination, Collaboration Characteristics, and Author Characteristics. We conduct horizontal and longitudinal analyses of five subfields (AI, CV, ML, NLP, Web\&IR) from 2000 to 2024. Using CSRankings classification and a dataset of 106,622 papers, we apply violin plots, chord diagrams, and sankey diagrams to characterize structural features and evolutionary paths. Results show that these subfields have entered high-intensity knowledge diffusion: academic impact increased, knowledge dissemination accelerated, external disciplinary reliance grown, and knowledge production shifted from closed accumulation to open, interdisciplinary, multi-actor networks. On this basis, subfields exhibit significant structural differentiation: CV leads in academic impact with a task-oriented trajectory; ML shows shrinking industry collaboration but concentrated international collaboration with a relatively dispersed structure; Web\&IR is strongly industry-driven with a stable collaboration network; AI shows continuous growth; NLP remains relatively stable. Overall, this study reveals artificial intelligence evolving from unified diffusion to structural differentiation, constructs an extensible multidimensional framework, and provides a quantitative approach for understanding complex technological field evolution.}


\maketitle
\section{Introduction}\label{sec1}

Artificial Intelligence, a transformative technological paradigm, has catalyzed various disciplinary and industrial transformations worldwide—encompassing medicine, finance, and autonomous driving (Yuan et al., 2020)\cite{Yuan2020ScienceBehindAI}. Bibliometric analyses of artificial intelligence profound insights into its evolutionary trends, academic influence, and interdisciplinary integration , laying a robust theoretical and empirical foundation for scholarly advancement and policy formulation (Fiala \& Tutoky, 2017)\cite{Fiala2017ComputerSciencePapers}.

Current bibliometric analyses of artificial intelligence center on three interrelated dimensions: first, the macro-evolution of artificial intelligence, encompassing technological advancements, evolving research priorities, and emerging trends. This shift not only reflects the gradual maturation of artificial intelligence technology but also indicates that research objectives in this field are shifting from "theoretical innovation" toward "practical implementation," particularly in subfields such as machine learning, natural language processing, and computer vision(Yuan et al., 2020)\cite{Yuan2020ScienceBehindAI} . Second, the influence of artificial intelligence across various disciplines. Bickley, Chan, and Torgler (2022)\cite{Bickley2022AIEconomics}explored artificial intelligence applications in economics, and demonstrated that technologies including machine learning, deep learning, and neural networks substantially enhance the predictive performance of economic models—particularly for financial market analysis and economic forecasting. Third, the development status and comparative analysis of artificial intelligence across nations and institutions. Liu et al. (2021)\cite{Liu2021TrackingDevelopmentsAI} noted that the global landscape of artificial intelligence research is gradually shifting from being dominated by a single hub toward a new phase characterized by both multipolar competition and cooperation.

Despite the rich insights offered by existing literature on artificial intelligence, certain limitations persist. Most investigations treat artificial intelligence as a monolithic field, neglecting in-depth analysis of inter-subfields differences and thus resulting in an oversimplified understanding of individual subfields. Concurrently, longitudinal studies on artificial intelligence's temporal evolution remain scarce, failing to delve into the uniqueness and evolutionary trajectories of individual subfields—resulting in an incomplete understanding of the subfields. As Bickley et al. (2022)\cite{Bickley2022AIEconomics} emphasized, the growing interdisciplinary applications of artificial intelligence, coupled with the field's increasing diversity and complexity, demand research conducted from a refined perspective. Accordingly, this study investigates the following research questions from three dimensions: Impact and Dissemination, Collaboration Characteristics and Author Characteristics.

Q1 (Horizontal Comparison): What are the differences among artificial intelligence subfields in various dimensions over the past twenty years? For example, how do different subfields perform in academic output, citation counts, author mobility and international collaboration?

Q2 (Longitudinal Comparison): How do artificial intelligence subfields develop longitudinally over time in various dimensions, and what are the differences between them? For example, how do the cooperation patterns, number of authors, and citation speed of each subfield change?

To address these research questions, this study employs a multi-dimensional comparative analysis framework grounded in twelve bibliometric indicators. First, we categorized artificial intelligence into five subfields based on CSRankings, and compiled a dataset of 106,622 articles from 13 top-tier conferences spanning 2000 to 2024, which were partitioned into five-year time intervals. Next, we clarified the above three dimensions through demand interview analysis with experts in the computer field. Integrating considerations of article volume, data support, and other practical factors, we systematically measured, screened, and categorized indicators under each dimension via a comprehensive literature review. Specifically, for Q1, we established mathematical models, counted representative articles, and conducted horizontal comparisons through violin plots and chord diagrams to reveal differences among subfields; for Q2, we calculated data for each time period through formulas, conducted joint analysis of results from five time periods, and used slope charts to explore the longitudinal evolution of subfields. Finally, we interpreted the visualization results in detail, discussed the underlying reasons, and validated and proofread our conclusions through deliberations with experts in the artificial intelligence field.

The findings show that the five subfields of artificial intelligence exhibit both several common trends and significant structural differences in their knowledge evolution.Overall, the academic influence and knowledge dissemination capacity of the five subfields have continued to increase, the speed of knowledge diffusion has significantly accelerated, dependence on closed internal knowledge circulation has weakened, and the absorption of external disciplinary resources has continued to strengthen; meanwhile, collaboration networks have become increasingly complex.Further comparison shows that there is clear heterogeneity across subfields in the pathways through which influence is formed, the patterns of interdisciplinary connection, and the structures of collaboration: AI shows sustained growth in academic influence and collaboration networks; CV has the strongest academic influence and displays a task-centered structure of interdisciplinary diffusion; ML shows a structure characterized by relatively dispersed industry collaboration but more concentrated international collaboration; NLP develops relatively steadily, but its technological iteration increasingly reflects a firm-led model of large-scale collaboration; and Web\&IR exhibits the most prominent industry-led and application-oriented characteristics, together with a relatively concentrated and stable structure of interdisciplinary collaboration.

The above findings suggest that artificial intelligence should not be simply understood as a set of parallel technical branches, but rather as an open knowledge system that continuously absorbs external knowledge, depends on the collaboration of multiple actors, and exhibits substantial internal heterogeneity. Through a bibliometric analysis of different AI subfields, this study reveals differences in knowledge evolution and structural characteristics, thereby deepening understanding of AI's overall development logic and internal structure. Methodologically, it offers a systematically organized, transferable bibliometric framework that can be applied to other emerging fields. Empirically, the curated dataset and longitudinal subfield comparisons provide a reliable foundation for future quantitative research in conference-driven disciplines. These contributions not only support evidence-based decisions in research organization, talent allocation, and innovation investment but also foster cross-community dialogue between bibliometrics and computer science.


\section{Previous Studies}
This study focuses on the cross-sectional comparison and longitudinal evolution of the artificial intelligence field as its research questions. Existing academic works centered on "Bibliometrics on Artificial Intelligence", "Construction of Bibliometric Indicators", and "Bibliometrics on Comparison among Research fields" have provided research background and methodological support for this study.

\subsection{Bibliometrics on Artificial Intelligence}
Bibliometric analysis targeting artificial intelligence primarily focuses on the evolution of the artificial intelligence field, the expansion of academic influence, and the comparison of research capabilities across nations and institutions. By tracing technological advancements, the evolution of research foci, and collaboration patterns, the research shows the current development status of the artificial intelligence field.

\subsubsection{Evolution and Development}

The macro-evolution of the artificial intelligence field encompasses technological advancements, the evolution of research foci, and emerging trends. For instance, Yuan et al. (2020)\cite{Yuan2020ScienceBehindAI} documented a theoretical-to-applied shift in artificial intelligence research—prominent in subfields including machine learning, natural language processing, and computer vision, where applied investigations now prevail. Fiala \& Tutoky (2017)\cite{Fiala2017ComputerSciencePapers} noted that the continuous advancement of artificial intelligence technologies in fundamental algorithms, computational models, and automated tools has propelled the development of the entire artificial intelligence field.Research by Al- Marzouqi \& Arabi (2024)\cite{AlMarzouqi2024ComparativeAnalysis} indicates that the global scientific impact of artificial intelligence has steadily grown, with the rapid development of China and India in the artificial intelligence field in particular establishing them as pivotal players in global scientific research competition.

\subsubsection{Dissemination and Influence}
With the continuous expansion of artificial intelligence, it has been widely applied in multiple disciplinary fields.Bickley, Chan, and Torgler (2022)\cite{Bickley2022AIEconomics} examined artificial intelligence applications in economics—specifically the proliferation of technologies including machine learning, deep learning, and neural networks across economic subfields. They noted that the introduction of artificial intelligence has substantially enhanced the predictive performance of economic models, particularly for financial market analysis and economic forecasting.Concurrently, Arencibia-Jorge, Vega-Almeida and Carrillo-Calvet(2022)\cite{ArencibiaJorge2022EvolutionaryStagesAI} explored artificial intelligence’s pivotal role in advancing the Sustainable Development Goals (SDGs), analyzing how artificial intelligence fosters technological innovation in fields such as social sciences and environmental protection while facilitating global collaboration.Within the cybersecurity field, research by Naeem Abbas et al. (2019)\cite{Abbas2019AICyberSecurity} demonstrates that artificial intelligence serves a critical function in mitigating data breaches, detecting cyberattacks, and safeguarding privacy, thereby providing robust technical support for global cybersecurity.

\subsubsection{Competition and Cooperation}

Furthermore, competition and collaboration among nations and institutions have also propelled the advancement of artificial intelligence. Liu et al. (2021)\cite{Liu2021TrackingDevelopmentsAI} showed that the United States took a leading position in the early artificial intelligence field, while China has ranked first in the number of artificial intelligence-related papers published in recent years, with increasing international cooperation networks; countries such as the United States, China and the United Kingdom play a core role in artificial intelligence field cooperation.Cai et al. (2024)\cite{cai2024} proposed that the generation mechanism of research leadership in international cooperation is particularly critical. Specifically, in China-US collaborations the distribution of leadership positions exerts a decisive influence on the global dissemination and application of artificial intelligence technologies.

\subsection{Construction of Bibliometric Indicators}

Bibliometric indicators hold substantial significance for research evaluation and academic advancement (Leibel \& Bornmann, 2024)\cite{Leibel2024DisruptionIndex}. They enable research institutions, governments, and funding bodies to quantify academic output, impact, and research collaboration (Binmakhashen et al., 2022)\cite{AlJamimi2022BibliometricsEmergingMarkets}, furnish data support for research policies, resource allocation, and academic rankings, and facilitate global research collaboration and academic innovation (Leibel \& Bornmann, 2024)\cite{Leibel2024DisruptionIndex}.

\subsubsection{Classification of Existing Indicators}

Currently, numerous bibliometric indicators can be categorized into four dimensions: research output, research impact, research innovation, and research development (Ma et al., 2008)\cite{Ma2008WorldUniversitiesCS}. In terms of research output, João M. Fernandes et al. (2016)\cite{Fernandes2017AuthorsCS}, through an analysis of the number of authors and papers, found that the number of authors of papers in the computer science field has increased annually, with research collaboration continuously intensifying.Porter and Rafols (2009)\cite{Porter2009InterdisciplinaryScience} analyzed citation counts and the number of authors per article, revealing that over time, citation numbers and co-author in six disciplinary fields dominated by computer science has increased significantly. In terms of research impact, Chakraborty et al. (2014)\cite{Chakraborty2014CitationInteractionsCS} analyzed the "inwardness" indicator and found that the status of artificial intelligence, algorithms and networks in computer science is constantly improving. Chavarro, Pérez-Taborda and Ávila (2022)\cite{Chavarro2022AIforSustainableDevelopment} analyzed patent citations and showed that the innovation of artificial intelligence technology has not only advanced research progress in academia but also yielded extensive applications in industry. In terms of research innovation, Abramo, D'Angelo, and Caprasecca (2009)\cite{Abramo2009GenderProductivityItaly}, by analyzing metrics including Contribution Intensity and Fractional Scientific Strength, found that the intensity of research innovation is closely associated with field maturity and the collaboration patterns of research teams. Singh, Uddin, and Pinto (2015)\cite{Singh2015Top100InstitutionsCS}, through an analysis of India's interdisciplinary collaborations and research hotspots, indicated that the global impact of its research innovation is relatively modest. In terms of research development, Abramo, D'Angelo and Caprasecca (2009)\cite{Abramo2009GenderProductivityItaly} analyzed the distribution of professional titles and changes in research output, revealing the differences in career development among researchers of different genders. Chakraborty et al. (2014)\cite{Chakraborty2014CitationInteractionsCS} analyzed shifts in global research focus, noting that certain fields have gradually fallen from the forefront, while emerging fileds such as big data and cloud computing have begun to gain prominence. Chavarro, Pérez-Taborda, and Ávila (2022)\cite{Chavarro2022AIforSustainableDevelopment} assessed global collaboration trends in artificial intelligence research through an analysis of cross-regional collaborations and collaborations among different income groups.

\subsubsection{Alternative Metrics}

Beyond metrics derived from bibliographic data, "altmetrics" have emerged as a crucial complement to research evaluation in recent years. Ding \& Du (2023)\cite{Ding2023PublicationDelaysPLOS} noted that altmetrics primarily deliver real-time feedback on the impact of academic outputs via social networks and news dissemination, addressing the limitation that traditional citation metrics fail to reflect research achievements promptly (Mingers, J., \& Leydesdorff, L. (2015))\cite{Mingers2015ReviewScientometrics}. Academic social network metrics proposed by Wiechetek \& Pastuszak (2022)\cite{Wiechetek2022AcademicSocialNetworks} highlight the potential of dynamic indicators on platforms like ResearchGate for assessing university research performance, expanding applications in academic interaction and collaboration networks, and boosting the societal visibility of research outputs. By introducing altmetrics such as social media interactions, download volumes, and citation deposits, Mingers, J., \& Leydesdorff (2014)\cite{Mingers2015ReviewScientometrics} provided real-time and multi-dimensional feedback, reflecting the dissemination of academic achievements outside the academic circle.

\subsubsection{Multi-Dimensional Indicator System}

A single metric has limitations. Contemporary research evaluation employs a multi-dimensional indicator framework that integrates citation data, collaborative dynamics, and academic activities to enable a comprehensive assessment of academic outputs.For instance, Bornmann et al. (2015)\cite{Bornmann2015HIndexSocialSciencesJournals} developed 17 bibliometric indicators, categorizing them into three groups—publication-based metrics, citation-based metrics, and hybrid metrics—and proposed a comprehensive scholar evaluation framework that holistically assesses scholars' impact and productivity.Wildgaard et al. (2015)\cite{Wildgaard2015AuthorLevelIndicators} evaluated the performance of scholars in disciplines such as astronomy, environmental science, philosophy, and public health by adopting indicators such as the number of publications, citation counts, and their combinations (e.g., h-index).Franceschet (2009)\cite{franceschet2009} analyzed the differences between Web of Science and Google Scholar data sources, selected 13 bibliometric indicators to measure the performance of scholars in the computer science field, and promoted research evaluation methods with diverse data sources. Vahdati et al. (2021)\cite{Vahdati2021QualityAssessmentScientificEvents} proposed a new scientific event quality evaluation system, designed a comprehensive method covering event-related, personnel-related, and publication-related metrics to solve the problem that traditional evaluation methods fail to fully assess the quality of scientific events.

Our study develops a three-dimensional research evaluation framework grounded in Impact and Dissemination, Collaboration Characteristics and Author Characteristics. By leveraging classical bibliometric methods and innovatively designing in response to research evaluation requirements, we propose 12 quantitative metrics to more accurately reflect the multi-dimensional nature of academic outputs.

\subsection{Bibliometrics on Comparison among Research Fields}

Through the comparison across different disciplines and the comparison of computer subfields, we can obtain a comprehensive perspective, thereby gaining in-depth insights into the research dynamics and academic influence of various disciplines.

\subsubsection{Comparison of Different Disciplines}

Research differences across disciplines are prominent in various aspects. Kim et al. (2016)\cite{Kim2016InfoVisDataVis} compared the knowledge landscape, development trends, and research topics of information visualization and data visualization over the past 15 years (2000–2014). While both focus on data visualization, information visualization primarily handles abstract data, whereas data visualization emphasizes presenting numerical or statistical data. Bordons et al. (2023)\cite{Bordons2023OpenAccessFundingLeadership} explored the relationship between open access publishing, research funding, international cooperation, and scientific leadership. Results indicated significant differences in OA practices across fields (e.g., biology \& biomedicine, materials science, social sciences), and external funding plays a key role in promoting OA publishing.Fathalla et al. (2020)\cite{Fathalla2020ScholarlyEventCharacteristics} analyzed the characteristics of academic events in four fields (computer science, physics, engineering, and mathematics) from the perspective of academic event characteristicsTheir study revealed that some academic events (such as CVPR and ICSE) have significant influence, and traditional academic events show unbalanced characteristics in terms of geographical distribution, publishing influence and participation.

\subsubsection{Comparison of Computer Subfields}

Subfields within computer science exhibit distinct differences in terms of knowledge structure, rate of development, and regional distribution. For instance, subfields of this discipline exhibit complex dynamic changes in citation relationships, self-citation ratios, and the number of cited regions. By analyzing 15 years of citation data, Zhu et al. (2015)\cite{Zhu2015DynamicSubfieldAnalysis} revealed the interdependence among different subfields and found that knowledge diffusion in these subfields demonstrates distinct dynamic trends. Additionally, over the past 60 years, the number of authors of academic papers in computer science has shown a rapid growth trend in emerging fields including bioinformatics, multimedia, and computer graphics. In contrast, traditional fields such as computer theory and operations research have exhibited relatively modest growth (Fernandes \& Monteiro, 2017)\cite{Fernandes2017AuthorsCS}. Subfields of computer science also exhibit disparities in regional distribution, and such differences affect the research dynamics and academic output of each subfield to a certain extent (Arruda et al., 2009)\cite{Arruda2009BrazilianCSGenderRegional}.

\vspace{\baselineskip}

\section{Data Collection}
\subsection{Data Processing}

\begin{figure}[H]
    \centering
    \includegraphics[width=\textwidth]{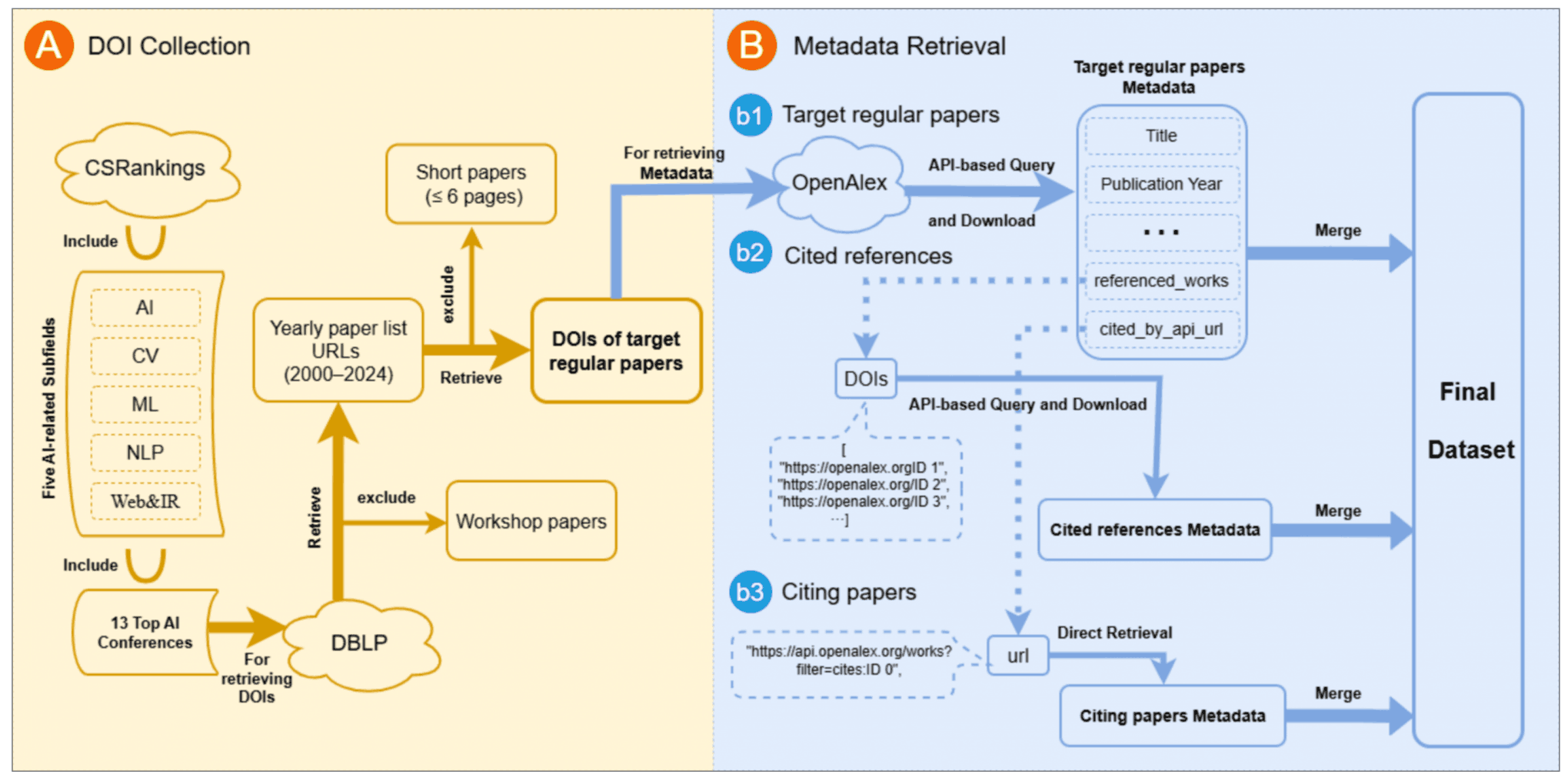}
    \caption{Pipeline of Data Collection}
    \label{fig:data_collection_pipeline}
\end{figure}

Since 2000, with the continued growth of computing power and the rapid accumulation of Internet data, artificial intelligence research has gradually entered a critical stage of transition from theoretical exploration to large-scale application. Therefore, this study sets 2000–2024 as the time window of analysis.Given that the field of computer science has long placed great emphasis on leading international conferences, and that prior research suggests that top conferences are comparable to high-quality journals in terms of influence and academic visibility, this study therefore selects papers published in top artificial intelligence conferences as its research objects (Freyne et al., 2010)\cite{Freyne2010}.It should be noted that traditional concerns regarding the data completeness and evaluative comparability of conference papers have been substantially alleviated with the improvement of relevant scholarly data infrastructures.With the transparent selection of top conferences by CSRankings, the high-quality bibliographic coverage of computer science conference papers by DBLP, and the open metadata and citation links provided by OpenAlex, conference papers are now able to provide a reliable data foundation for bibliometric research.Given that this study focuses on the diffusion and influence of artificial intelligence knowledge, and that citation relations are an important carrier of knowledge diffusion, in addition to basic metadata such as titles, keywords, abstracts, authors, and publication dates, it is also necessary to collect the citing documents, references, and related metadata of the target literature.Since no single database is currently able to export all of the above information completely in one step, this study adopts a two-stage data collection strategy: first, it identifies the target top conferences through CSRankings and obtains the DOIs of the corresponding papers from DBLP; second, using the DOIs as identifiers, it retrieves the metadata and citation information of the target literature from OpenAlex, and further obtains the citing documents, references, and their metadata on this basis.Figure \ref{fig:data_collection_pipeline} presents the specific procedure, which is described in detail below.

The first step is the selection of target literature and the collection of DOIs.This study references the authoritative classification of the artificial intelligence field by CSRankings\footnote{https://csrankings.org}. CSRankings evaluates the academic performance of institutions and researchers based on the number of papers published in top-tier conferences; the conferences selected by this platform are widely recognized by experts in computer science and are commonly used for international research assessment and academic statistical analysis. As shown in Table \ref{tab:ai_subfield_papers}, CSRankings divides the artificial intelligence field into five subfields: Artificial Intelligence (AI), Computer Vision (CV), Machine Learning (ML), Natural Language Processing (NLP), and The Web \& Information Retrieval (Web\&IR), encompassing 13 top-tier conferences with broad international influence (e.g., AAAI, IJCAI, CVPR). To avoid cumbersome expressions in the following sections, this paper uses AI, CV, ML, NLP, and Web\&IR as abbreviations for the five subfields mentioned above. It should be specially noted that this paper strictly distinguishes between the overall field of "artificial intelligence" and the AI subfield within it: when referring to the entire field, the term artificial intelligence is used uniformly; when referring to the Artificial Intelligence subfield among the five subfields, it is abbreviated as AI, so as to avoid conceptual confusion.This paper takes the papers published in these 13 conferences as the research objects.

Since conference papers lack a unified indexing system, we used the DBLP database to obtain DOIs (unique identifiers for academic literature) of papers from each conference annually. DBLP\footnote{https://dblp.org} is a renowned indexing platform for computer science publications, recognized for its high data accuracy and timeliness, and is widely utilized in research evaluation and academic analysis. As depicted in Figure \ref{fig:data_collection_pipeline}: First, we crawled URLs of paper lists for each conference from 2000 to mid-2024, filtering out workshop papers by excluding links containing the term "Workshop". Subsequently, DOIs were extracted from the retained URLs of main conference proceedings, and further screening was conducted to exclude "Short Papers" and "Posters", retaining only full-length papers exceeding 6 pages. Ultimately, a set of 106,622 high-quality conference paper DOIs was obtained, with the number of papers per subfield presented in Table \ref{tab:ai_subfield_papers}.

\begin{table}[h]
\caption{Number of Papers Based on Representative Conferences}\label{tab:ai_subfield_papers}%
\begin{tabular}{@{}llr@{}}
\toprule
Artificial Intelligence Subfields (Abbr.) & Conferences & Papers \\
\midrule
Artificial intelligence (AI) & AAAI, IJCAI & 24248 \\
Computer vision (CV) & CVPR, ECCV, ICCV & 28934 \\
Machine learning (ML) & ICLR, ICML, NeurIPS & 23135 \\
Natural language processing (NLP) & ACL, EMNLP, NAACL & 7280 \\
The Web \& information retrieval (Web\&IR) & SIGIR, WWW & 23025 \\
\bottomrule
\end{tabular}
\end{table}

The second step is to obtain bibliographic metadata based on the DOIs.This study leverages OpenAlex to accurately locate and retrieve comprehensive metadata for the target papers. OpenAlex\footnote{\url{https://openalex.org/}} is an emerging open citation database characterized by rich fields, standardized structures, and timely data updates (Culbert et al., 2025)\cite{culbert2025reference}, with support for efficient retrieval and automated collection via Application Programming Interfaces (APIs). Following the acquisition of high-quality DOIs, we batch-downloaded metadata of target papers from OpenAlex using its open API. Within the metadata of target papers, the $referenced\_works$ field provides a set of DOIs for reference papers, while the $cited\_by\_api\_url$ field offers URLs for all citing papers. Using the same method, metadata for reference papers and citing papers was further collected. Collectively, target papers, reference papers, and citing papers form a complete dataset, enabling subsequent interdisciplinary analysis and knowledge diffusion modeling.

\subsection{Data Overview}

\begin{figure}[H]
    \centering
    \includegraphics[width=\textwidth]{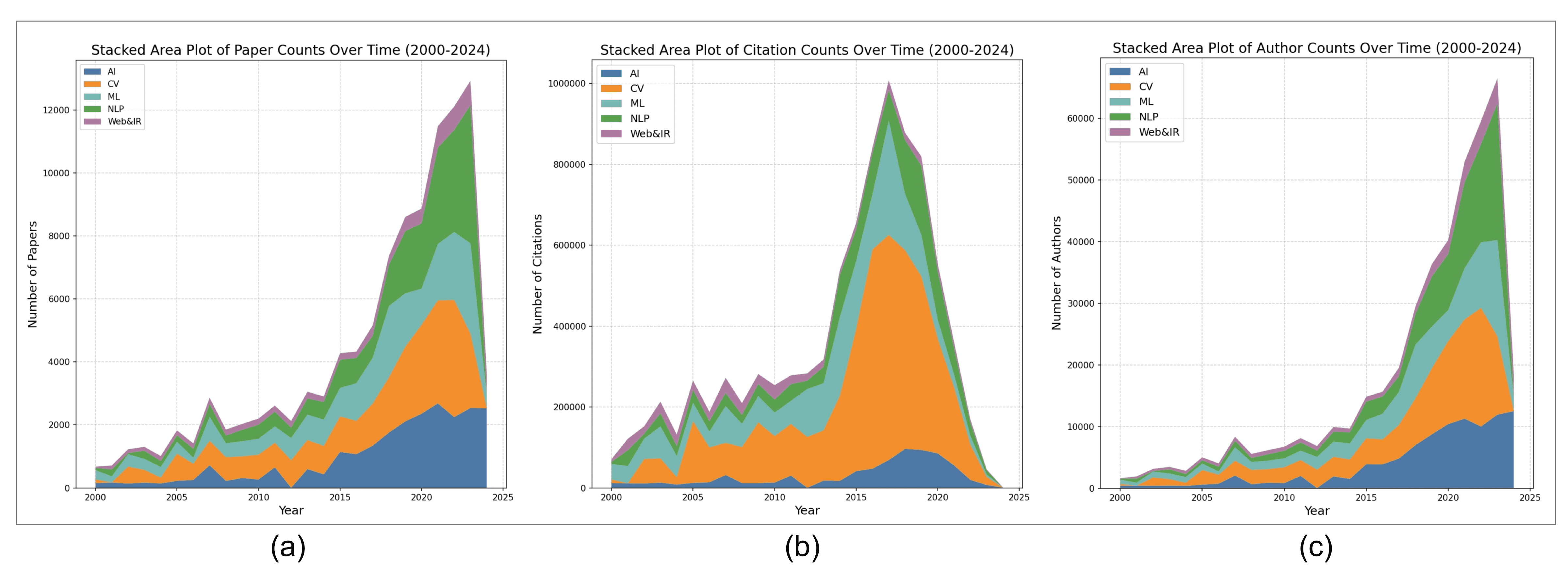}
    \caption{Scale Changes of Each Subfield}
    \label{fig:science}
\end{figure}

To establish a preliminary understanding of the collected data and lay a consistent foundation for subsequent analyses, this section presents an overview statistical analysis of literature data from 13 top artificial intelligence conferences in the artificial intelligence field between 2000 and 2024. As illustrated in Figure \ref{fig:science}, the analysis primarily includes trends in annual paper publication volume, changes in the number of authors, temporal evolution of citation counts, and structural characteristics of authors' affiliated countries or regions.

By demonstrating temporal trends in publication volume, citation counts, and author numbers across the five artificial intelligence subfields, this study intuitively reveals the evolutionary trajectory of the artificial intelligence field. These preliminary analyses and visualizations construct a foundational cognitive framework, providing essential conditions for subsequent in-depth horizontal and vertical comparative analyses, as well as a clear perspective for understanding the macro developmental trends of the field.

In Figure \ref{fig:science}(a), we plot the temporal evolution of publication volume across the five subfields, with the total annual number of papers representing the sum of papers across all subfields. As shown, the number of papers published in all five subfields exhibits an overall steady growth trend over the entire period. However, a decline was observed in the artificial intelligence subfield in 2012, attributed to the non-holding of relevant conferences that year, leading to data collection bias and an underestimation of the number of papers for that year. After 2015, the growth rate of paper publications accelerated significantly across all subfields—a phenomenon likely driven by the rapid advancement of artificial intelligence technologies such as deep learning, the exponential growth of big data, and the proliferation of open-access datasets, which collectively boosted the output of research findings and the number of published papers. Even amid the global COVID-19 pandemic in 2020, the number of artificial intelligence-related papers maintained a growth momentum. By 2023, with the further maturation of artificial intelligence technologies and their widespread adoption across diverse application scenarios, the number of papers in each subfield reached a new peak.

Figure \ref{fig:science}(b) depicts the temporal evolution of citation counts across the five subfields. As indicated, citation counts for all subfields show a steady upward trend over the entire period. The slow growth in citations during the early phase may be attributed to the fact that artificial intelligence, as a research field, had not yet gained widespread recognition in academia, resulting in relatively limited citations of research findings. Citation growth began to accelerate after 2013 and became particularly prominent post-2015, driven by the successful application of relevant technologies, which significantly advanced artificial intelligence research and attracted substantial attention and citations. Following a peak in 2020, citation counts began to decline—potentially due to lockdowns and social distancing measures during the pandemic, which hindered academic exchange, collaboration, and the dissemination/citation of research findings. Alternatively, this decline may reflect the inherent citation lag effect: due to constraints on the research timeframe, papers published after 2020 have not had sufficient time to be fully cited by the academic community.

Figure \ref{fig:science}(c) illustrates the temporal evolution of the number of authors across the five subfields. As shown, the number of authors exhibits a marked overall growth trend. Starting from 2015, the number of artificial intelligence researchers began to increase—a trend likely fueled by the widespread application of artificial intelligence technologies across various industries, which heightened industrial demand for artificial intelligence research and promoted increased academic investment in the field. Concurrently, significant increases in government policy support and funding for artificial intelligence research provided additional resources for research activities, encouraging more researchers to engage in artificial intelligence-related work. By 2023, the number of authors in the artificial intelligence field reached its peak. This phenomenon may be attributed to the maturation of artificial intelligence technologies and their extensive application across multiple domains, which attracted a growing number of researchers to the field. The continuous emergence of research findings during this period further expanded the author base of artificial intelligence research, ultimately reaching a historical high in 2023.

\subsection{Science Maps}

Science maps are tools for visualizing the structure and dynamic relationships of academic knowledge. They enable researchers to identify and track changes and developments in research hotspots within academic fields, thereby providing profound insights into the evolution of academic activities (Petrovich, 2021)\cite{Petrovich2021ScienceMapping}.

\begin{figure}[H]
    \centering
    \includegraphics[width=\textwidth]{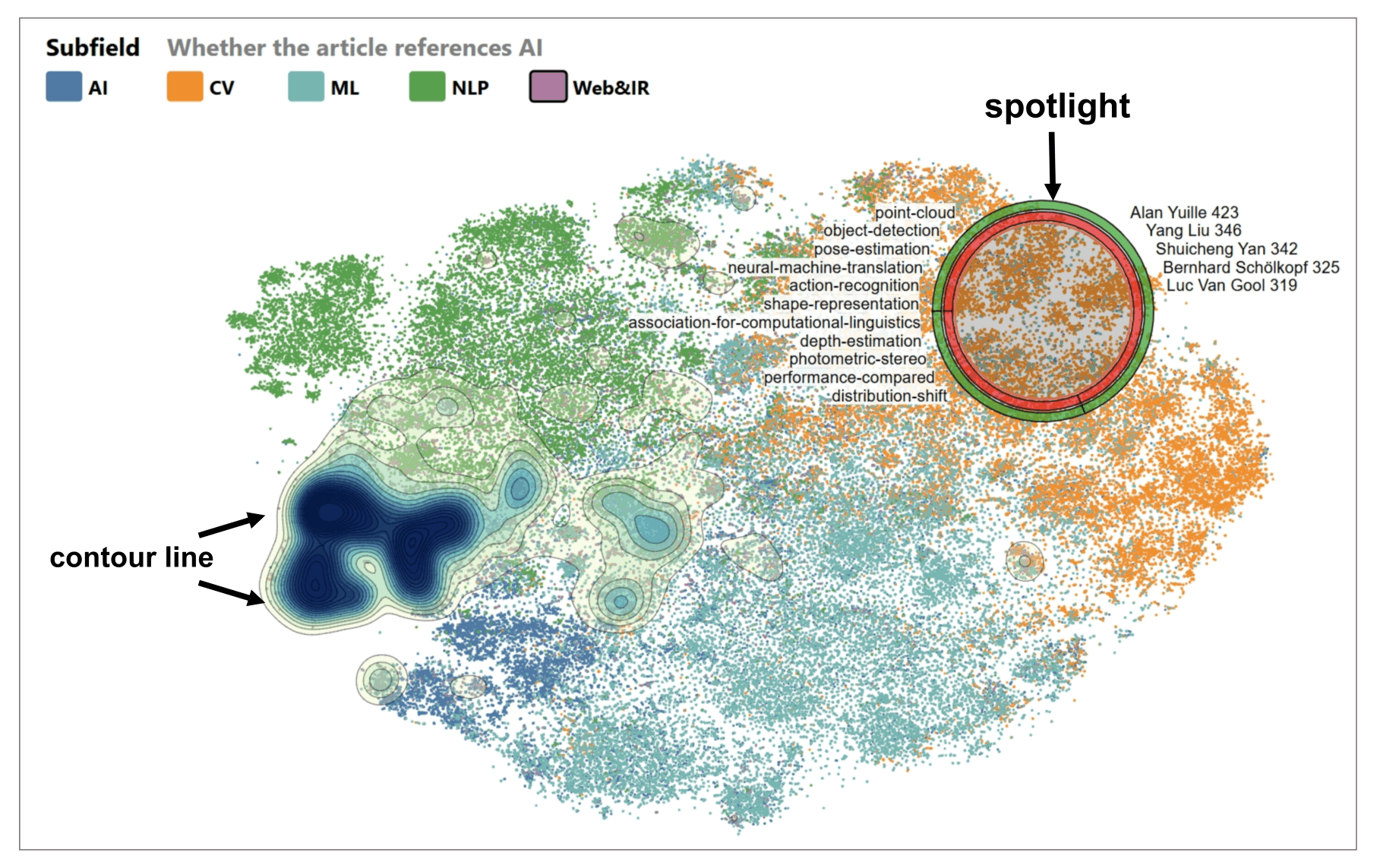}
    \caption{Science Mapping of Research Topics}
    \label{fig:science_map}
\end{figure}

We first employed natural language processing techniques such as SentenceTransformer to convert textual data from the dataset into high-dimensional vectors, where each vector represents the theme and core content of a paper. We then reduced the dimensionality of the data to 30 dimensions using Principal Component Analysis, followed by further dimensionality reduction to 2 dimensions via t-distributed Stochastic Neighbor Embedding (t-SNE). The final result is presented as a 2D scatter plot, where each point represents a paper, and the position of each point is determined by the similarity between papers. After dimensionality reduction, papers were clustered based on semantic features, grouping similar papers into clusters, with five themes distinguished by different colors. Additionally, density contour plots generated by Kernel Density Estimation were integrated to visualize paper aggregation regions, where the depth of color indicates the density of papers (darker colors represent higher paper concentration). The dark-shaded region on the left side of Figure~\ref{fig:science_map} denotes the high-density distribution of Web\&IR studies, providing visual evidence of their spatial concentration within the semantic map.Furthermore, users can adjust the view size through zoom and pan operations, and drag a selection circle to highlight regions of interest in the map, thereby accessing detailed information about papers within the selected region (e.g., keywords, lead authors, number of papers published by authors).

The science map reveals that the AI subfield is relatively scattered; nonetheless, it exhibits significant overlap with other subfields, indicating that AI encompasses multiple research themes, plays a foundational role across various domains, and possesses distinct interdisciplinary characteristics. The CV subfield is mainly distributed on the right side of the figure and forms a large orange cluster, indicating that computer vision research has developed a relatively mature and active thematic group.The ML subfield is mainly located in the central and lower regions of the figure and shows considerable overlap with the AI and CV subfields.This distribution feature shows that machine learning plays an important methodological supporting role in the artificial intelligence research system. In contrast, the NLP and Web\&IR subfields have weaker connections with other domains, presenting relatively independent research directions. Specifically, NLP focuses primarily on language processing and text analysis (Khurana et al., 2017)\cite{Khurana2017NLPStateOfTheArt}, while Web\&IR concentrates on search engines, data retrieval, and Web-related artificial intelligence applications (Kuyoro et al., 2012)\cite{Kuyoro2012InformationRetrievalOverview}.

\begin{figure}[H]
    \centering
    \includegraphics[width=\textwidth]{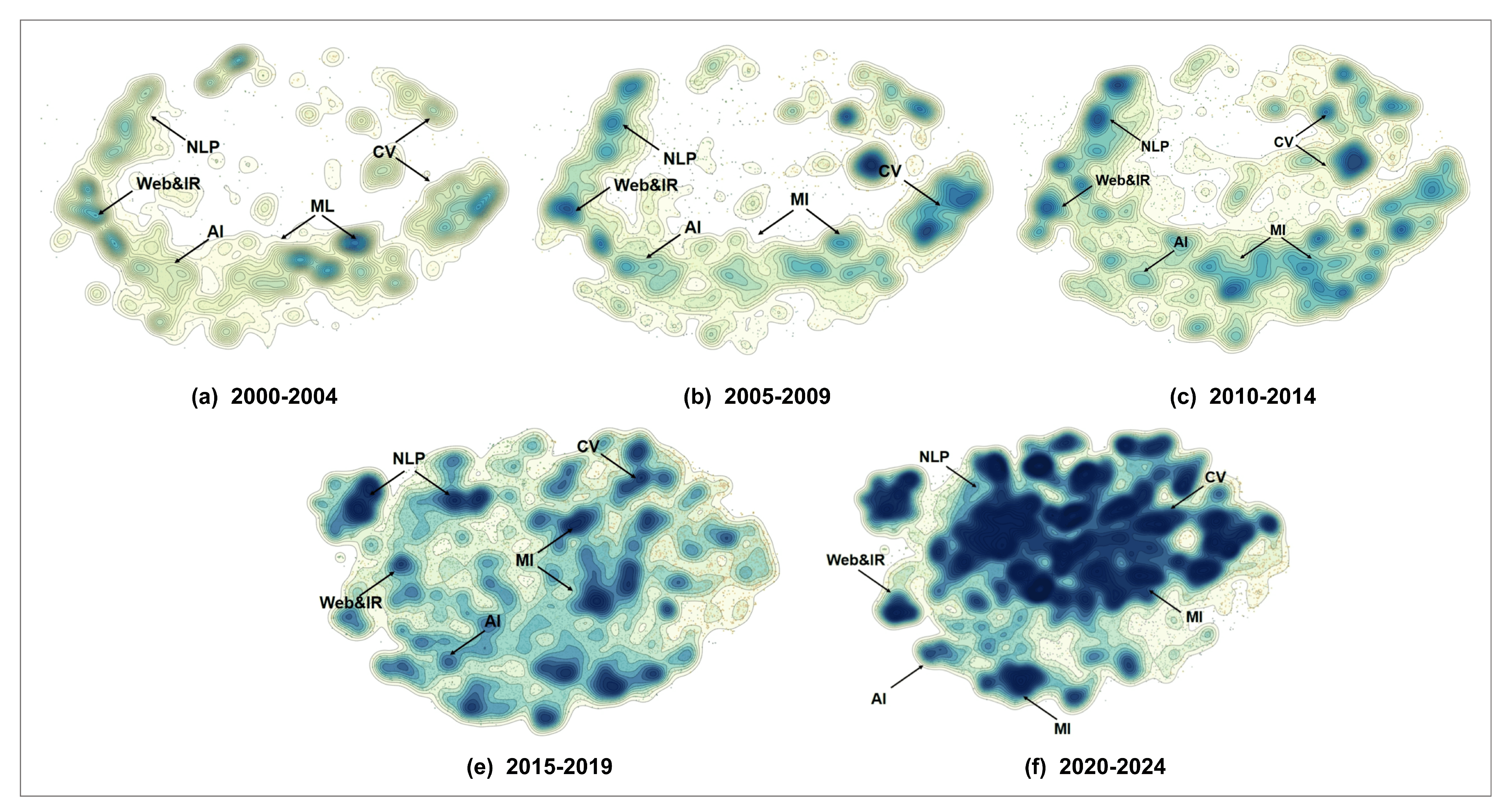}
    \caption{Evolution of Research Hotspots in Artificial Intelligence Subfields}
    \label{fig:placeholder}
\end{figure}

Figure \ref{fig:placeholder} shows that the evolution of artificial intelligence subfields shows a trend from local exploration to deep integration.In 2000–2009, the distribution of subfields was discrete with clear boundaries, reflecting independent early research.Since 2010, the heat centers converged to the center, and ML expanded, becoming the basis connecting fields.By 2020–2024, the map formed a whole structure with a high-density core, boundaries blurred, showing full integration.

The evolution of the spatial structure of the science map reveals the changes and integration of disciplinary attributes among subfields.The general AI field is always at the center, and its connection density with CV and NLP increased significantly over time.Reflecting its functional shift from a "technological source" to a "cross-task integration center."Meanwhile, the CV region in the upper right and the NLP region in the upper left connected together in the later stage.This overlapping of spatial structure shows the explosion of multimodal research and the unification of vision and language models\cite{Du2022VisionLanguageSurvey}.

The densification of the citation network reflects the systematic unification of artificial intelligence research paradigms.The color change from light green to dark blue represents growth in literature volume and increased citation closeness.This high-density closed structure shows that artificial intelligence research has evolved from scattered early topics to a collaborative stage based on large frameworks.Forming a highly integrated and jointly driven system.


\section{Methods}\label{sec4}

This study analyzes the horizontal differences among artificial intelligence subfields across multiple dimensions and their longitudinal evolutionary trends over time through a constructed indicator system. As quantitative tools for academic evaluation, indicators comprehensively and systematically reflect the core characteristics of each subfield\cite{franceschet2009}. This chapter first outlines the development of the two-level indicator system, followed by detailed descriptions of each metric's purpose, data source, selection rationale, measurement approach, and corresponding formula.

\subsection{Construction of the Indicator System}\label{sec4_1}
To comprehensively quantify the core characteristics of each subfield, we first conducted needs-assessment interviews with experts in the field of computer science to identify three dimensions. We then collected potential indicators under each dimension through a literature review, and screened these indicators by considering factors such as space limitations and data availability. Finally, we constructed a two-level indicator system consisting of three dimensions and twelve indicators.The detailed process of the construction is described as follows.

\textbf{Needs Analysis:} Four computer science experts (three specializing in artificial intelligence) were consulted through a 40-minute focus group discussion. We clarified the research objectives and solicited recommendations on dimensions and indicators. Three core dimensions were finalized: Influence and Dissemination, Collaboration Characteristics and Author Characteristics. Expert priorities included: Expanding beyond traditional metrics (e.g., h-index) to include citation distribution, growth rate, and cross-domain dissemination for Influence and Dissemination; Integrating collaboration scale, industry-academia partnerships, and relationship sustainability for Collaboration Characteristics; Focusing on research output, topic migration, and institutional mobility for Author Characteristics.

\textbf{Preliminary Collection:} We systematically reviewed the indicators used in the existing literature to characterize disciplines and subfields.A comprehensive analysis shows that existing studies have developed relatively mature measurement frameworks across three dimensions—Impact and Dissemination, Collaboration Characteristics, and Author Characteristics—and have accumulated a rich set of operational indicators.Specifically, the Impact and Dissemination dimension is used to measure the academic impact and dissemination breadth of research outputs, with indicators including citation counts, h-index, g-index, impact factor, acceptance rate, and publication venues.The Collaboration Characteristics dimension reflects the scale, scope, and structure of collaboration in the knowledge production process, with indicators including Authors per Paper, International Collaboration, and Collaboration Frequency.The Author Characteristics dimension is used to describe the academic productivity and cross-domain mobility of research actors, with indicators including Author Productivity and Topic Mobility of Authors.

\textbf{Screening:} We considered three factors in selecting the indicators.First, we prioritized indicators with clear definitions, ease of interpretation, and a certain degree of generalizability and popularity, while excluding methods that are only applicable in specific contexts or rely on complex assumptions.For example, in the Collaboration Characteristics dimension, we selected Authors per Paper and International Collaboration rather than complex constructed indicators that require full-network modeling or composite weighting.Second, we evaluated the data availability and practical operability of each indicator to ensure that the selected indicators could be effectively measured on the basis of the available data.For example, although the "cross-disciplinary collaboration coefficient" is useful for evaluating Collaboration Characteristics, it requires additional collection of each author’s disciplinary background information and involves a substantial workload; therefore, it was not adopted.Finally, due to space limitations, this study retains only 2 to 5 core indicators for each dimension.For some indicators that show limited differentiation across subfields or contribute little to the study’s conclusions, such as Citing Ratios, they were not included in the final indicator system.

\begin{figure}[H]
    \centering
    \includegraphics[width=0.8\textwidth]{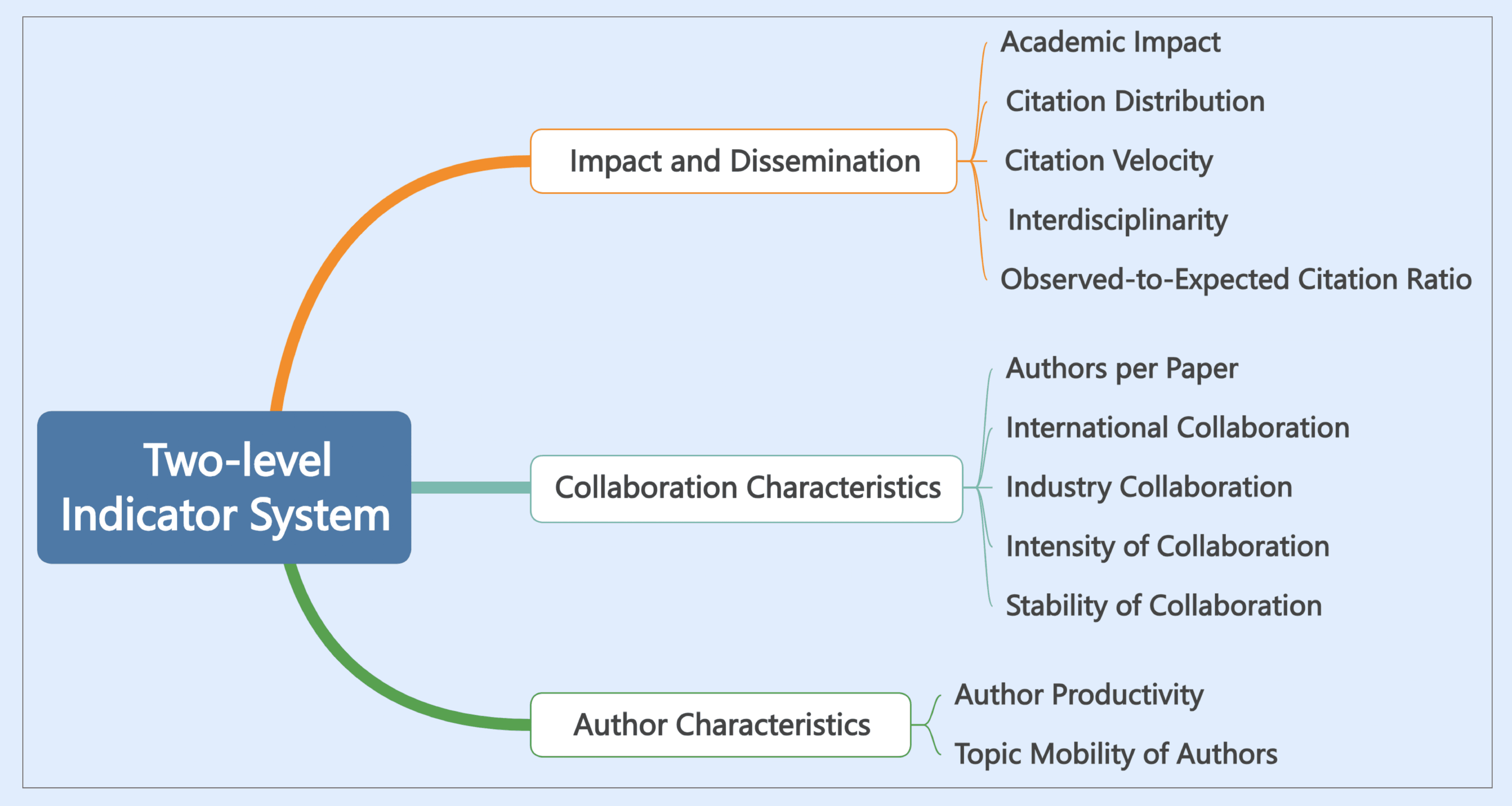}
    \caption{Two-level Indicator System}
    \label{fig:placeholder1}
\end{figure}

\subsection{Metrics}\label{sec4_2}
\subsubsection{Influence and Dissemination}\label{sec4_2_1}
\textbf{1. Academic Impact}

The H-index is a composite indicator commonly used to measure academic output and academic impact, taking into account both the quantity of publications and their quality (as reflected by citations).We measured the H-index of the 13 top artificial intelligence conferences in our dataset, defined as having at least $h$ papers each cited more than $h$ times.Clearly, a larger $h$ indicates greater influence of the conference.

\textbf{2. Citation Distribution}

Compared with indicators that rely only on total citations or averages, the Citation Distribution better reflects the knowledge dissemination characteristics of a field.McCain (1990)\cite{mccain1990} and Todeschini. (2016)\cite{todeschini2016} point out that citation networks in computer science generally follow a long-tail distribution (i.e., a power-law distribution), where a small number of papers receive a large number of citations, while most papers receive few or no citations.To analyze the citation intensity and concentration of this distribution, we fit a power-law distribution to the Citation Distribution of a field.
\begin{equation}
y = C x^{-\alpha} \label{eq:powerlaw}
\end{equation}
where $y$ = citation count of a paper, $x$ = rank by citations, $C$ = scale constant (citations of the top-ranked paper), and $\alpha$ = power-law exponent (governing distribution concentration). Larger $\alpha$ indicates more concentrated citations. $C$ and $\alpha$ are estimated via log-transformed linear regression, with goodness-of-fit evaluated by $R^2$:
\begin{equation}
R^2 = 1 - \frac{\sum_{i=1}^{n} (y_i - \hat{y}_i)^2}{\sum_{i=1}^{n} (y_i - \bar{y})^2}
\end{equation}
( $y_{i}$: actual value, $\hat{y}_{i}$: predicted value, $\bar{y}$: mean value.A higher $R^2$ (close to 1) indicates a good fit of the power-law model).

\textbf{3. Citation Velocity}

Citation Velocity is used to evaluate the efficiency of knowledge dissemination within a field and reflects its dynamism.Tahamtan and Bornmann (2019)\cite{tahamtan2019} point out that there are systematic differences in citation growth rates across topics, document types, and collaboration patterns, and that rapid citation is often associated with research hotspots, high visibility, and interdisciplinary content.In general, Citation Velocity can be measured in two ways: (i) the time required to obtain the first citation, and (ii) the time required to accumulate $n$ citations (in days).Due to the rapid development of the artificial intelligence field and the use of the preprint platform arXiv, where papers often receive citations before formal publication, the time to first citation is often substantially shortened. Therefore, we adopt the second approach by counting citations within a fixed time window after publication and construct two complementary indicators.
\begin{enumerate}[label=(\arabic*), nosep]
    \item Overall Citation Velocity: We calculate the number of days required for papers in the field with at least 25 citations to reach 25 citations. The velocity is defined as 25 divided by this number of days. The threshold of 25 is chosen because it is close to the median level of the citation distribution.
    \item High-impact Citation Velocity (as highly cited papers represent the most dynamic part of a field): We calculate the number of days required for the top 20\% most cited papers to reach 100 citations. The velocity is defined as 100 divided by this number of days.
\end{enumerate}

\textbf{4. Interdisciplinarity}

Interdisciplinarity measures the diversity and unevenness of citation sources within a field.As a discipline with broad applications, artificial intelligence exhibits significant interdisciplinarity\cite{porter2007}\cite{kusters2020}.We use Shannon entropy to characterize the breadth and balance of this distribution. Based on the subject classification of papers in the dataset, we categorize both citing and cited references according to $subfield\ level$ discipline names ($display\ name$), and calculate the proportion of each category to form a proportion vector $p = (p_1, \ldots, p_K)$.We compute the Shannon entropy of the proportion vector for each paper, defined as follows:
\begin{equation}
H(p) = -\sum_{k=1}^{K} p_k \log p_k
\end{equation}
A higher value of $H(p)$ indicates a more uniform distribution of citation sources and thus a higher degree of interdisciplinarity.

\textbf{5. Observed-to-Expected Citation Ratio }

Observed-to-Expected Citation Ratio is used to measure whether directional citation relationships between subfields exceed the expected level determined by their size, and to reveal the intensity of knowledge flows between subfields.Following Sinatra et al. (2015)\cite{sinatra2015}, we adopt a random independent baseline (null model) as a reference, decomposing each citation into two "half-links", and allowing these half-links to be randomly matched while preserving the total out-degree and in-degree of each subfield, thereby defining the expected number of cross-domain citations.That is, in the absence of any cross-domain preference, the expected number of citations from subfield \textit{i} to subfield \textit{j} is:
\begin{equation}
E_{ij} = \frac{N_i}{N} \times \frac{N_j}{N} \times C
\end{equation}
( $N_{i}$: papers in $i$, $N$: total papers, $C$: total citations). Standardized intensity:
\begin{equation}
R_{ij} = \frac{C_{ij}}{E_{ij}}
\end{equation}
 $R_{ij}>1$ indicates stronger-than-expected citation flow from $i$ to $j$.

\subsubsection{Collaboration Characteristics}\label{sec4_2_2}

\textbf{1. Authors per Paper }

Average number of authors per paper, reflecting team size and collaboration prevalence\cite{guan2004}:
\begin{equation}
CI = \frac{\sum_{j=1}^{A} j f_j}{N}
\end{equation}
 $f_{j}$= number of papers with $j$ authors, $A$=maximum number of authors, $N$=total papers. Higher $CI$ indicates more widespread collaboration.

\textbf{2. International Collaboration}

International Collaboration plays an important role in the global dissemination of scientific outputs\cite{gomezcaridad2007}.Following Rousseau \& Zhang (2021)\cite{rousseau2021}, we adopt a pair-based co-authorship perspective and characterize the International Collaboration of a paper by the proportion of cross-country author pairs among all author pairs.For a paper with $n$ authors, the total number of author pairs is $\binom{n}{2}$. If $m$ pairs consist of authors from different countries, then the International Collaboration is defined as $m / \binom{n}{2}$.The closer this ratio is to 1, the higher the level of International Collaboration within the author team.

\textbf{3. Industry Collaboration}

Industry Collaboration is defined as the proportion of articles that involve collaboration with industry among all articles, and is regarded as a key dimension for measuring the strength of academia–industry linkages within national, institutional, and field-level research ecosystems (Garousi, 2015)\cite{garousi2015}.We use the $"raw\_affiliation\_string"$ field under authorships in the dataset to determine whether each author's affiliation contains industry-related keywords, such as $"company", "tech", "inc", "ltd", "corp", "enterprise"$, or common company names such as $"Huawei", "IBM", "Tencent", "Google", and "Microsoft"$.We collected a total of 171 keywords.If at least one author meets the industry criteria, the article is labeled as involving Industry Collaboration.A higher proportion indicates a closer connection between the field and industry.

\textbf{4. Intensity of Collaboration}

Stability of Collaboration measures the closeness of collaboration among researchers within a field.When the nodes in a field’s collaboration network are broadly connected, the field exhibits strong collaboration.Based on this idea, we construct a weighted collaboration network of the top 1000 most productive authors in the field, where edge weights represent the number of collaborations.Following the weighted clustering coefficient defined by Opsahl and Panzarasa (2009)\cite{opsahl2009}, and extending the traditional clustering coefficient that only considers the presence of edges, we use a function $g(x, y)$ to combine the weights of two edges centered on a node, thereby computing the weighted closure of triangles.The local weighted clustering coefficient for node $i$ is defined as follows:
\begin{equation}
{C_i^{(w)}} = \frac{\sum_{j<k} g(\hat{w}_{ij}, \hat{w}_{ik}) a_{ij} a_{ik} a_{jk}}{\sum_{j<k} g(\hat{w}_{ij}, \hat{w}_{ik}) a_{ij} a_{ik}}
\end{equation}
Here, $a_{ij}$ indicates whether an edge exists between nodes $i$ and $j$ (1 if present, 0 otherwise), and $w_{ij}$ denotes the normalized edge weight. When the node degree $k_i < 2$ or the denominator equals $0$, $C_i^{(w)}$ is defined as $0$. Common choices for the function $g(x,y)$ include the arithmetic mean and the geometric mean. We adopt the geometric mean $g(x,y) = \sqrt{xy}$ to balance the strengths of the two edges.The global weighted clustering coefficient is defined as the average of the local weighted clustering coefficients across all nodes.A higher weighted clustering coefficient indicates that collaboration relationships within the field are more concentrated in locally closed structures.

\textbf{5. Stability of Collaboration}

Intensity of Collaboration is used to measure whether research collaboration relationships remain stable over a longer period or whether collaborators are changed frequently.Following Bird et al. (2009)\cite{bird2009} and Ibáñez et al. (2012)\cite{ibanez2012}, this study uses set similarity based on the Jaccard coefficient for measurement.We select the top 3,000 authors in the field by publication volume, construct collaboration networks among them across multiple time periods, examine each author’s set of collaborators, and use the Jaccard coefficient to calculate set similarity between adjacent periods.For any author $i$, let the collaborator sets in two adjacent periods $t$ and $t+1$ be $C_i^t$ and $C_i^{t+1}$, respectively; then the Jaccard similarity coefficient is defined as follows:
\begin{equation}
J_i^t = \frac{\left| C_i^t \cap C_i^{t+1} \right|}{\left| C_i^t \cup C_i^{t+1} \right|}
\end{equation}
Taking the average across all authors yields the overall collaboration stability for that time period:$S\mathrm{=}\frac{1}{N}\sum_{i\mathrm{=}1}^{N}{J_i^t}$ ( $N$ is the total number of authors involved in the two time periods.)A higher value of $S$ indicates more stable collaboration relationships.

\subsubsection{Author Characteristics}\label{sec4_2_3}

\textbf{1. Author Productivity}

Author Productivity is used to measure researchers’ capability and activity in scholarly output.We adopt the classic H-index, which takes into account both the number of publications and the number of citations. A score of $n$ means that an author has published at least $n$ papers, each of which has been cited at least $n$ times.

\textbf{2. Topic Mobility of Authors}

Topic Mobility of Authors is a key indicator for measuring the "openness" of a field, and can effectively reflect the efficiency of knowledge circulation and the innovation potential of the field (Bird et al., 2009)\cite{bird2009}.We select the top 3,000 authors in each field by publication volume, compile their complete publication records, and divide them into several consecutive time periods according to publication year.For author $i$, the distribution of research topics in time period $t$ (defined by the subfield corresponding to the primary topic field of each paper) is represented as a vector:

\begin{equation}
N_i^t = \left( N_{i1}^t, N_{i2}^t, \dots, N_{iK}^t \right)
\end{equation}

Among them, ${{N_{ik}^t}}$ represents the number of papers on the kth topic by author $i$ during time period $t$, and $K$ is the total number of topics. Normalize the vector to obtain the proportion distribution of research topics for author $i$ during this time period:

\begin{equation}
p_{ik}^t = \frac{N_{ik}^t}{\sum_{j=1}^{K} N_{ij}^t}
\end{equation}
We calculated the frequency of changes in the research topic of author $i$ between adjacent time periods $t$ and $t+1$ using $JS$ divergence as follows:

\begin{equation}
f_{i,t} = JS\left(p_{it} \parallel p_{i(t+1)}\right).
\end{equation}

The overall Topic Mobility of Authors for author $i$ is defined as the average of $f_{i,t}$ across multiple time periods.Finally, the mobility of the field is obtained by averaging the overall mobility values of all authors in the field.A higher value indicates more frequent changes in research topics overall.


\section{Result}\label{sec5}

\subsection{Impact and Dissemination}\label{sec5_1}

\subsubsection{Academic Impact}\label{sec5_1_1}

\begin{figure}[H]
    \centering
    \includegraphics[width=\textwidth]{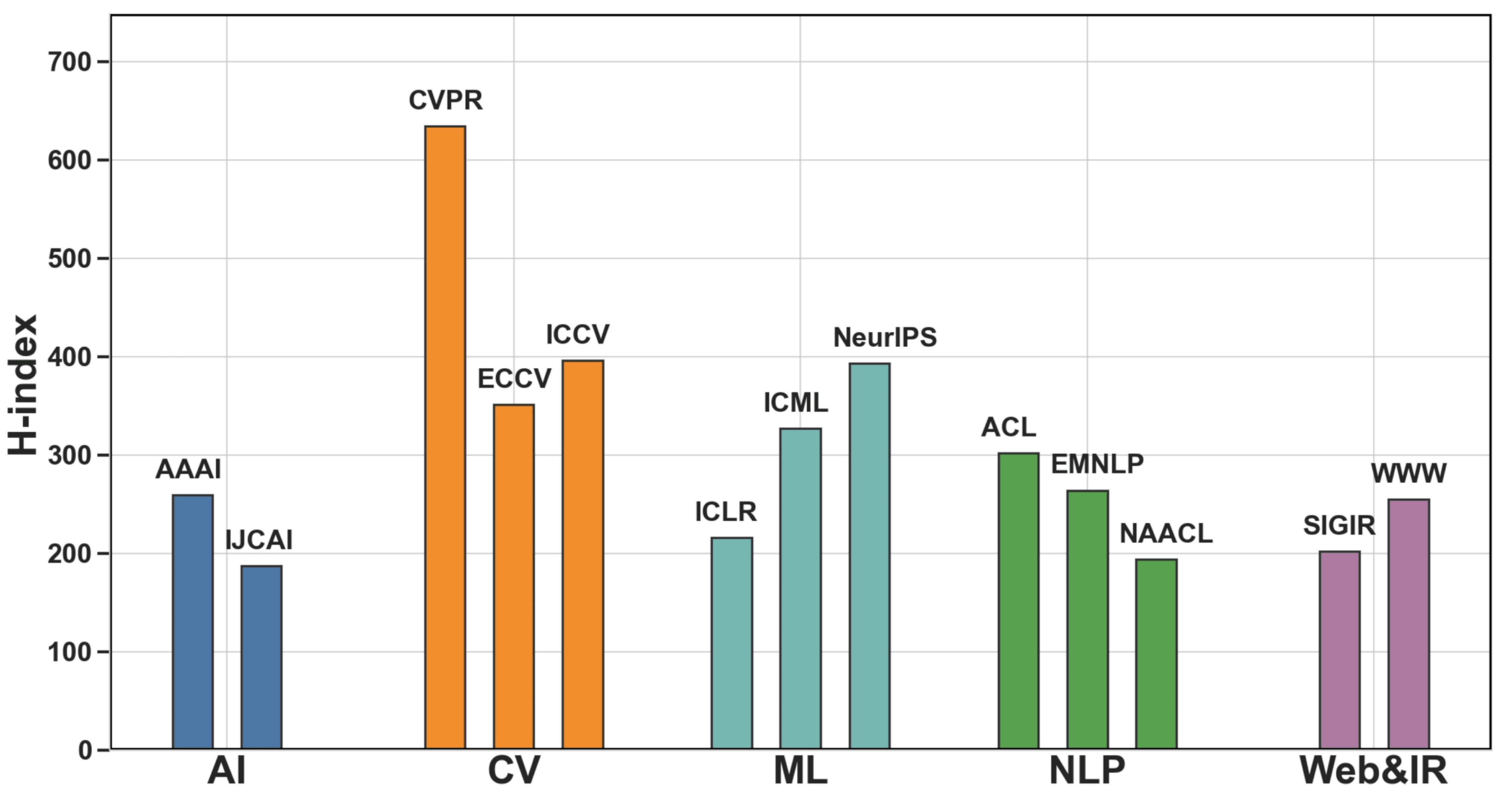}
    \caption{H-index of Conferences of Each Subfield}
    \label{fig:hindex}
\end{figure}

We used the H-index metric described in Section \ref{sec4_2_1} to measure the academic impact of 13 top artificial intelligence conferences. Figure \ref{fig:hindex} shows the H-index for conferences in each subfield; a higher value indicates stronger overall academic impact. Overall, top conferences across different subfields exhibit a distinct hierarchical structure in terms of academic influence. Among them, the CV field has the highest overall H-index, with all three conferences exceeding 350; CVPR, in particular, reaches 634, significantly higher than other conferences. ML field follows, with both ICML and NeurlPS having H-indexes in the 300–400 range. The leading position of top conferences in CV and ML in terms of H-index may be related to their knowledge dissemination mechanisms and standardized evaluation systems centered around these conferences. This structure facilitates the concentrated dissemination of research results and the continuous accumulation of citations. At the same time, methods in these fields possess strong generalizability (e.g., GANs, Transformers) and can be widely applied across multiple tasks, thereby further enhancing their academic influence. In contrast, the H-indexes of top conferences in other subfields—such as IJCAI in AI, ICLR in ML, NAACL in NLP, and SIGIR in Web\&IR—are relatively low, hovering around 200. This may be attributed to factors such as their relatively short history (ICLR was founded in 2013, and NAACL was initially held biennially) and smaller conference scales.

\subsubsection{Citation Distribution}\label{sec5_1_2}

\begin{figure}[H]
    \centering
    \includegraphics[width=\textwidth]{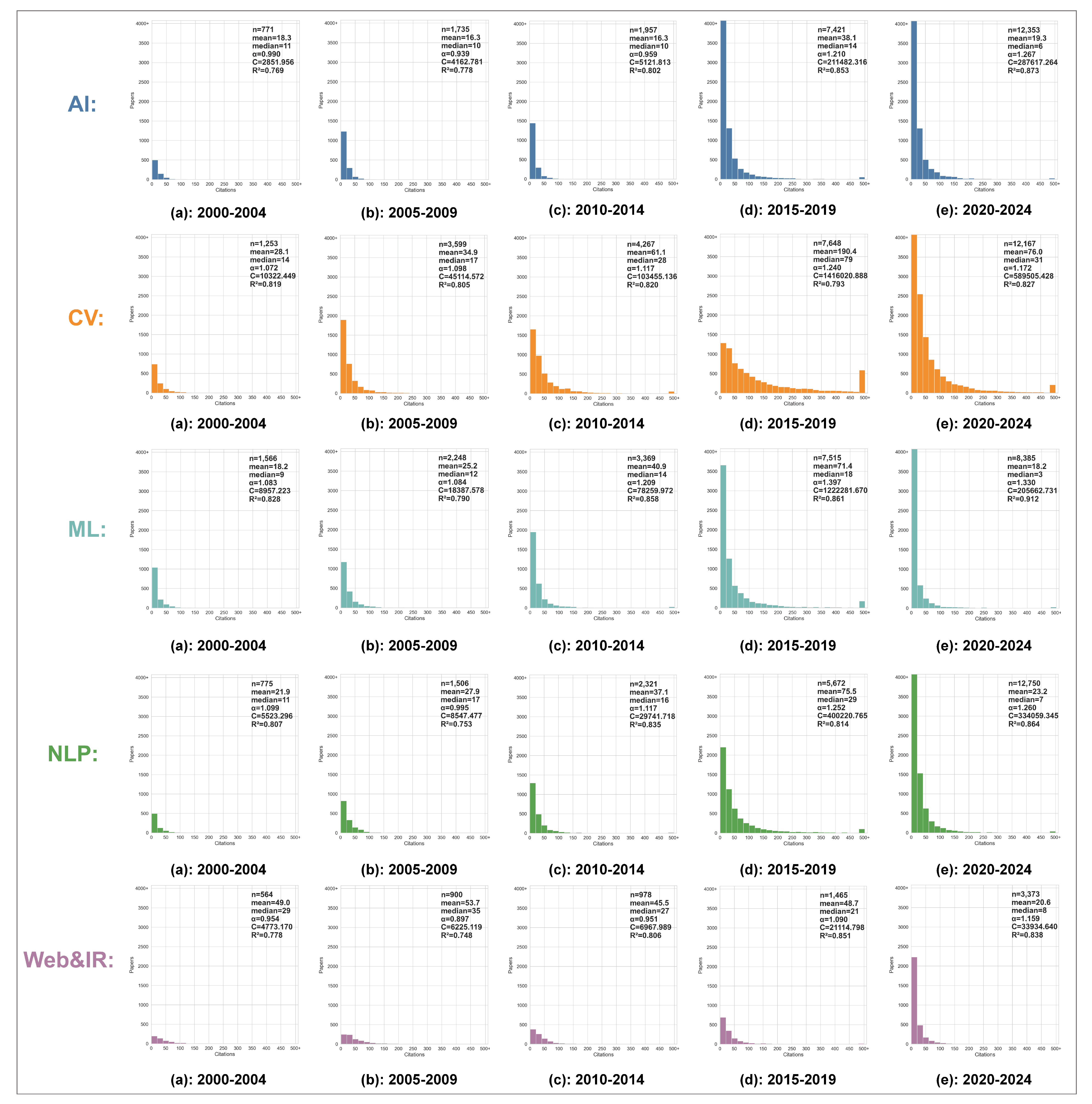}
    \caption{Citation Distribution Histogram}
    \label{fig:citation_powerlaw}
\end{figure}

To analyze the distribution of citation counts, we applied a power-law model to fit the citation counts of papers across each subfield over five time periods. The effectiveness of the model fit is measured by the coefficient of determination $R^2$; a higher value indicates a more pronounced power-law characteristic. Figure \ref{fig:citation_powerlaw} illustrates the citation distributions and parameter values for each subfield. Overall, the $R^2$ values for all subfields are around 0.8, indicating a good fit of the power-law model. Subfields generally exhibit a long-tail distribution, characterized by a small number of highly cited papers and a large number of low-cited papers, reflecting an uneven distribution of academic influence. This phenomenon can be partially attributed to the "Matthew Effect," whereby highly cited papers are more likely to receive subsequent citations, thereby creating a cumulative advantage. From a temporal perspective, the long-tail distribution characteristics in each subfield became more pronounced between 2010 and 2019, with the highest citation count rising from approximately 100 to over 500. This trend may be related to the expansion of research scale, the continuous accumulation of citations, and the emergence of methodological breakthroughs with broad impact (such as the Transformer). Over the past decade, the number of high-impact papers with over 500 citations in each subfield has grown from zero to over 100. Table \ref{tab:top5_cited}  indirectly confirms this phenomenon: the most highly cited papers in each subfield appeared within the past decade, and all have been cited over 1,000 times. It is worth noting that the long-tail distribution characteristic has consistently been less pronounced in Web\&IR, with the highest citation count remaining around 150.

\begin{figure}[H]
    \centering
    \includegraphics[width=\textwidth]{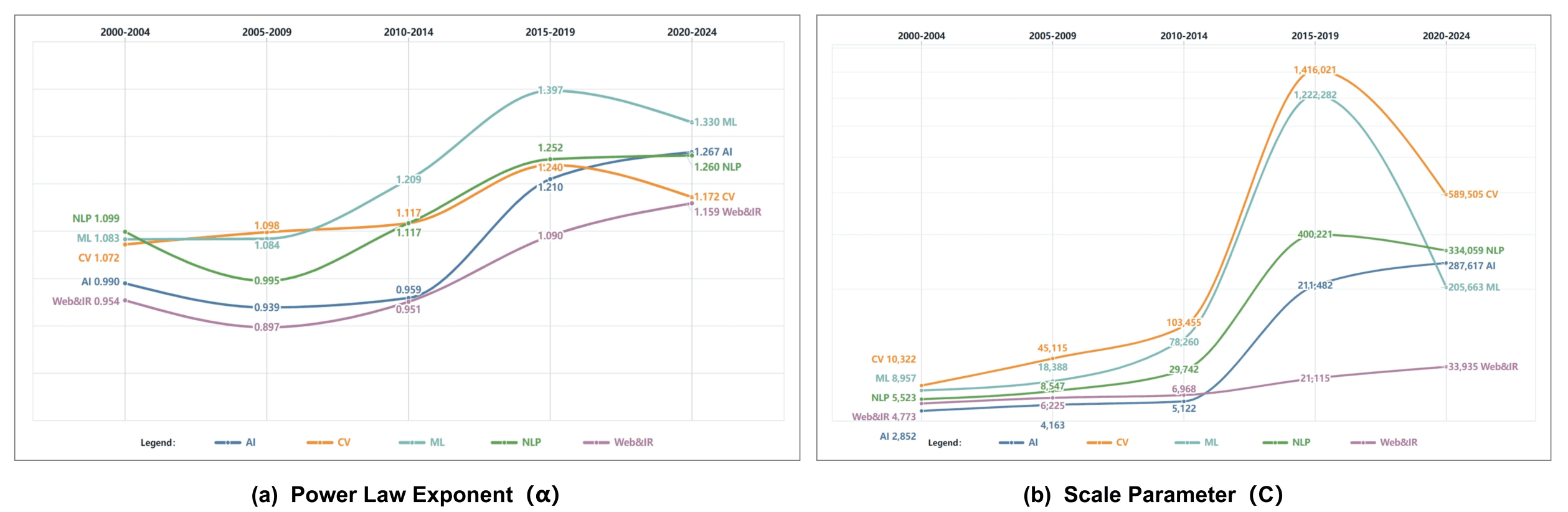}
    \caption{Power Law Exponent and Scale Parameter of Each Subfield}
    \label{fig:powerlaw_params}
\end{figure}

Figure \ref{fig:powerlaw_params} further illustrates the changes in the power-law exponent ($\alpha$) and scale parameter ($C$) of the power-law model across different time periods. A higher value of $\alpha$ indicates a greater disparity between highly cited and low-cited papers, resulting in a more pronounced long-tail effect; a higher value of $C$ indicates that the top influential papers in the field receive a greater number of citations. Overall, the $\alpha$ values across subfields generally follow a "decline followed by rise" trend, typically peaking between 2015 and 2019 (with a maximum of 1,397). Meanwhile, the $C$ values increased significantly after 2010, reaching a peak between 2015 and 2019 (with a maximum of 1,416,021). As shown in Figure (a), ML generally had the highest $\alpha$ values across all periods (averaging 1,200), while Web\&IR consistently had the lowest (averaging around 900). Combined with Figure \ref{fig:citation_powerlaw}, it can be seen that the citation distribution in the Web\&IR field is more balanced, indicating a more dispersed academic influence. As shown in Figure (b), the $C$ values for CV and ML are significantly higher than those of other subfields, averaging 400,000, while the peak values for most other fields do not exceed this level. Data in Table \ref{tab:top5_cited} confirms that the top five most-cited papers in the CV and ML fields dominate the citation landscape (e.g., \textit{"Attention Is All You Need,"} published at NeurlPS, has been cited over 20,000 times, and the most-cited papers in the CV field average 8,000 citations). From a longitudinal perspective, the $C$-values of all subfields rose significantly between 2010 and 2020, with an average increase of 600,000. In particular, the $C$-value in the CV field surged rapidly between 2015 and 2019, with an increase exceeding 1,300,000. In contrast, the Web\&IR field saw the smallest increase in $C$-value, at only 15,000. We speculate that this is because progress in these fields largely occurs within established frameworks, achieved through algorithmic fine-tuning, feature engineering, or system optimization—representing cumulative advancements. It should be noted that the citation data in Table \ref{tab:top5_cited} are drawn from OpenAlex, whose statistics may underestimate the absolute citation counts of some highly cited papers because different versions are counted separately; however, under a unified data source and measurement framework, this bias usually does not substantially alter the relative comparison results across subfields. A detailed discussion of this issue can be found in Section \ref{sec7_1} "Limitations of Data Sources."
\begin{landscape}
\scriptsize
\setlength{\tabcolsep}{4pt}

\begin{longtable}
{p{6.5cm}p{3.0cm}p{1cm}p{2.5cm}p{1.3cm}p{2.2cm}}
\caption{Top Five Most-Cited Papers of Each Subfield}\label{tab:top5_cited} \\
\toprule
\textbf{Title} & \textbf{Authors} & \textbf{Year} & \textbf{Inst-1} & \textbf{Citations} & \textbf{Subfield} \\
\midrule
\endfirsthead

\multicolumn{6}{c}{\tablename\ \thetable\ (continued)} \\
\toprule
\textbf{Title} & \textbf{Authors} & \textbf{Year} & \textbf{Inst-1} & \textbf{Citations} & \textbf{Subfield} \\
\midrule
\endhead

\bottomrule
\multicolumn{6}{p{15.4cm}}{\footnotesize \textit{Note}. First author only (et al.). Inst-1 = first author’s primary affiliation ("University" $\rightarrow$ Univ.).} \\
\endfoot

Inception-v4,Inception-ResNet and the Impact of Residual Connections on Learning & Christian Szegedy, et al.& 2017 & Google & 5433 & AI \\
Informer: Beyond Efficient Transformer for Long Sequence Time-Series Forecasting & Haoyi Zhou, et al. & 2021 & Beihang Univ. & 3194 & AI \\
Distance-IoU Loss: Faster and Better Learning for Bounding Box Regression & Zhaohui Zheng, et al. & 2020 & Tianjin Univ. & 2782 & AI \\
Random Erasing Data Augmentation & Zhun Zhong, et al. & 2020 & Xiamen Univ. & 2213 & AI \\
Regularized Evolution for Image Classifier Architecture Search & Esteban Real, et al. & 2019 & Google & 2173 & AI \\

Swin Transformer: Hierarchical Vision Transformer using Shifted Windows & Ze Liu, et al. & 2021 & Microsoft Research Asia & 9998 & CV \\
Momentum Contrast for Unsupervised Visual Representation Learning & Kaiming He, et al. & 2020 & Meta & 9010 & CV \\
End-to-End Object Detection with Transformers & Nicolas Carion, et al. & 2020 & Univ.Paris Dauphine--PSL & 8756 & CV \\
High-Resolution Image Synthesis with Latent Diffusion Models & Robin Rombach, et al. & 2022 & Ludwig-Maximilians-Universität München & 6910 & CV \\
YOLOv7: Trainable Bag-of-Freebies Sets New State-of-the-Art for Real-Time Object Detectors & Chien-Yao Wang, et al. & 2023 & Institute of Information Science & 6819 & CV \\

GAN (Generative Adversarial Nets) & Ian Goodfellow, et al. & 2017 & Univ.de Montréal & 9420 & ML \\
PyTorch:An Imperative Style, High-Performance Deep Learning Library & Adam Paszke, et al. & 2019 & N/A & 7370 & ML \\
Attention Is All You Need & Ashish Vaswani, et al. & 2017 & Google & 21210 & ML \\
A Simple Framework for Contrastive Learning of Visual Representations & Ting Chen, et al. & 2020 & Google & 6025 & ML \\
Faster R-CNN: Towards Real-Time Object Detection with Region Proposal Networks & Shaoqing Ren, et al. & 2016 & Univ.of Science and Technology of China & 5611 & ML \\

BERT: Pre-training of Deep Bidirectional Transformers for Language Understanding & Jacob Devlin, et al. & 2019 & N/A & 9186 & NLP \\
Deep Contextualized Word Representations & Matthew E. Peters, et al. & 2018 & University of Washington & 7993 & NLP \\
BART: Denoising Sequence-to-Sequence Pre-training for Natural Language Generation, Translation, and Comprehension & Mike Lewis, et al. & 2020 & N/A & 6154 & NLP \\
Transformers: State-of-the-Art Natural Language Processing & Thomas Wolf, et al. & 2020 & Hugging Face & 5919 & NLP \\
Sentence-BERT: Sentence Embeddings using Siamese BERT-Networks & Nils Reimers, et al. & 2019 & Ubiquitous Energy & 5079 & NLP \\

Neural Collaborative Filtering & Xiangnan He, et al. & 2017 & National Univ. of Singapore & 2549 & Web\&IR \\
LightGCN & Xiangnan He, et al. & 2020 & Univ. of Science and Technology of China & 2456 & Web\&IR \\
What is Twitter, a social network or a news media? & Haewoon Kwak, et al. & 2010 & Korea Advanced Institute of Science and Technology & 2162 & Web\&IR \\
Neural Graph Collaborative Filtering & Xiang Wang, et al. & 2019 & National Univ. of Singapore & 1962 & Web\&IR \\
Heterogeneous Graph Attention Network & Xiao Wang, et al. & 2019 & Beijing Univ.of Posts and Telecommunications & 1615 & Web\&IR \\

\end{longtable}
\normalsize
\end{landscape}

\subsubsection{Citation Velocity}\label{sec5_1_3}

\textbf{1. Overall Citation Velocity}

\begin{figure}[H]
    \centering
    \includegraphics[width=\textwidth]{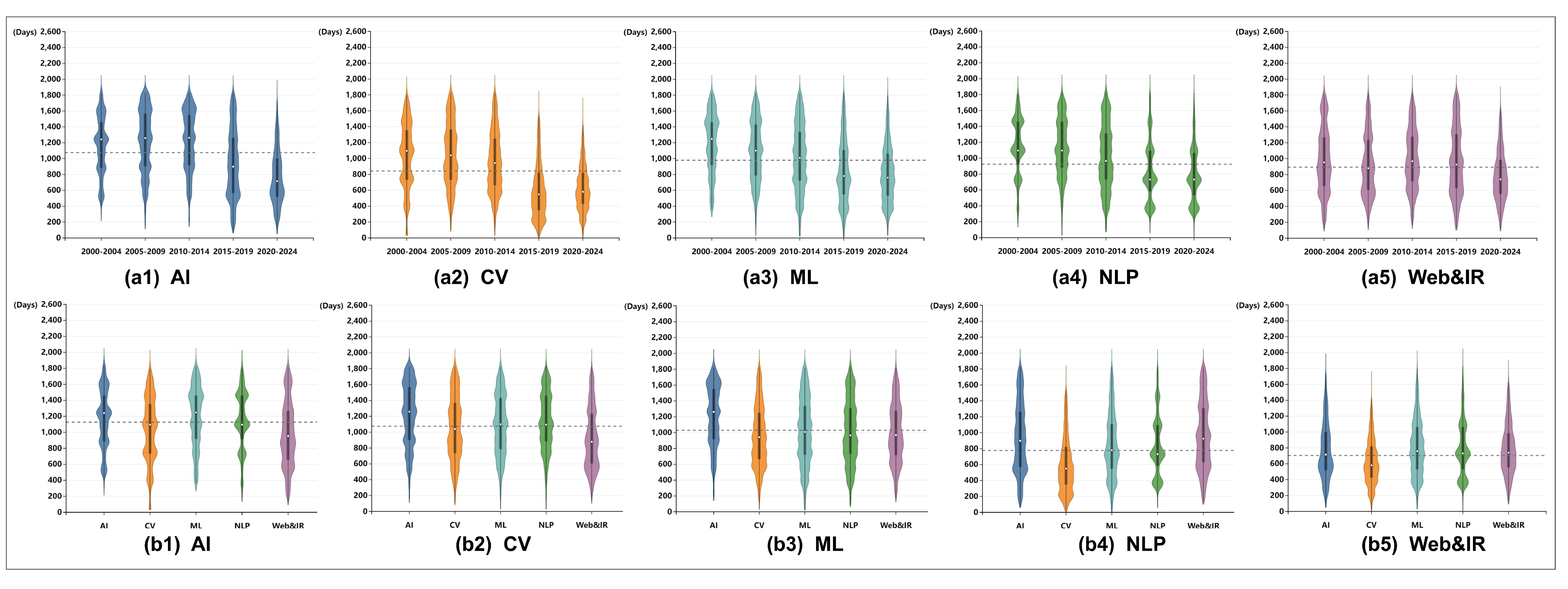}
    \caption{Distribution of Days Needed for 25 Citations}
    \label{fig:citation_velocity}
\end{figure}

To compare the efficiency of knowledge dissemination, we use the overall citation velocity metric defined in Section \ref{sec4_2_1} to calculate the number of days required for papers in a given field with 25 or more citations to reach 25 citations. The fewer the days required, the higher the overall citation velocity. All violin plots were generated using the "kde" function in the Matplotlib library, and all interquartile ranges (IQRs) were automatically determined using Silverman’s IQR method. Figure \ref{fig:citation_velocity} shows the distribution of the number of days required for papers in each field to reach 25 citations. Specifically, Figures (a1)–(a5) illustrate changes across fields over time, while Figures (b1)–(b5) present comparisons between fields across different time periods. Figures (a1)–(a5) reveal that the time required for each subfield to reach 25 citations has decreased from an average of 1,150 days to approximately 700 days, indicating a significant reduction in the time required and suggesting that citation velocity has generally increased over time. The downward trend was most pronounced between 2010 and 2019, a shift likely driven by the new dissemination paradigm shaped by the combination of arXiv preprints and GitHub code-sharing platforms. Since 2012, when AlexNet was released early via arXiv and sparked a breakthrough in the field, it has become standard practice for papers at top artificial intelligence conferences to simultaneously publish preprints and code during the review process. This model has reduced the time to global visibility of research findings to 24 hours, significantly accelerating the speed of knowledge dissemination. As shown in Figures (b1)–(b5), the citation velocity in the CV field is significantly higher than in other fields, with the average time required being around 800 days. The Web\&IR field exhibits relatively minor fluctuations, with the required number of days decreasing from 1,000 to 800, indicating a relatively stable pace of knowledge dissemination. Notably, the citation network in the NLP field consistently displays an abnormal nodular structure, reflecting the persistent presence of "top-tier" papers in this field whose citation rates far exceed the average. We speculate that this may be due to the rapid accumulation of citations for infrastructure-related papers—such as those on standard corpora and evaluation tasks—which has created a dissemination pattern significantly different from that of conventional papers.

\textbf{2. High-Impact Citation Velocity}

\begin{figure}[H]
    \centering
    \includegraphics[width=\textwidth]{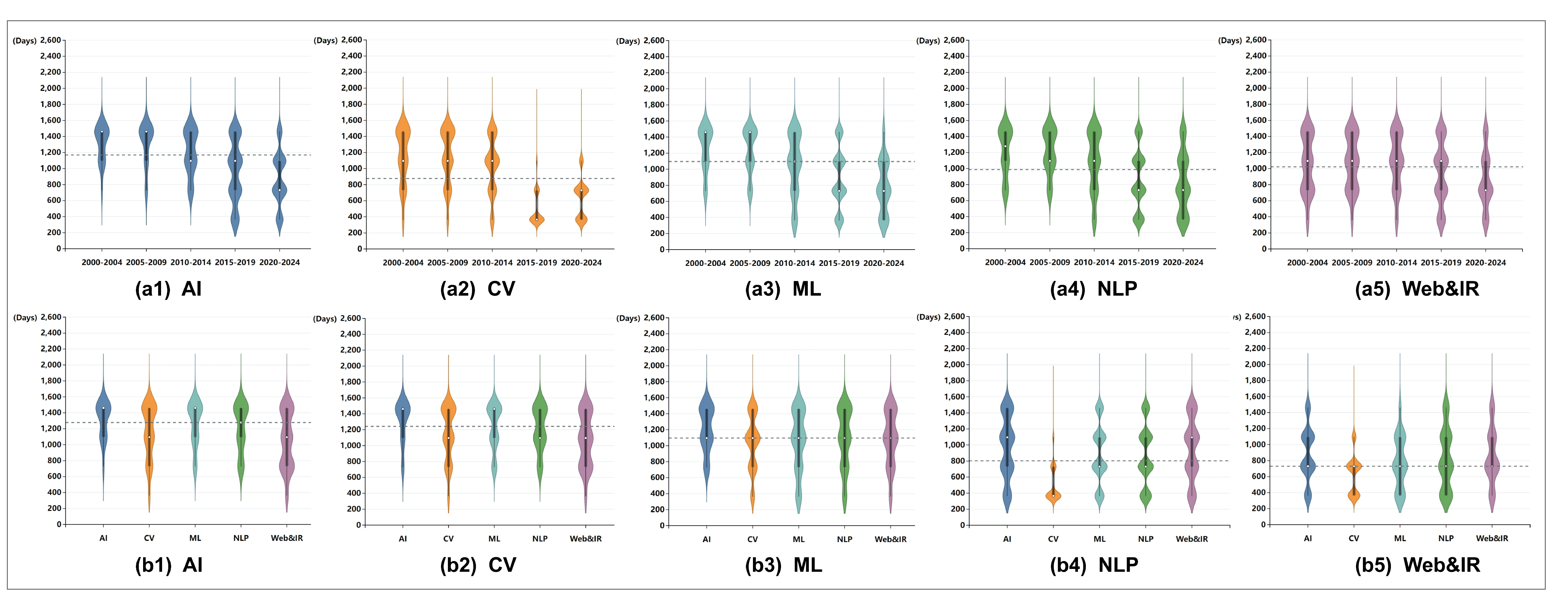}
    \caption{ Distribution of Days Needed for 100 Citations}
    \label{fig:highcited_velocity}
\end{figure}

To further compare the citation rates of highly cited papers, we calculated the number of days required for papers in the top 20\% by total citations to reach 100 citations; this group represents the most dynamic segment within a given field. Figure \ref{fig:highcited_velocity} shows the distribution of the number of days required for papers in the top 20\% by total citations across various fields to reach 100 citations. As shown in Figures (a1)–(a5), the overall high-citation rate across subfields has shown an upward trend over time, reflecting the gradually increasing vitality of highly cited papers in each field. Notably, the time required to reach 100 citations decreased from an average of 1,100 days to 800 days between 2010 and 2019, indicating a significant shortening of the time required to achieve high citation rates—a trend that is most pronounced. During the same period, the average number of days required for papers in the CV field to receive 100 citations decreased from 1,100 to 400. We speculate that this is because CV can be evaluated using standard benchmark datasets, resulting in a relatively short reproduction cycle. As shown in Figures (b1)–(b5), prior to 2015, the rates at which highly cited papers emerged across subfields did not vary significantly, with an average of 1,000 days required and a difference of approximately 400 days between the maximum and minimum values. Among these, the disparity between the CV and AI fields was the greatest, with a gap of 700 days in the time required. Notably, all fields exhibited four distinct nodal structures after 2005. We speculate that there may be a hierarchical structure within the top papers: breakthrough papers, such as those involving revolutionary open-source frameworks (e.g., AlexNet, BERT), can surpass 100 citations within a short period; hot-topic follow-up papers, typically published in large numbers on the same subject within a short timeframe, exhibit slightly slower but concentrated citation growth; and steadily accumulating papers, as well as relatively lagging papers, are often theoretical or foundational achievements. For reference, Appendix Tables \ref{tab:fastest_spreading} and \ref{tab:fastest_spreading_100} report the five fastest-spreading articles within five years under citation thresholds of at least 25 and at least 100 citations, respectively.

\subsubsection{Interdisciplinarity}\label{sec5_1_4}

\textbf{1. Interdisciplinarity of Cited Literature}

\begin{figure}[H]
    \centering
    \includegraphics[width=\textwidth]{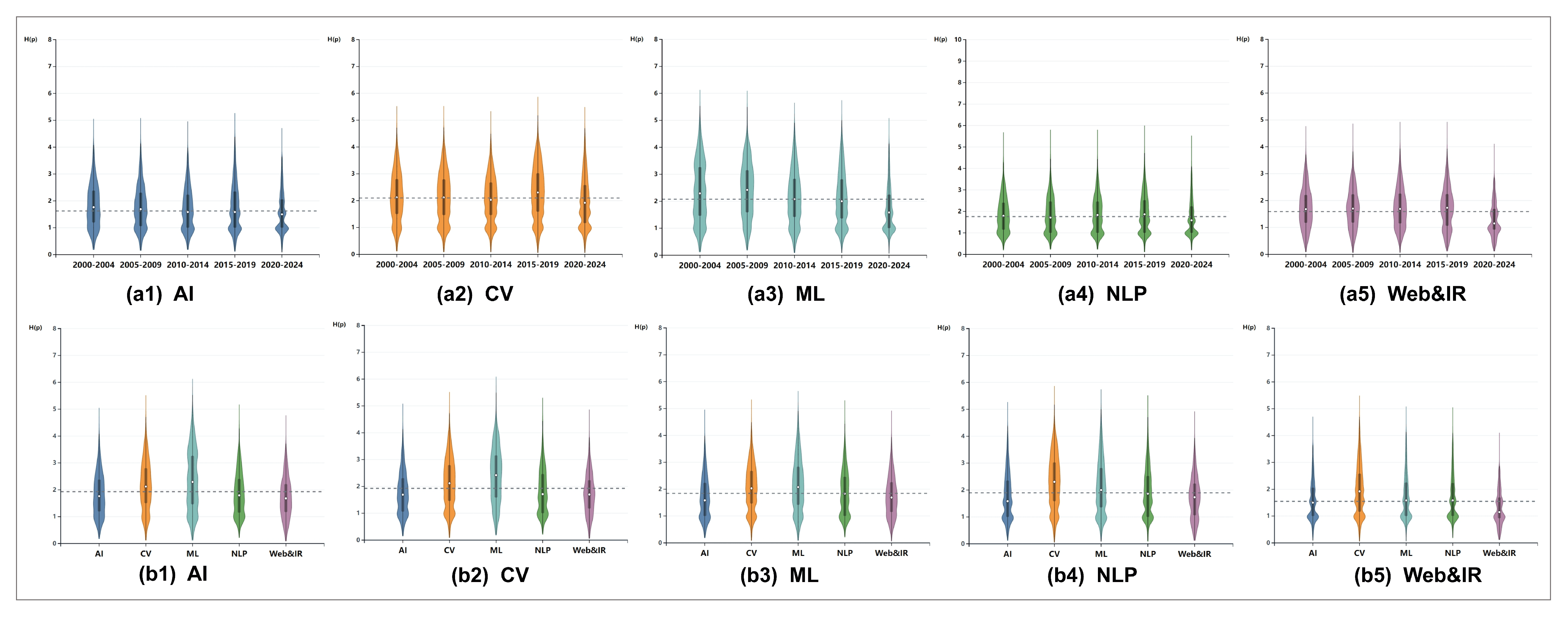}
    \caption{Interdisciplinarity of Cited Literature}
    \label{fig:interdisciplinarity_input}
\end{figure}

We use the Shannon entropy metric defined in Section \ref{sec4_2_1} to measure the disciplinary diversity of cited literature (i.e., knowledge output) in each subfield, represented by the $H(p)$ value; a higher value indicates greater interdisciplinarity. Figure \ref{fig:interdisciplinarity_input} shows the distribution of $H(p)$ values across subfields at different time periods. As shown in Figures (a1)–(a5), there are significant differences in the overall interdisciplinary nature of each subfield. The CV and ML fields have the highest overall values, averaging around 2.0. This may be because their core technologies (such as ResNet and Vision Transformer) have relatively broad application scenarios. The Web\&IR field has the lowest overall values, fluctuating around 1.4. We propose the following hypothesis: the core problem addressed by Web\&IR is "information retrieval efficiency." Since this objective is relatively focused, the knowledge generated by it, while having a profound impact along specific pathways, has a relatively narrow scope of lateral influence across disciplines, resulting in a lower Shannon entropy. As shown in Figures (b1)–(b5), the overall level of interdisciplinarity across subfields has remained largely unchanged over time. However, it is worth noting that the tails of the violin plots for all subfields exhibit an anomalous nodular structure. This indicates the presence of a stable set of papers within each subfield that produce relatively limited knowledge output. We speculate that such research focuses on more mature technologies, with core contributions lying in engineering integration and optimization rather than transferable new knowledge.

\textbf{2. Interdisciplinarity of Citing Literature}

\begin{figure}[H]
    \centering
    \includegraphics[width=\textwidth]{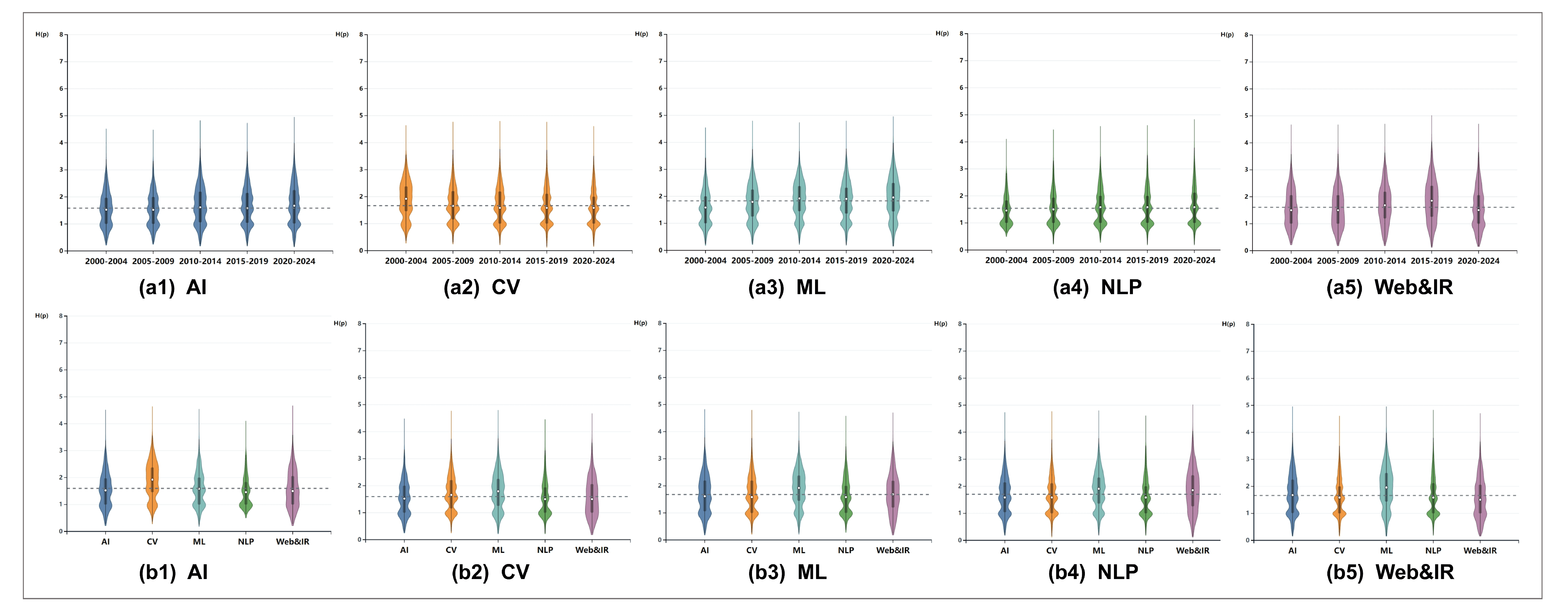}
    \caption{Interdisciplinarity of Citing Literature}
    \label{fig:interdisciplinarity_output}
\end{figure}

\begin{figure}[H]
    \centering
    \includegraphics[width=\textwidth]{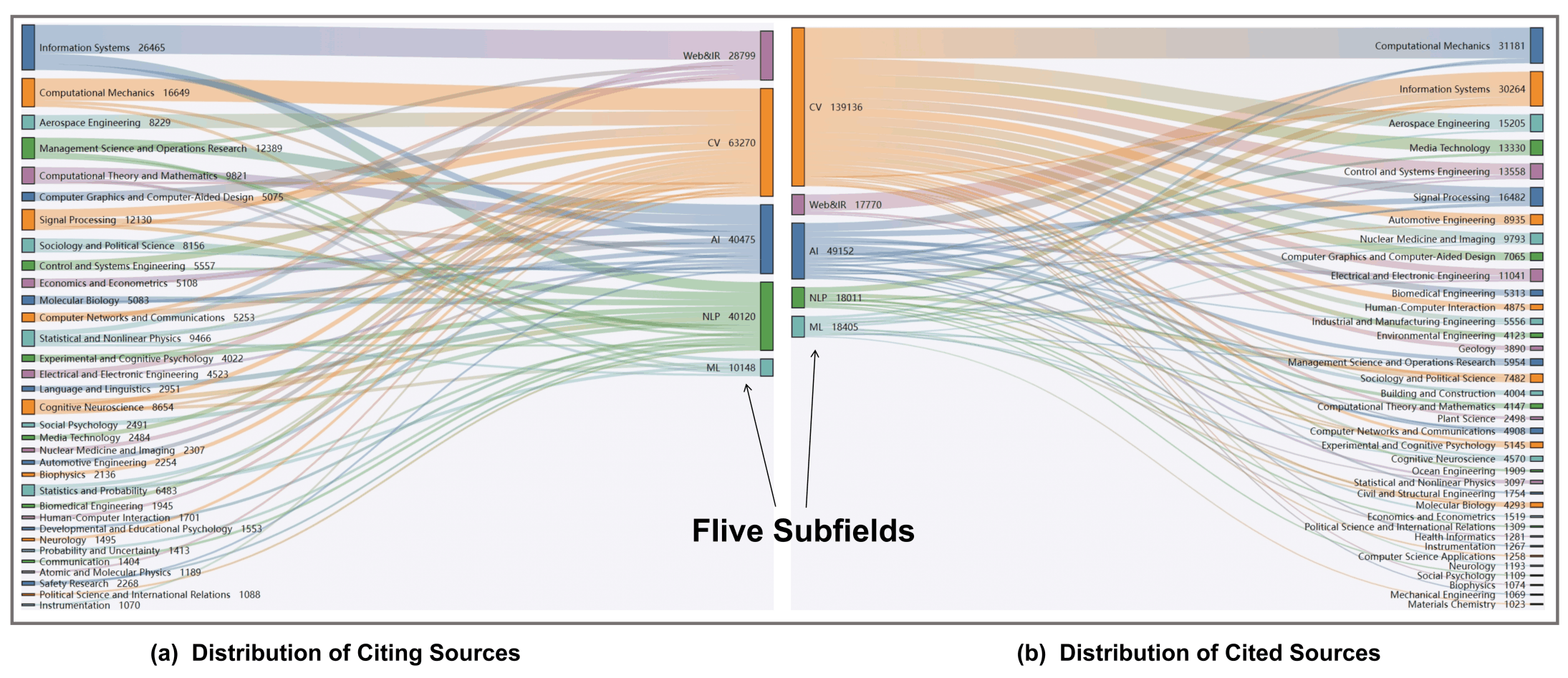}
    \caption{Citing Sources and Cited Sources}
    \label{fig:sankey}
\end{figure}

We use the Shannon entropy metric defined in Section \ref{sec4_2_1} to measure the disciplinary diversity of cited literature (i.e., knowledge input) in each subfield. The $H(p)$ value indicates that the higher the value, the stronger the interdisciplinary nature. Figure \ref{fig:interdisciplinarity_output} shows the distribution of the interdisciplinary nature of citing literature across different time periods for each subfield. As shown in (a1)–(a5) of Figure \ref{fig:interdisciplinarity_output}, the ML subfield exhibits the highest level of interdisciplinarity, averaging around 1.9, while the differences among other subfields are minimal, with values generally stabilizing around 1.5. We hypothesize that, as the methodological core, ML must extensively draw upon foundational disciplines such as statistics to develop general theories and algorithms, resulting in the most diverse input. This hypothesis is indirectly corroborated by Figure \ref{fig:sankey}. Combining Figures \ref{fig:interdisciplinarity_input} and \ref{fig:interdisciplinarity_output}(b1)–(b5), we observe that the overall interdisciplinary nature of both cited and citing references across subfields remains largely unchanged over time. We speculate that each subfield has established relatively stable disciplinary dependencies, with fixed pathways for knowledge input. However, it is worth noting that, with the exception of Web\&IR, the tails of the violin plots for other subfields exhibit anomalous nodular structures. We speculate that after the mainstream paradigm matures, some research focuses on optimization within established frameworks. The references in such papers mostly originate from a few core papers within the field, resulting in extremely narrow knowledge input, abnormally low $H(p)$ values, and the formation of tail knots.

We further constructed a sankey diagram based on the disciplinary classifications of citing and cited sources for papers of each subfield. Figure \ref{fig:sankey} illustrates the knowledge flow (citing sources and cited sources) between the five subfields and external fields. Knowledge input into the CV field primarily originates from engineering disciplines such as Computational Mechanics, Signal Processing, and Aerospace Engineering, accounting for over 40\% of the total. Knowledge output, on the other hand, is heavily concentrated in Information Systems, accounting for approximately 20\%. For Web\&IR, citations from Computational Mechanics and Information Systems account for over 80\% and 50\%, respectively. We speculate that this is because its core research (information retrieval) also falls within the scope of Information Systems. Meanwhile, the core algorithms relied upon to accomplish these tasks (such as ranking learning and graph neural networks) depend on computational theories from the field of Computational Mechanics. Other subfields exhibit a richer, more basic-science-oriented interdisciplinary nature. Twenty percent of cited AI and ML literature originates from engineering disciplines such as Control and Systems Engineering and Electrical and Electronic Engineering. Thirty percent of ML citations come from statistical disciplines, such as Statistics and Probability and Statistical and Nonlinear Physics. This may reflect that modern machine learning is, at its core, computable, high-dimensional statistical inference. Bayesian theory, hypothesis testing, and nonparametric statistics form the foundation of ML model interpretability, generalization theory, and algorithm design. At the same time, the knowledge output of both fields extends widely into numerous engineering and natural science disciplines, such as Biomedical Engineering and Environmental Engineering, confirming their foundational status as general-purpose modeling and inference methods. The citations and cited sources of NLP are primarily concentrated in the humanities, such as Language and Linguistics, Cognitive Neuroscience, and Experimental and Cognitive Psychology. We speculate that this stems from the fact that its theoretical development is grounded in linguistic ontology and cognitive mechanisms. Overall, the knowledge flow network exhibits distinct structural characteristics: tight bidirectional flows have formed between Information Systems and Web\&IR, Computational Mechanics and CV, and Control and Systems Engineering and AI and ML. At the same time, foundational methodological disciplines such as Statistics and Probability and Statistical and Nonlinear Physics primarily serve as sources of knowledge, continuously supplying methodological approaches to fields like ML, while reverse flow remains relatively limited. This asymmetrical structure clearly reveals the research landscape across subfields in terms of macro-level knowledge flow, ranging from foundational theory and core algorithms to vertical applications.

\subsubsection{Observed-to-Expected Citation Ratio}\label{sec5_1_5}

\begin{figure}[H]
    \centering
    \includegraphics[width=0.6\linewidth]{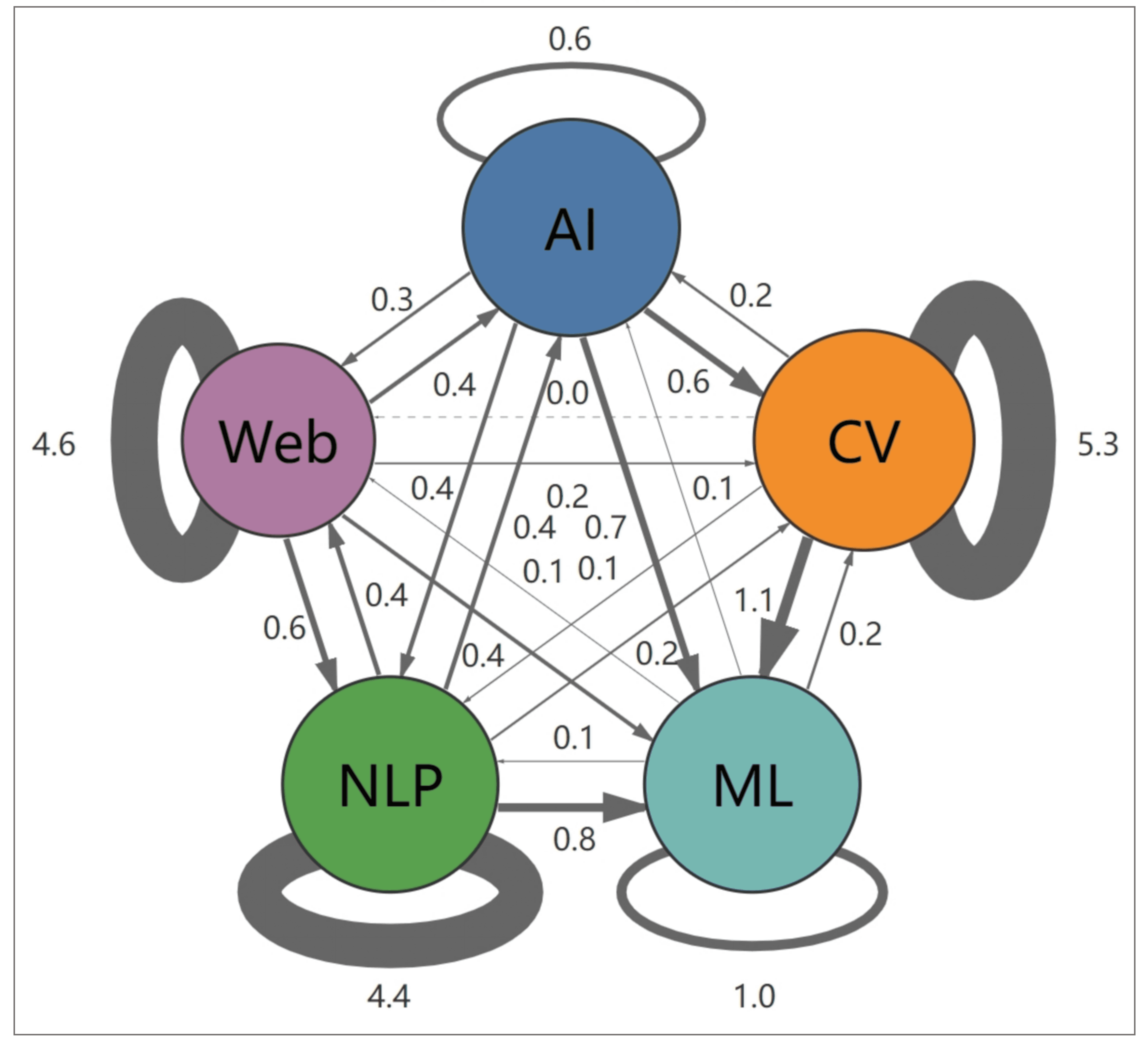}
    \caption{Network Diagram of Observed-to-Expected Citation Ratio}
    \label{fig:rel_citation_strength}
\end{figure} 

We used the relative citation strength metric described in Section \ref{sec4_2_1} to measure the intensity of knowledge flow among the five subfields. Figure \ref{fig:rel_citation_strength} shows the relative citation strength network for each subfield from 2000 to 2024, where the thickness of the lines reflects the relative citation strength values between nodes. As shown in the Figure \ref{fig:rel_citation_strength}, CV, NLP, and Web\&IR exhibit high levels of internal knowledge flow, with overall citation strengths of approximately 5.3, 4.6, and 4.4, respectively—far higher than the random baseline of 1.0. This may be attributed to the formation of relatively stable technical pathways within these fields, such as Convolutional Neural Networks (CNNs) in CV and Transformers in NLP. Subsequent research, while building upon and fine-tuning these pathways, generates intensive internal knowledge transfer and citations. This phenomenon is corroborated by the interdisciplinary nature observed in Figures \ref{fig:interdisciplinarity_input} and \ref{fig:interdisciplinarity_output}. Regarding cross-domain knowledge flow, Figure \ref{fig:rel_citation_strength}  reveals non-bidirectional knowledge flow. The relative citation strength of CV to ML is 1.1, making it the only cross-subdomain citation relationship that significantly exceeds the random baseline. This may be because CV is highly dependent on ML technologies, particularly as deep learning algorithms (such as CNNs and GANs) have seen rapid expansion and scenario-specific applications within CV. In contrast, the citation strength of ML to other fields does not exceed 0.3. Combined with the sankey in Figure \ref{fig:sankey}, we speculate that ML knowledge absorption stems more from external foundational disciplines such as statistics.

\begin{figure}[H]
    \centering
    \includegraphics[width=\textwidth]{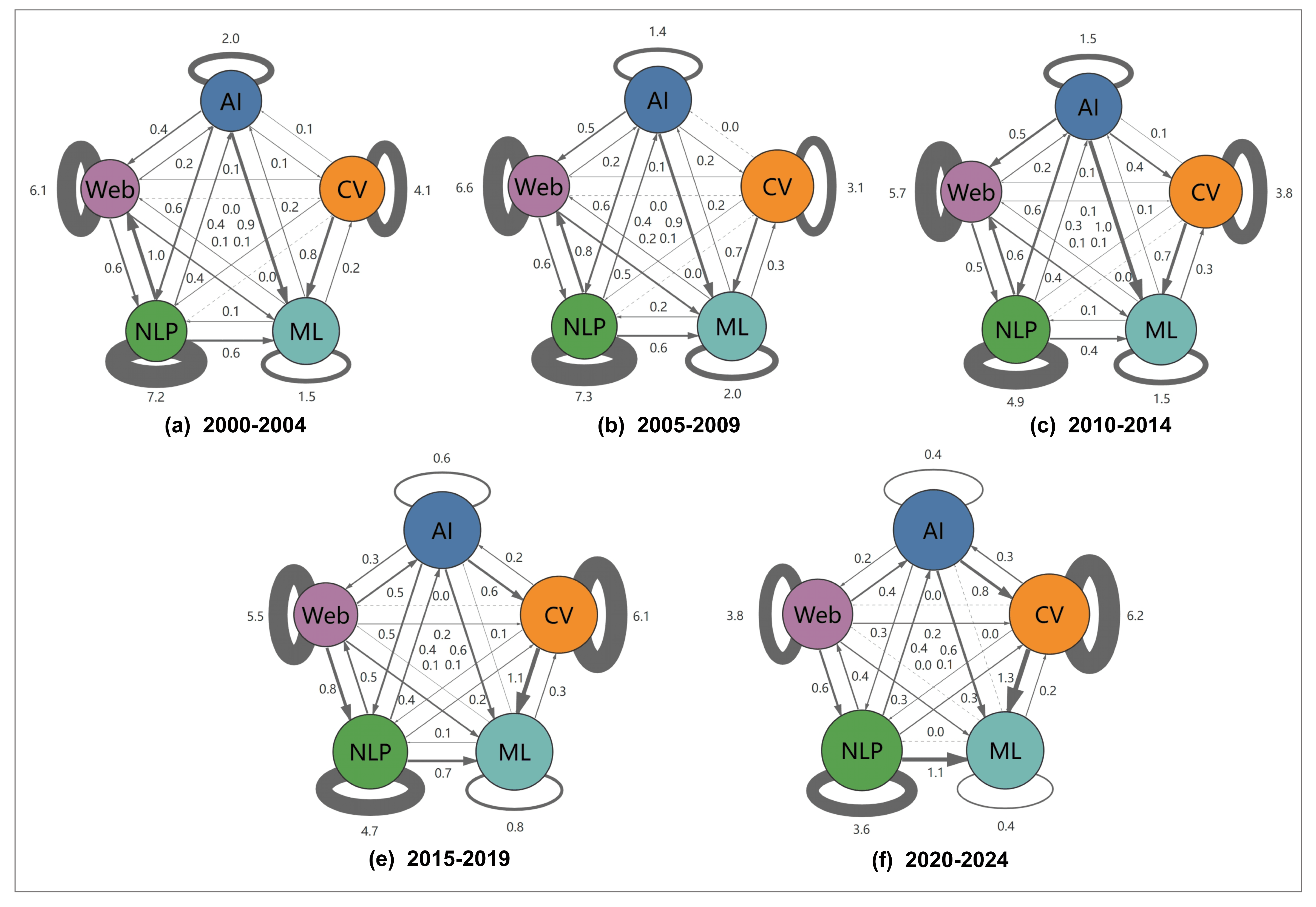}
    \caption{Network Diagram of Citation Ratio of Each Period}
    \label{fig:rel_citation_time}
\end{figure} 

Building on Figure \ref{fig:rel_citation_strength}, we further measured changes in knowledge flow among the five subfields over time. Figure \ref{fig:rel_citation_time} shows a network diagram of relative citation strength by time period. Overall, the relative citation strength within each subfield gradually declined over time; for example, AI’s relative citation strength dropped from 2.0 to 0.4, indicating that most subfields have gradually transformed from relatively independent disciplines into open disciplines characterized by interdisciplinary collaboration. Notably, the relative citation strength of the CV subfield rose from 4.1 to 6.2. We hypothesize that the CV subfield rapidly established a standardized research paradigm, leading to most relevant literature originating from within the field. At the same time, there are relatively stable directions of knowledge flow between subfields. The citation strength of CV and NLP toward ML gradually increased (lower right of (a)–(f)), rising from 0.8 and 0.6 to 1.3 and 1.1, respectively. The mutual citation relationship between Web\&IR and NLP has consistently remained above 0.5, likely due to their natural overlap in text information processing. The distribution of mutual citation strength between AI and other subfields is relatively balanced, generally ranging between 0.2 and 0.6, without forming fixed dependencies. We speculate that research topics in the AI field intersect directly with those of various subfields, but AI does not serve as the core technology provider for any single field.

\subsection{Collaboration Characteristics}\label{sec5_2}

\subsubsection{ Authors per Paper}\label{sec5_2_1}

\begin{figure}[H]
    \centering
    \includegraphics[width=\textwidth]{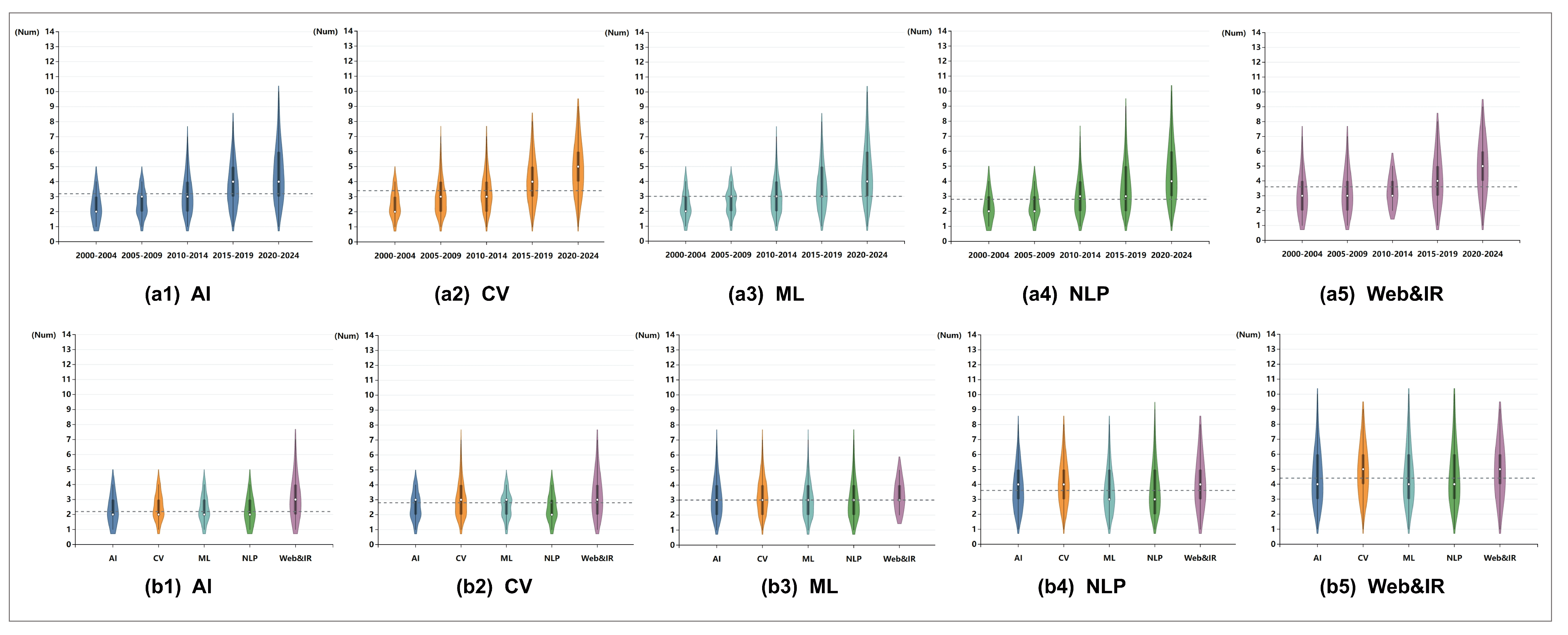}
    \caption{Distribution of Authors per Paper of Each Subfield}
    \label{fig:author_distribution}
\end{figure}

Figure \ref{fig:author_distribution} shows the distribution of the number of authors per paper across five subfields. Figures (a1)–(a5) illustrate how the distribution in each field changes over time, while Figures (b1)–(b5) present a comparison of the distributions across fields during different time periods. The gray dashed lines in the figures represent the average of the medians of the five subfields. As shown in Figures (a1)–(a5), the average number of authors per paper in each field increases over time. At the same time, the range of the violin plots for each field continues to widen, the data distribution becomes increasingly dispersed, and peaks appear at the top. This indicates that the variance in the number of authors across fields has increased, with the emergence of some papers featuring a large number of authors. For example, "Why is the Winner the Best?" (CV 2022; OpenAlex: W4386071488), with 100 authors, is a challenge summary paper that typically lists all participating team members as co-authors, resulting in an extremely large author list. "The BigScience ROOTS Corpus: A 1.6TB Composite Multilingual Dataset" (Web\&IR 2023; OpenAlex: W4307225507), with 54 authors, is a typical open, collaborative multilingual large-scale corpus project. It brings together universities and industry institutions from multiple countries, reflecting the collaborative paradigm in the field of computer science. GEMv2: Multilingual NLG Benchmarking in a Single Line of Code (NLP 2022; OpenAlex: W4385573234), with 77 authors, features a large team size due to the significant computational engineering efforts required for multilingual benchmark development, including data collection and cleaning, evaluation protocols and platforms, and baseline systems. As shown in Figures (b1)–(b5), when viewed horizontally across different time periods, the differences between fields are relatively small. Among them, the median for Web\&IR is slightly higher than the average median across all fields, while NLP is slightly lower than this average. Taking the period from 2020 to 2024 (Figure (b5)) as an example, the median for Web\&IR was 5 authors, slightly higher than the average median of 4.4 across all domains; the median for NLP was 4 authors, slightly lower than this average, indicating that Web\&IR holds a slight advantage in terms of author count.

\begin{figure}[H]
    \centering
    \includegraphics[width=\textwidth]{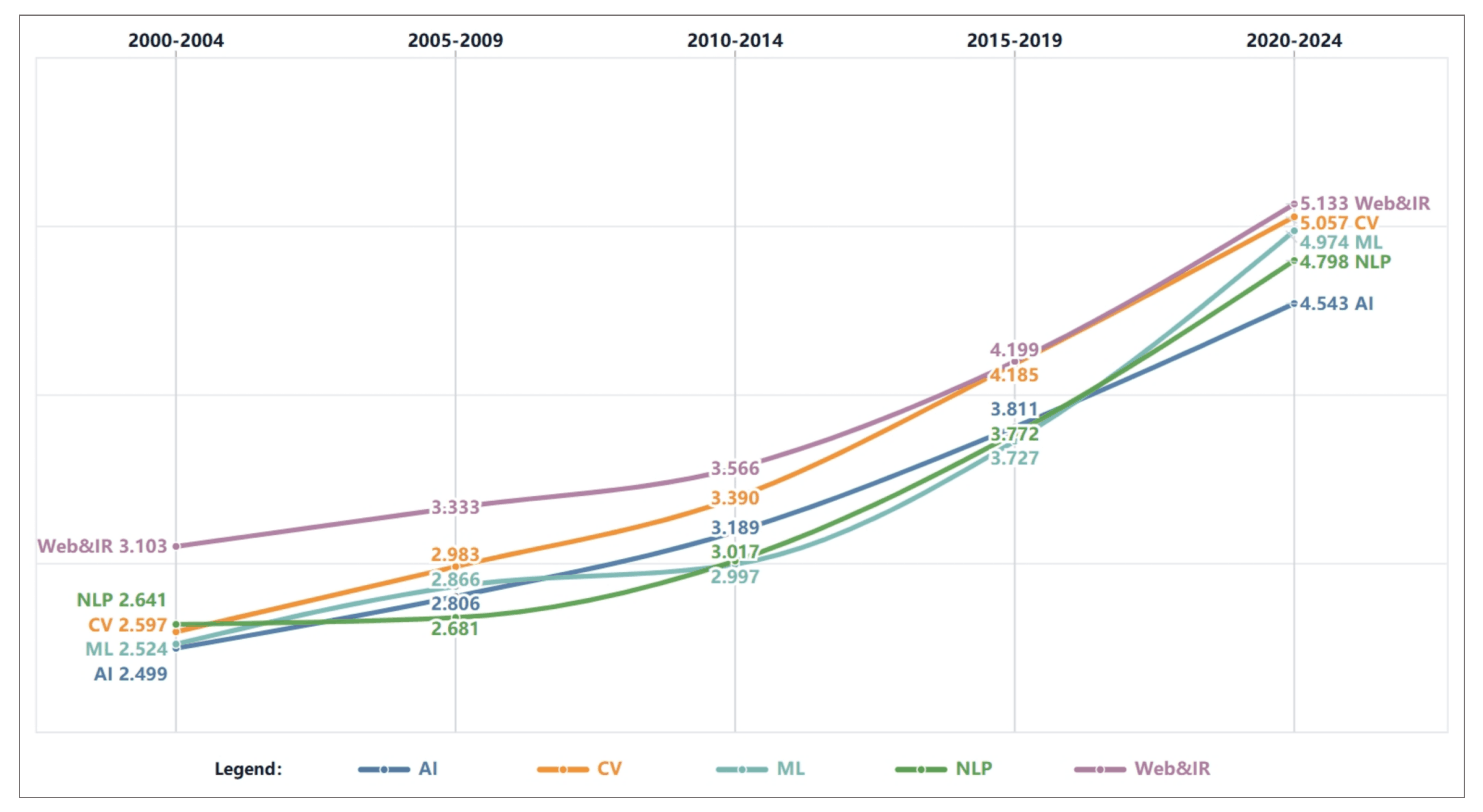}
    \caption{CI of Each Subfield}
    \label{fig:CI}
\end{figure}

Based on the distribution characteristics described above, we used the collaboration index ($CI$) defined in Section \ref{sec4_2_2}—specifically, the average number of authors per paper—to further analyze the breadth of academic collaboration across the five subfields. Figure \ref{fig:CI} illustrates the changes in $CI$ values for each subfield over time; a higher $CI$ value indicates a broader scope of academic collaboration. Overall, $CI$ values across all fields show an upward trend, and the gaps between fields are gradually narrowing. This indicates that as the complexity of research problems increases and the scale of data expands, multi-author collaboration is becoming the norm across all five subfields. Among them, Web\&IR increased from 3.103 in 2000–2004 to 5.133 in 2020–2024, maintaining the highest level across all time periods. We speculate that this is closely related to the interdisciplinary nature of research in this field—work in this area typically requires the integration of multiple technologies, such as algorithm design, data processing, and user analysis, and therefore relies more heavily on multi-author collaboration. In contrast, the $CI$ value in the AI field increased from 2.499 in 2000–2004 to 4.543 in 2020–2024, with a growth rate slightly lower than that of other fields. This may be attributed to the relatively independent nature of artificial intelligence research tasks, which rely less on cross-institutional collaboration, resulting in a more gradual expansion of the scale of collaboration.

\subsubsection{International Collaboration}\label{sec5_2_2}

\begin{figure}[H]
    \centering
    \includegraphics[width=\textwidth]{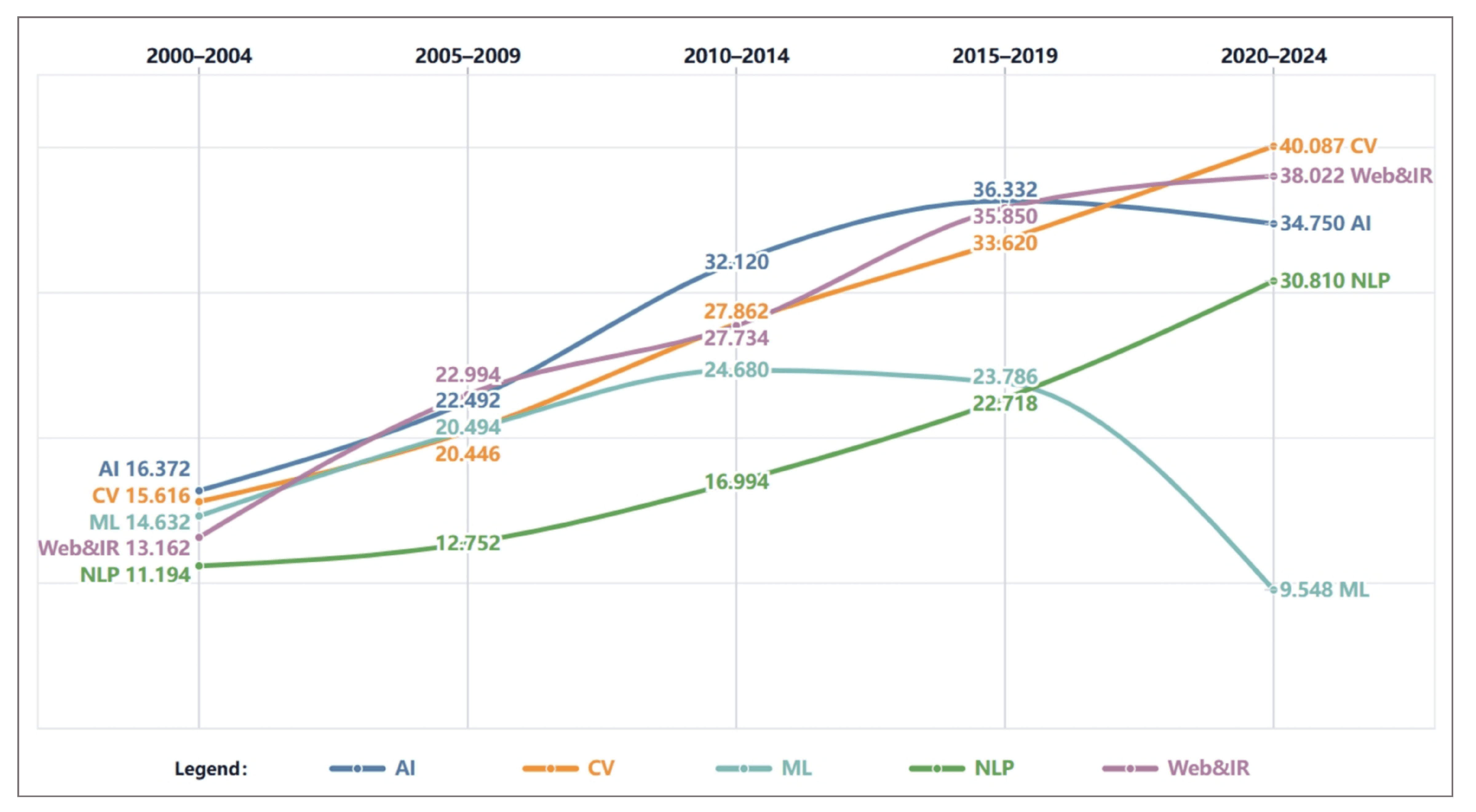}
    \caption{International Collaboration Rate of Each Subfield}
    \label{fig:intl_collab_rate}
\end{figure} 

First, we calculated the simple ratio—that is, the number of papers involving international collaboration divided by the total number of papers. Figure \ref{fig:intl_collab_rate} shows how this ratio has changed over time. Overall, the proportion of international collaboration in most fields shows an upward trend, reflecting the deepening level of international cooperation across various disciplines. As shown in the figure, the proportion of international collaboration in CV and Web\&IR has continued to grow, reaching 40.087\% and 38.022\%, respectively, in 2020–2024, indicating significant progress in international collaboration in these two fields. AI reached a peak of 36.332\% between 2015 and 2019, indicating that international collaboration in AI was particularly active during this period. In contrast, the proportion of international collaboration in ML declined sharply between 2020 and 2024, falling significantly below that of other fields. We speculate that the ongoing deterioration of U.S.-China relations in recent years may have had some impact. The proportion of international collaboration in NLP grew steadily from 11.194\% to 30.810\%, but remains significantly lower than in other fields. This may be due to the higher barriers to collaboration in NLP, as cross-border efforts must overcome challenges such as multilingual adaptation and data barriers.

\begin{figure}[H]
    \centering
    \includegraphics[width=\textwidth]{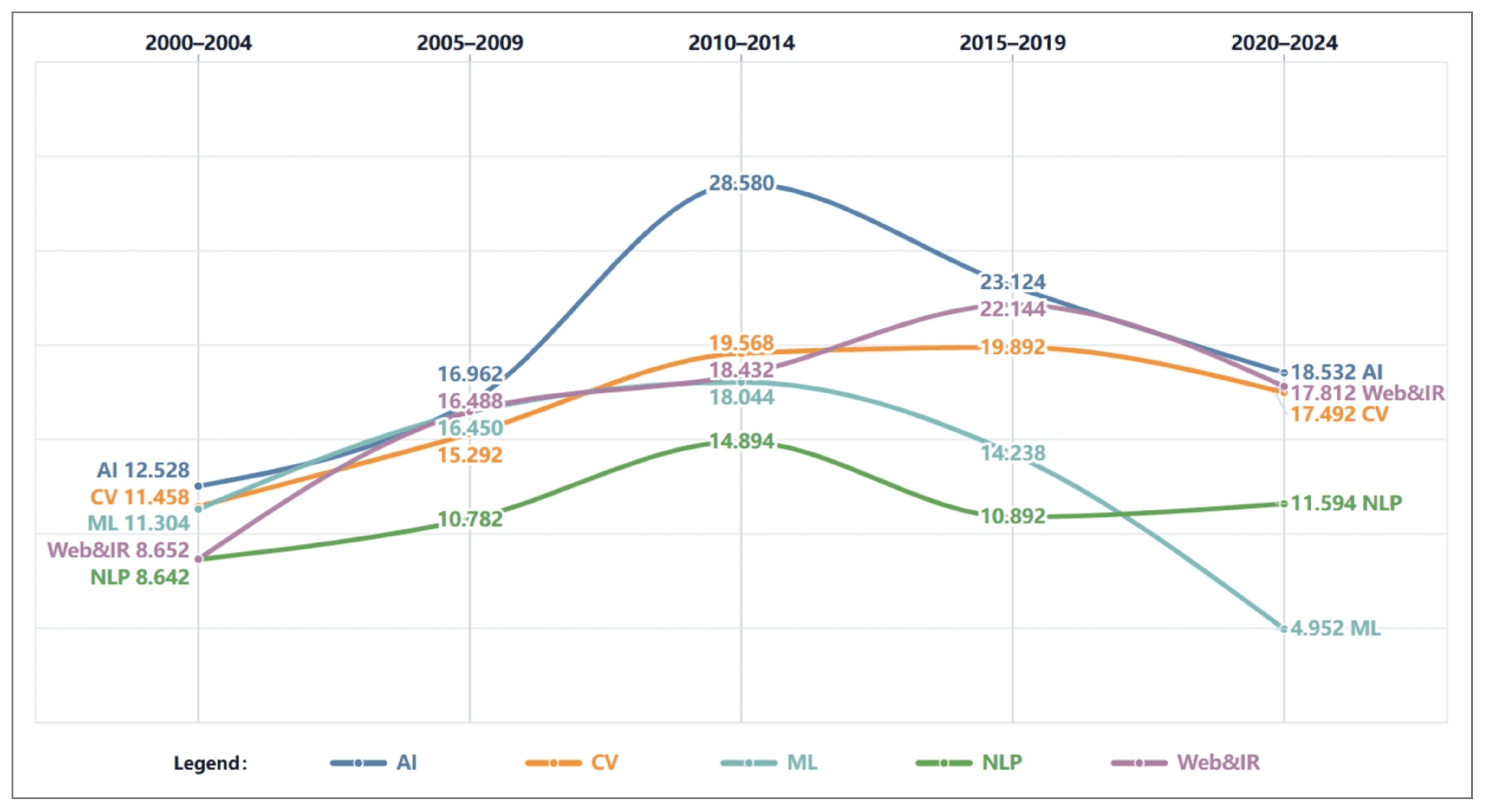}
    \caption{International Cooperation Rate of Each Subfield}
    \label{fig:intl_coop_subfield}
\end{figure} 

Second, using the proportion of international collaboration defined in Section \ref{sec4_2_2} and drawing on the pair-based co-authorship perspective of Rousseau \& Zhang (2021)\cite{rousseau2021}, we measured the degree of international collaboration across the five subfields by calculating the proportion of transnational author pairs out of all author pairs. Figure \ref{fig:intl_coop_subfield} illustrates how this proportion has changed over time across each subfield; values closer to 1 indicate a higher degree of international collaboration among author teams. Overall, the proportion of international collaboration in all fields followed an upward-then-downward trend, with most peaking between 2010 and 2014 before declining. We speculate that the primary reason for this phenomenon may be the recent decline in academic exchanges between China and the United States. Notably, the proportion for ML dropped to 4.952\% between 2020 and 2024, significantly lower than other fields, indicating that international collaboration in this field has been most severely constrained. The rate of international collaboration in AI fluctuated significantly, rising rapidly from 16.962\% in 2005–2009 to 28.580\% in 2010–2014, which was notably higher than in other fields. We speculate that this is closely related to deep learning research, exemplified by the Google Brain project\cite{Le2012GoogleBrain}. Jointly initiated by Google and Stanford University, this project brought together researchers from the United States, Canada, and other countries. Its breakthroughs in training large-scale neural networks sparked a global research boom and fostered frequent international collaboration. In Table \ref{tab:intl_collab_top5} of the Appendix, we summarize and list the top five papers with the highest degree of international collaboration in each subfield for reference.

\begin{figure}[H]
    \centering
    \includegraphics[width=\textwidth]{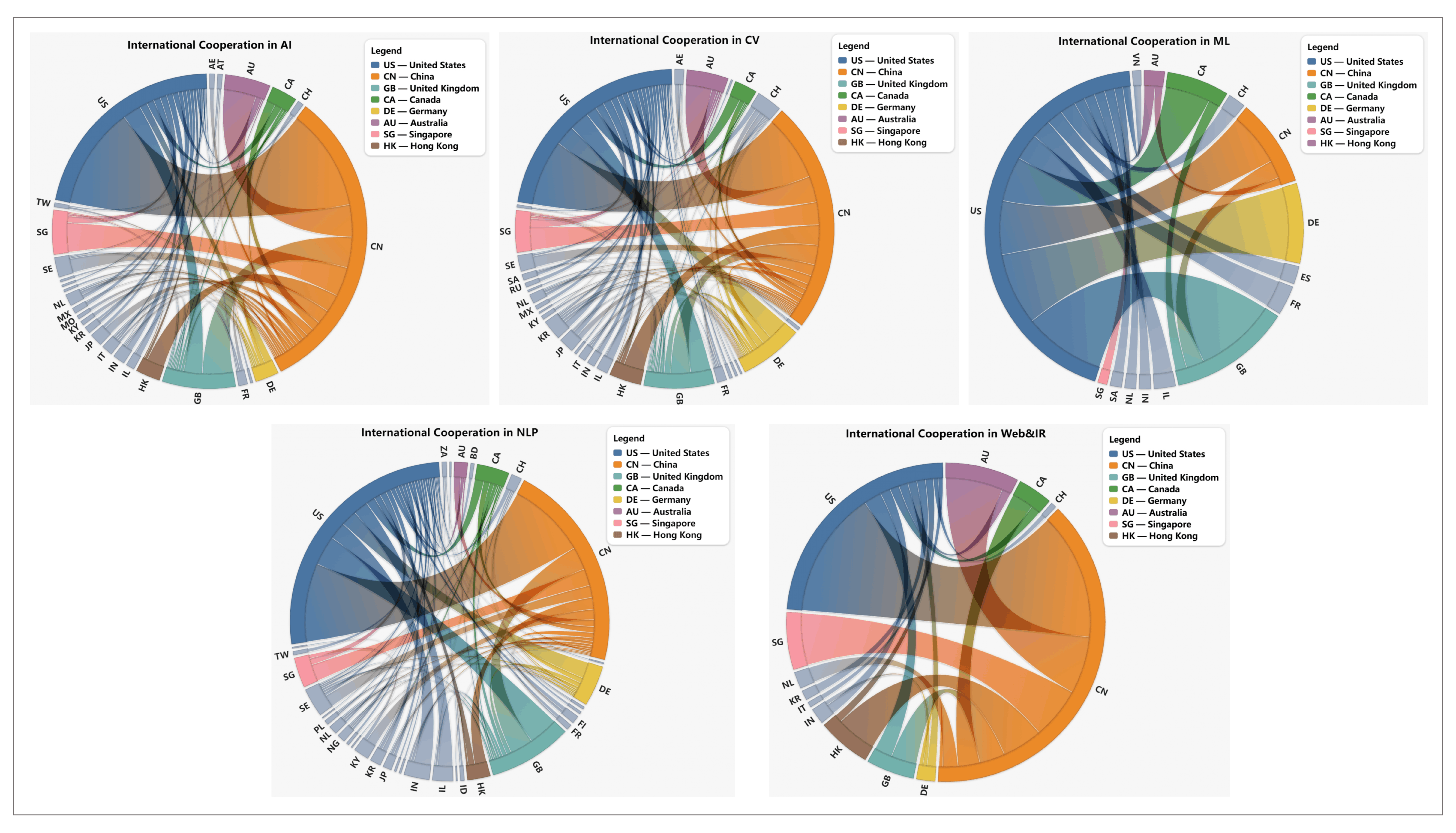}
    \caption{International Collaboration Patterns of Each Subfield}
    \label{fig:intl_collab_patterns}
\end{figure} 

Figure \ref{fig:intl_collab_patterns} further illustrates international collaboration across the five subfields from 2020 to 2024. The length of the outer arcs represents the proportion of each country’s participation in international collaboration, while the width of the chords reflects the intensity of bilateral collaboration. Overall, international collaboration is primarily concentrated among CN, US, and GB, with collaboration between China and the US accounting for the highest proportion, at approximately 12.4\%. The United States accounts for approximately 34.6\% of total international collaboration, while China accounts for approximately 29.8\%, which is generally consistent with the distribution of papers across these fields. In terms of the structural characteristics of the international collaboration network, in the fields of AI, CV, and NLP, the US and CN form the core of collaboration, with countries and regions such as GB, DE, CA, AU, SG, and HK also deeply involved, creating a multi-centered, open collaboration structure. In the ML and Web\&IR fields, collaboration is primarily concentrated among a few core countries. Apart from CN and US, the participation of other countries is significantly lower, resulting in a centralized structure dominated by a small number of nations. A comparison of these two collaboration structures reveals that the global collaboration networks in AI, CV, and NLP are more open and dispersed, whereas the collaboration patterns in ML and Web\&IR tend to be closed and centralized, with relatively limited diversity in international collaboration.

We found that in the Web\&IR field, CN holds a clear advantage, accounting for approximately 30\%; in the ML field, the US holds a prominent dominant position, accounting for approximately 35\%. With the exception of ML, the majority of international collaborations in other fields occur between China and the US. It is worth noting that in the Web\&IR field, where CN holds the advantage, the proportion of China-U.S. collaboration is the highest, at approximately 15\%; however, in the ML field, where the US holds the advantage, the proportion of China-U.S. collaboration is only about 8\%, lower than that of GB and DE. This is consistent with the results shown in Figures \ref{fig:intl_collab_rate} and \ref{fig:intl_coop_subfield} above, corroborating the hypothesis that the decline in the proportion of international collaboration in ML from 2020 to 2024 is related to a reduction in China-U.S. collaboration. Meanwhile, Stanford University’s {"2023 AI Index Report"}\cite{StanfordHAI2023AIIndex} notes that from 2010 to 2021, the number of collaborative AI publications between China and the US increased by approximately fourfold; however, from 2020 to 2021, the total number of collaborations grew by only 2.1\%, marking the smallest year-over-year growth rate since 2010. This further illustrates that the intensifying Sino-U.S. technological competition and strained international relations have already impacted the overall landscape of transnational collaboration in the AI field, leading to a significant contraction in the scale of international cooperation in subfields such as ML and exhibiting certain "anti-globalization" characteristics. This conclusion also validates our speculation in Figure \ref{fig:intl_collab_rate} regarding the reasons for the recent decline in the proportion of international collaboration.

\subsubsection{Industry Collaborationn}\label{sec5_2_3}

\begin{figure}[H]
    \centering
    \includegraphics[width=\textwidth]{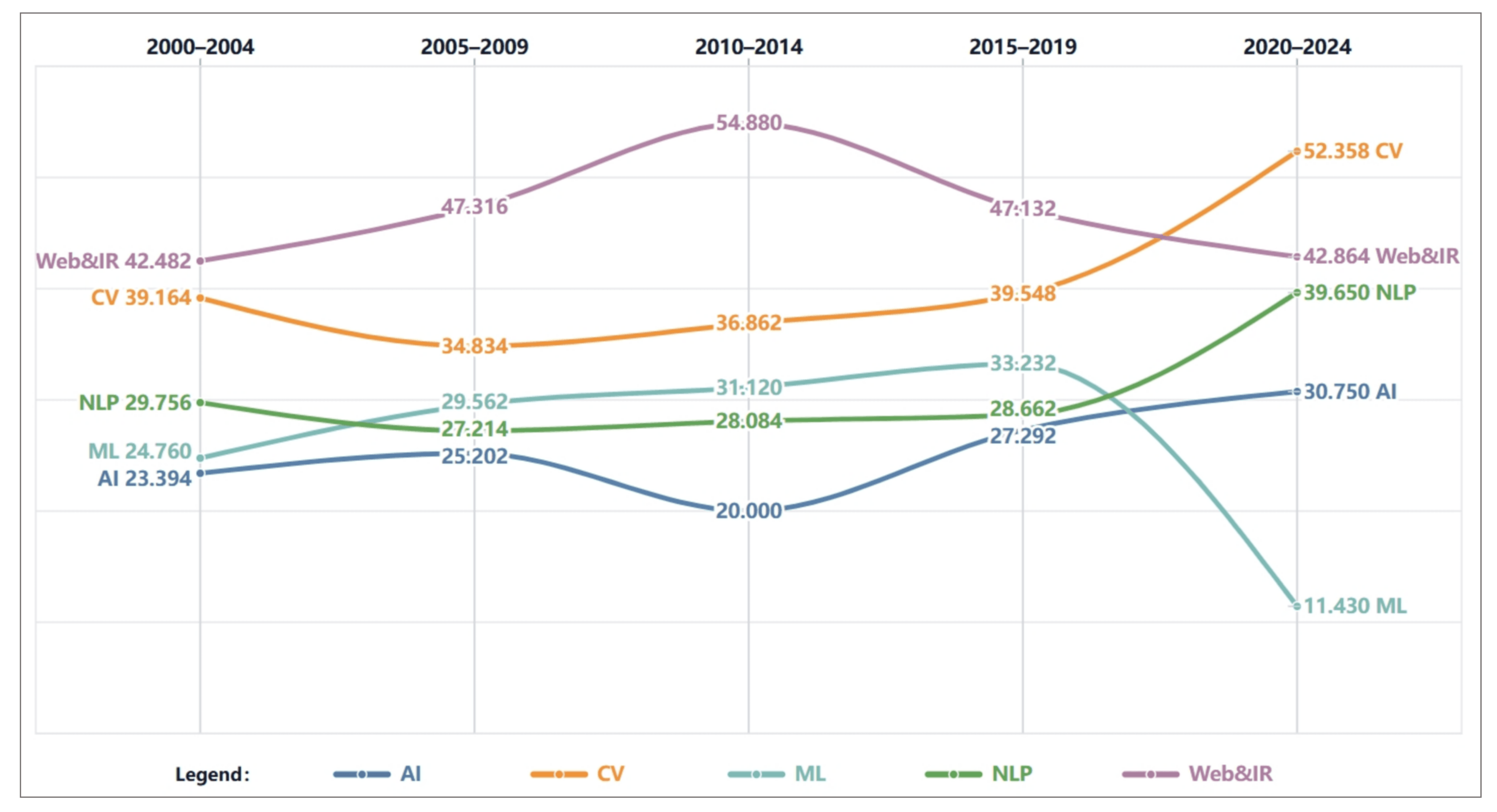}
    \caption{Industry Collaboration Rate of Each Subfield}
    \label{fig:industry_collab_rate}
\end{figure} 

We use the indicator "proportion of industry collaboration" from Section \ref{sec4_2_2} to measure the strength of industry-academia ties in each field by calculating the proportion of articles involving industry collaboration out of the total number of articles across five subfields. Figure \ref{fig:industry_collab_rate} illustrates how this proportion has changed over time across different fields; a higher proportion indicates closer ties with industry. Overall, the proportion of industry collaboration shows an upward trend in most fields. Among these, the collaboration ratio in the Web\&IR field has consistently remained within the 40\%–60\% range, peaking at 54.880\% between 2010 and 2014—significantly higher than other fields—demonstrating a long-term, stable partnership with industry. The collaboration ratio in the CV field rose rapidly between 2020 and 2024, aligning closely with the wave of industrialization driven by deep learning. The NLP field has shown generally gradual changes, with only slight growth from 2020 to 2024, corresponding to the advancement of commercial applications of language models and the gradual increase in corporate participation. The overall collaboration rate in the AI field remains relatively low, hovering around 25\%; the decline observed from 2010 to 2014 was primarily due to insufficient sample size resulting from the cancellation of IJCAI in 2012. The collaboration rate in ML dropped significantly to 11.430\% between 2020 and 2024. This may be attributed to companies shifting their collaborative focus toward LLMs, multimodal systems, and dataset construction, rather than foundational optimization algorithms. Additionally, the development of deep learning and large-scale models relies heavily on massive computational resources and data support, prompting companies to establish their own research teams (such as Google Brain and Baidu’s Deep Learning Lab), thereby reducing direct collaboration with the academic community. 

\begin{figure}[H]
    \centering
    \includegraphics[width=\textwidth]{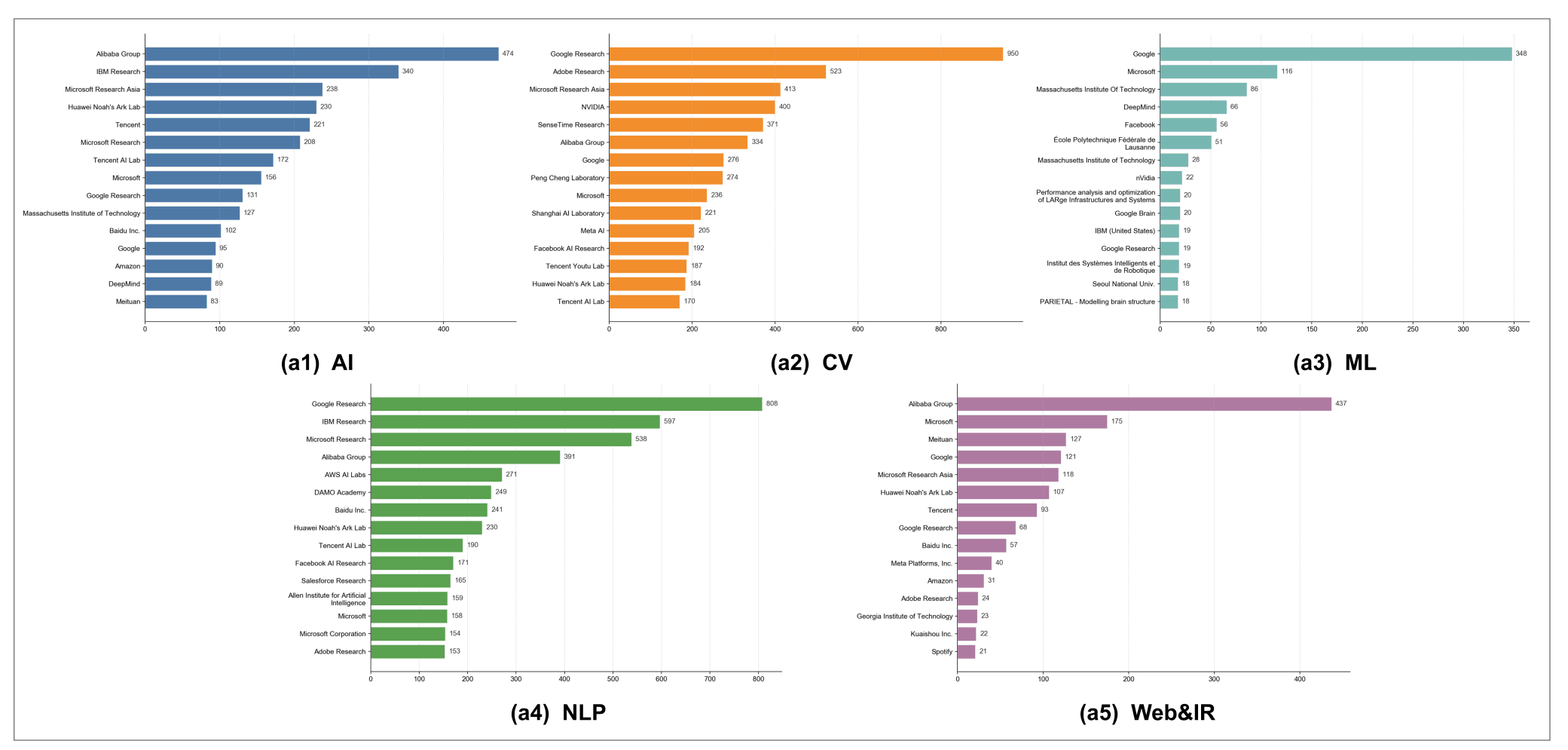}
    \caption{Top 15 Industry Collaborators of Each Subfield}
    \label{fig:top15_industry_collab}
\end{figure} 

To gain a deeper understanding of the specifics of collaboration between the five subfields and industry organizations, we have created bar charts showing the top 15 industry organizations by number of collaborations in each subfield (Figure \ref{fig:top15_industry_collab}). Overall, the list generally includes major companies such as Google, Microsoft, Alibaba Group, and Huawei Noah’s Ark Lab, along with their research institutes, indicating that they are common collaborators across multiple fields. In the AI field, collaborating institutions are primarily from China and the United States. Alibaba Group had the highest number of collaborations at 474, while IBM Research, Microsoft Research Asia, Huawei Noah’s Ark Lab, and Tencent were also major partners. In the CV field, Google Research ranked first with 950 collaborations, significantly ahead of second-place Adobe Research, with relatively stable gaps among the remaining companies. China’s SenseTime (371 collaborations) ranks among the top five. In the ML field, collaborations are highly concentrated around Google (348 collaborations); the up-and-coming company DeepMind ranks relatively high in terms of collaboration frequency, while no Chinese companies made it into the top 15. In the NLP field, Google Research had the highest number of collaborations, at 808. China’s Alibaba Group and DAMO Academy also maintained a significant collaboration frequency, with 391 and 249 collaborations, respectively. The Web\&IR field is dominated by Chinese institutions, with Alibaba Group leading with 437 collaborations, while Microsoft and Meituan also rank among the most frequent collaborators.

\subsubsection{Intensity of Collaboration}\label{sec5_2_4}

\begin{figure}[H]
    \centering
    \includegraphics[width=\textwidth]{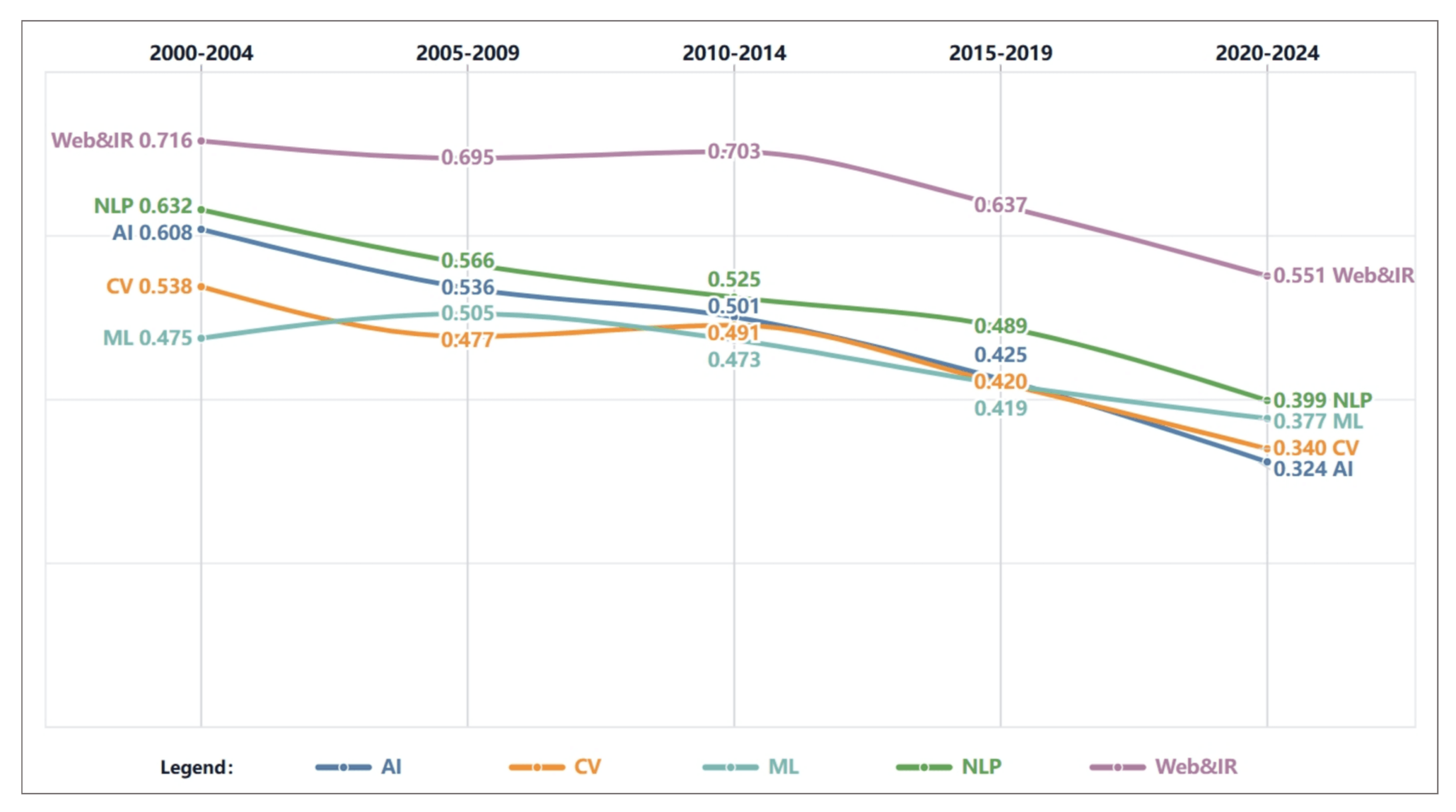}
    \caption{Weighted Average Clustering Coefficient of Each Subfield}
    \label{fig:weighted_clustering}
\end{figure} 

We utilized the collaboration intensity metric described in Section \ref{sec4_2_2}, following the definition of the weighted clustering coefficient by Opsahl \& Panzarasa (2009)\cite{opsahl2009}, which introduces a weight based on the number of collaborations to measure the density of collaboration among researchers. Figure \ref{fig:weighted_clustering} illustrates the changes in the weighted clustering coefficient across subfields over time; higher values indicate that collaboration within a field is more concentrated in localized, closed structures, suggesting closer collaborative relationships. Overall, the weighted clustering coefficients across all fields show a continuous downward trend, indicating that collaborative relationships are gradually shifting from closed, intra-team collaboration toward open, cross-team collaboration. We speculate that this may be related to the gradual maturation of technical ecosystems. For example, in the ML field, with the refinement of open-source deep learning frameworks such as TensorFlow and PyTorch, researchers are able to conduct independent research with less reliance on tight-knit collaborative networks. As shown in the figure, from 2000 to 2004, the overall clustering coefficients across the five fields remained at a relatively high level, with Web\&IR recording the highest value. We speculate that this may be because research during this period focused on a few core retrieval tasks (such as web page ranking and information retrieval model optimization), with researchers primarily concentrated within a small number of research groups, forming tight local collaboration networks. For example, the early optimization and application of the PageRank algorithm were jointly accomplished by research teams at Stanford University, engineers from companies such as Google, and some academic teams specializing in information retrieval theory. Over the subsequent 15 years, the clustering coefficients for all fields continued to decline, and the differences between fields gradually narrowed. By 2020–2024, the clustering coefficients for all five fields had dropped to their lowest values (AI 0.324, CV 0.340, ML 0.377, NLP 0.399, Web\&IR 0.551). Notably, the clustering coefficient for Web\&IR remained significantly higher than that of other fields across all five time periods, likely due to the high correlation of its research tasks and the concentrated nature of industrial demand.

\begin{figure}[H]
    \centering
    \includegraphics[width=\textwidth]{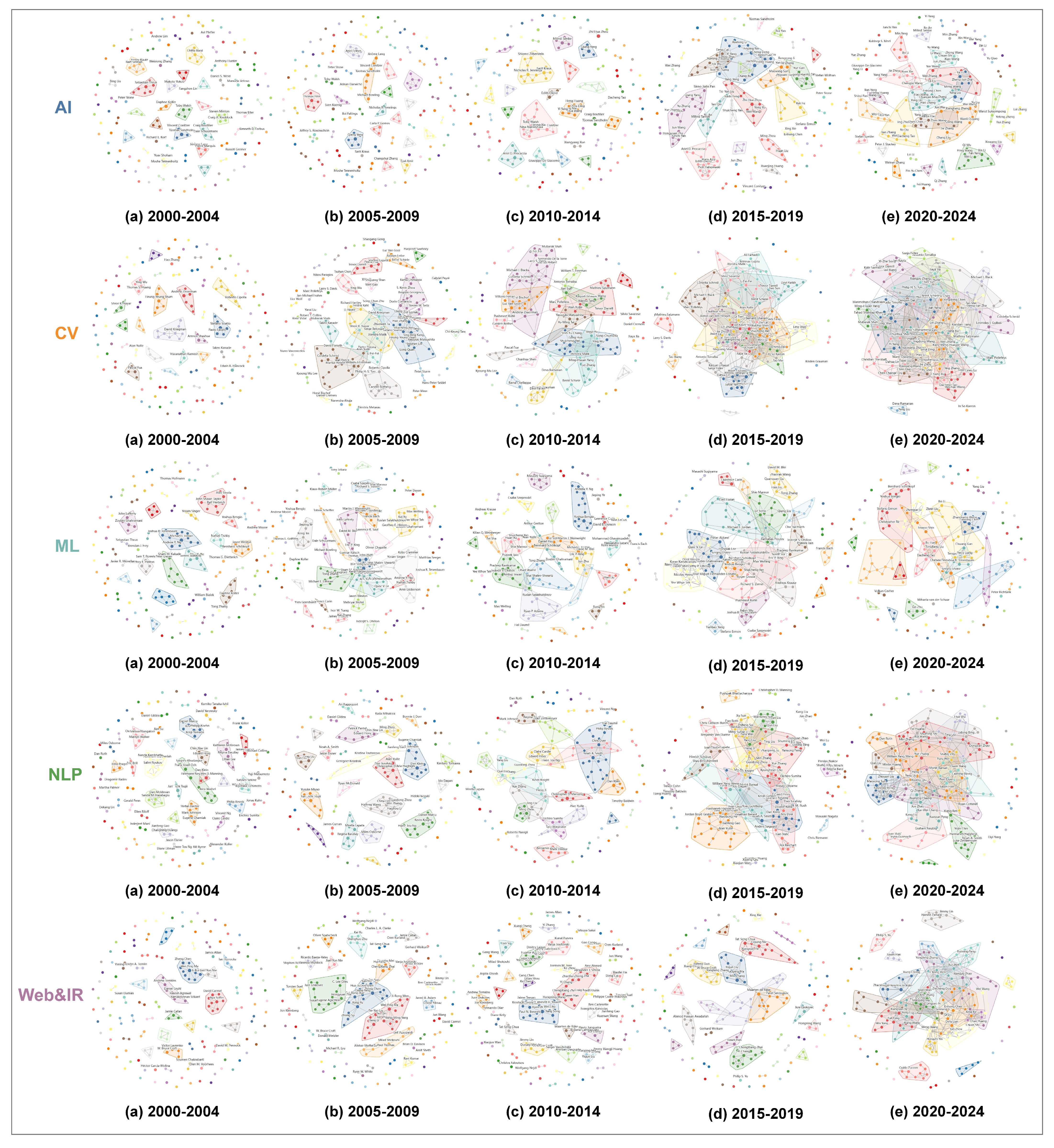}
    \caption{Collaboration Networks (Top-150 Authors) of Each Subfield}
    \label{fig:collab_networks}
\end{figure} 

To further elucidate the structural characteristics reflected by changes in the clustering coefficient, we constructed a collaboration network using the Louvain community detection algorithm and force-directed layout, based on data from the top 150 authors by publication volume in each of the five subfields across different time periods. As shown in Figure \ref{fig:collab_networks}, each cluster represents a group of high-frequency collaborators formed during that period; colors distinguish different collaborative communities, node size maps to author publication volume, and edge thickness reflects collaboration strength. Overall, the collaboration networks across the five subfields all exhibit an evolutionary pattern from sparse to dense, indicating that collaboration patterns have shifted from intra-team collaboration to cross-team collaboration. This result corroborates the trends shown in Figure \ref{fig:top15_industry_collab}. From 2000 to 2004, the collaboration networks in each field were dispersed, with sparse node connections and no core structures yet formed; only Web\&IR exhibited relatively concentrated collaboration clusters, consistent with its highest clustering coefficient. From 2005 to 2009, multiple independent and clearly defined collaborative clusters emerged in CV, ML, and Web\&IR, corresponding to the rise in top-conference submission volumes and the increase in the number of collaborative groups during this period. From 2010 to 2014, the collaborative networks of CV and ML began to exhibit cross-cluster connections, reflecting the initial formation of cross-team collaborative relationships; simultaneously, Web\&IR maintained the highest level of aggregation, with its core structural features becoming increasingly distinct. From 2015 to 2019, all five fields entered a phase of intensive collaboration, with a rapid increase in submissions to top-tier conferences. The cluster concentration in AI and Web\&IR further increased, while peripheral nodes in CV, ML, and NLP gradually established connections with core clusters and integrated into the core structure, significantly enhancing the integration of the collaboration network. From 2020 to 2024, the core cluster structures in each field stabilized, connections between clusters strengthened, and the collaboration network exhibited a continuous character. Among these, the collaboration networks in CV and NLP were particularly complex. We speculate that this change was driven primarily by two factors: first, the rise of interdisciplinary research such as multimodal learning has promoted cross-domain collaboration; second, the expansion of renowned research teams has strengthened connections between fields. For example, the team led by Fei-Fei Li in the CV field and the team led by Michael Collins in the NLP field—whose members, while focusing on their own research, actively collaborate with external research teams—have contributed to the increasing complexity of the collaborative network.

\subsubsection{Stability of Collaboration}\label{sec5_2_5}

\begin{figure}[H]
    \centering
    \includegraphics[width=\textwidth]{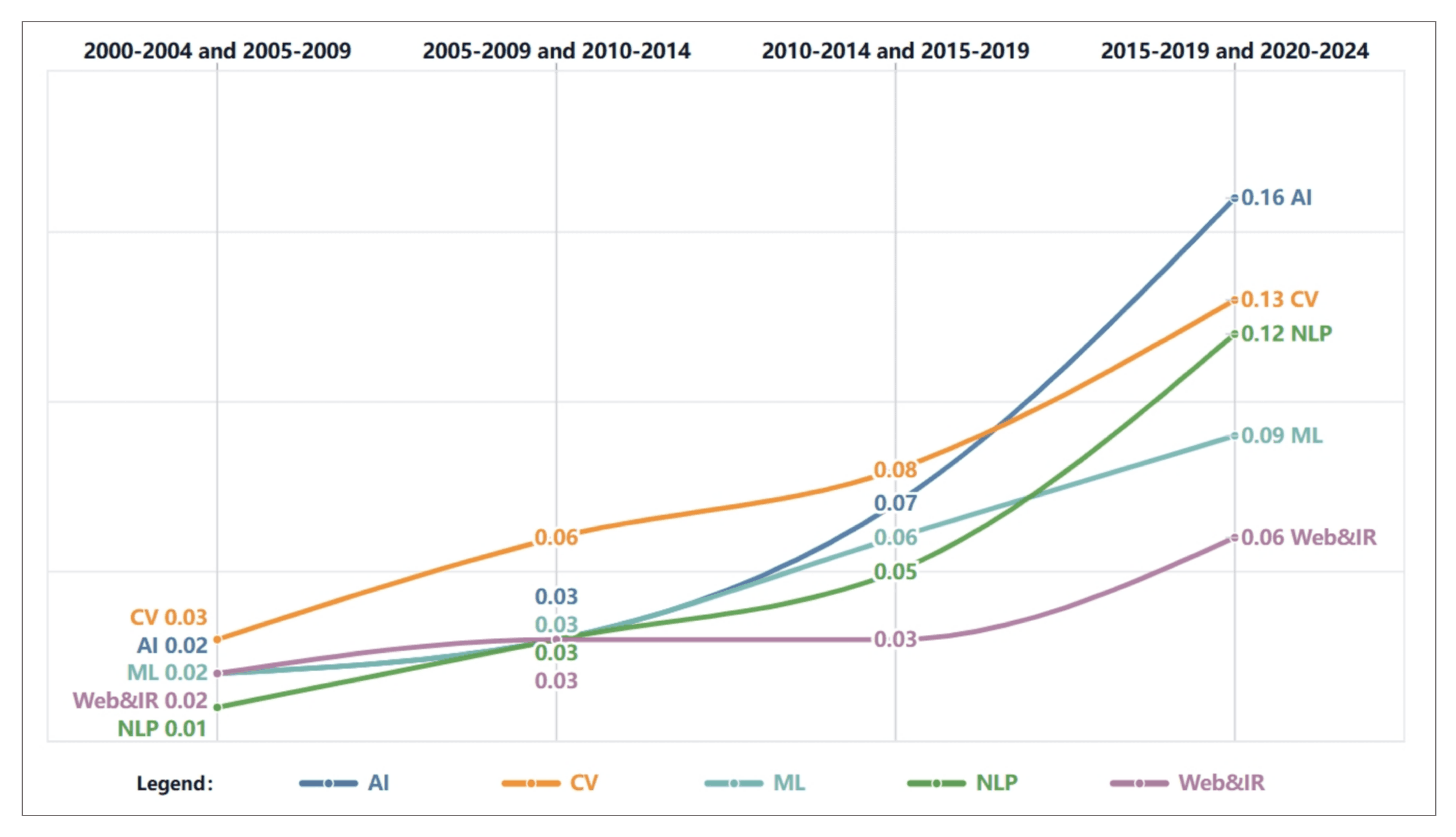}
    \caption{Collaboration Stability Measured by Jaccard Similarity}
    \label{fig:collab_stability}
\end{figure} 

We used the set similarity based on the Jaccard Similarity coefficient defined in Section \ref{sec4_2_2}\cite{bird2009}\cite{ibanez2012} to measure the stability of collaborative relationships across the five subfields by comparing the overlap in the sets of collaborators among authors across adjacent time periods. Figure \ref{fig:collab_stability} shows the changes in Jaccard values for each subfield over time; higher values indicate more stable collaborative relationships. Overall, the Jaccard values for all subfields show an upward trend, suggesting that the stability of collaboration is gradually improving, particularly over the past decade. At the same time, the disparity in stability among the subfields has gradually widened. As shown in Figure \ref{fig:collab_stability}, during the 2000–2004 and 2005–2009 periods, the stability of the five subfields was generally low, with small differences in values, all falling within the range of 0.01–0.03, indicating poor stability in scholarly collaboration across these fields during that time. During the 2005–2009 and 2010–2014 periods, stability improved across all fields, with the CV field rising to 0.06, leading among the five fields. We speculate that this is related to the launch of the ILSVRC competition in 2010, which fostered long-term collaboration among cross-institutional teams, thereby enhancing the stability of collaborative relationships within the field. Between 2010–2014 and 2015–2019, stability improved significantly across all fields except Web\&IR, with Jaccard values reaching 0.07, 0.08, 0.06, and 0.05 for AI, CV, ML, and NLP, respectively. The 2015–2019 and 2020–2024 periods represent the phases with the greatest improvement across all fields within the entire cycle. Among these, AI showed the fastest growth, reaching 0.16, and exhibited the highest level of collaboration stability. We speculate that this is due, on the one hand, to the formation of long-term collaborative mechanisms between large research institutions (such as OpenAI and Google DeepMind) and top universities (such as Stanford University and MIT) in the areas of large models and foundational algorithms, which has enhanced the similarity among collaborative groups; on the other hand, the integration of industrialization with the open-source ecosystem (such as the PyTorch community led by Meta AI) has driven sustained collaboration between academia and industry based on a unified technology stack, thereby enhancing the stability of cross-organizational collaboration.

\subsection{Author Characteristics}\label{sec5_3}

\subsubsection{Author Productivity}\label{sec5_3_1}

\begin{figure}[H]
    \centering
    \includegraphics[width=0.6\textwidth]{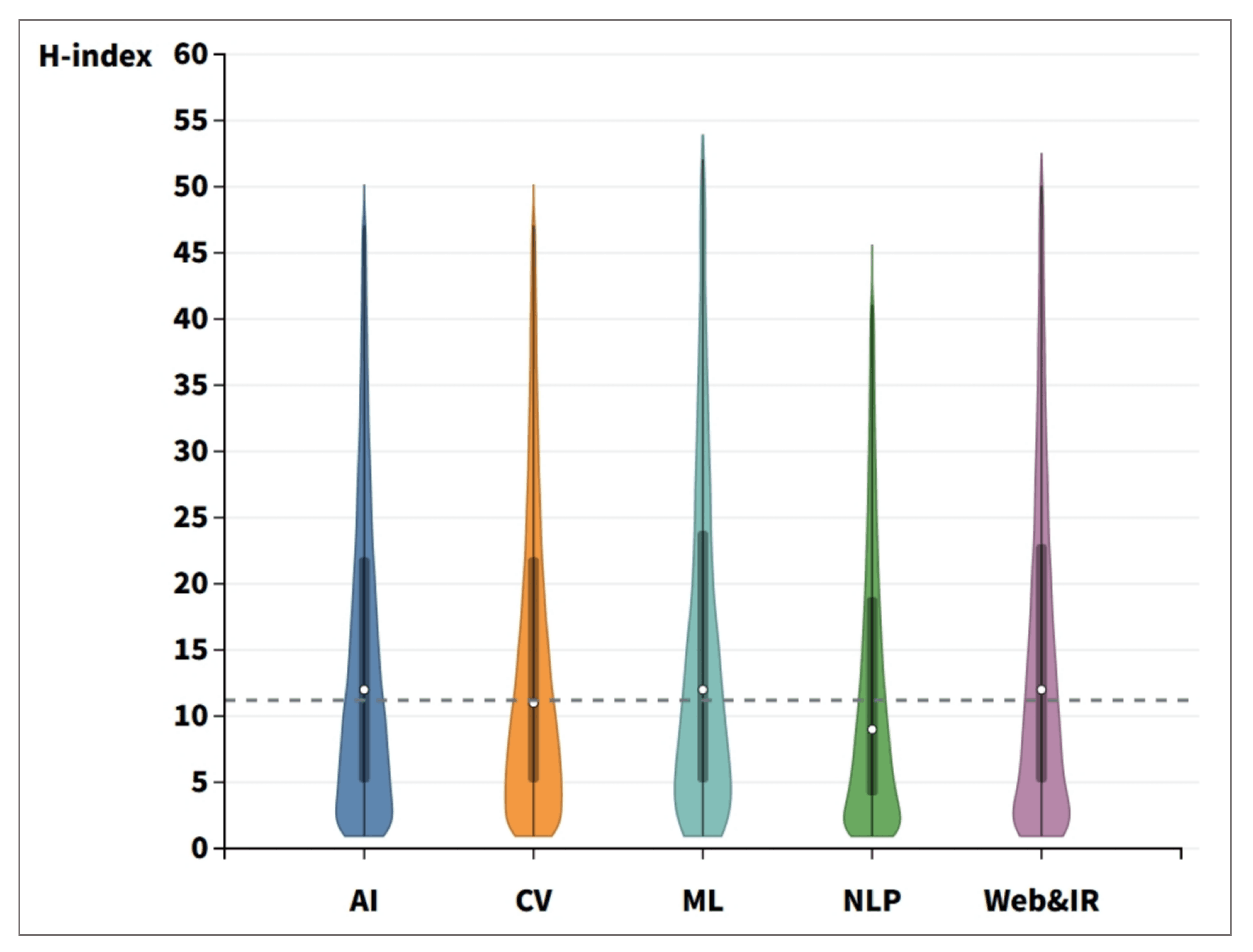}
    \caption{Distribution of Author Productivity}
    \label{fig:author_productivity}
\end{figure}

We use the author productivity metrics from Section \ref{sec4_2_2} and employ the classic H-index to measure researchers’ academic output. Figure \ref{fig:author_productivity} illustrates the H-index distribution across the five subfields, with the gray dashed line representing the average of the medians of the five subfields. Overall, the H-index distribution across the five subfields exhibits a distinct concentration pattern: the H-index of most researchers is low, concentrated in the 0–20 range, with only a small number of researchers reaching an H-index of 50 or higher. This reflects the "head effect" in academic output and influence. The median H-index for all five subfields is around 10, indicating that the overall academic productivity across these fields is at a similar level. Among them, the median for NLP is slightly lower than the overall average, possibly because much of the research in NLP focuses on theoretical depth or niche scenarios, resulting in a narrower citation scope and a relatively longer "research-to-citation" cycle.

\subsubsection{Topic Mobility of Authors}\label{sec5_3_2}

\begin{figure}[H]
    \centering
    \includegraphics[width=0.6\textwidth]{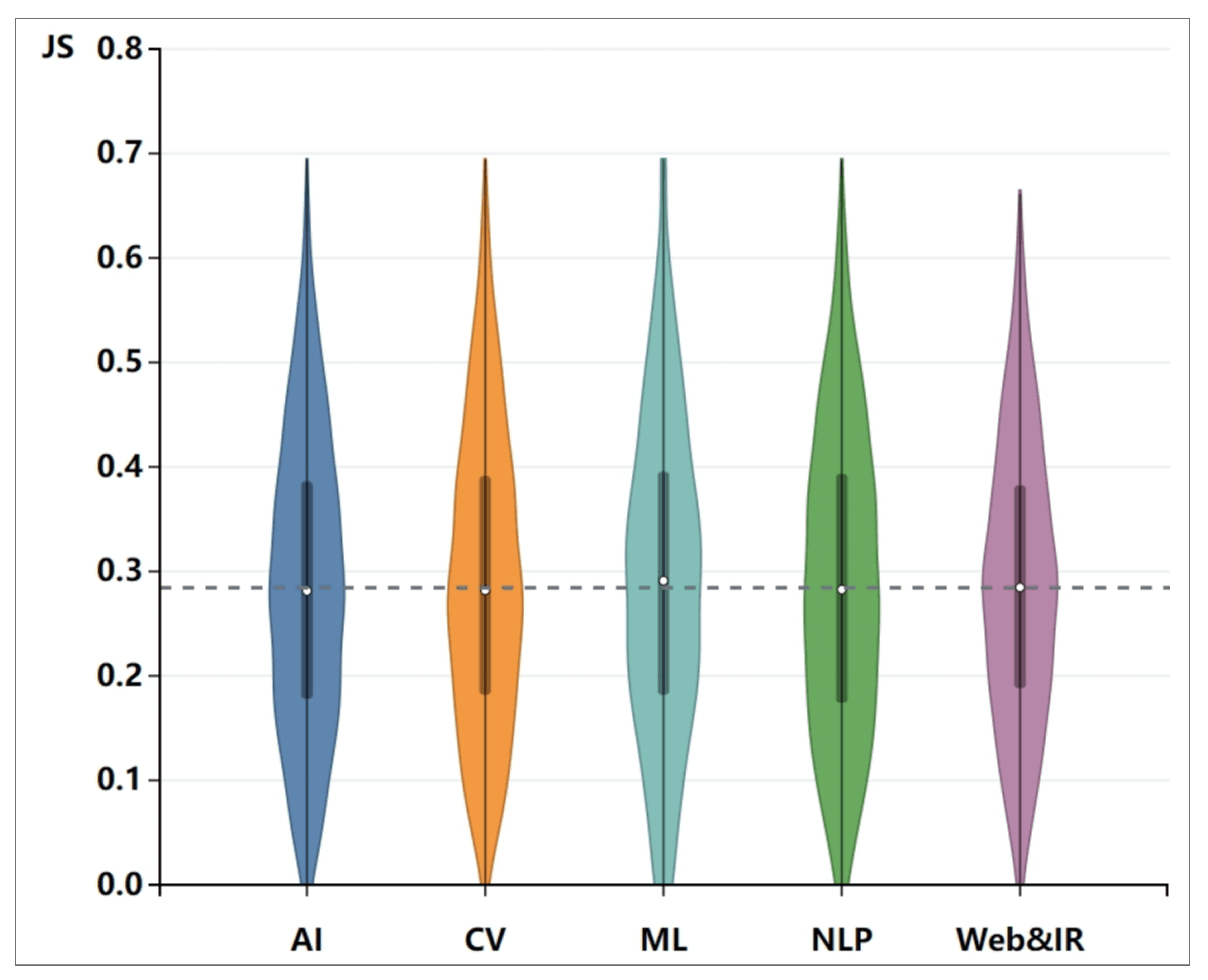}
    \caption{Research Topic Mobility Across Subfields}
    \label{fig:topic_mobility}
\end{figure}

We utilized the author research topic mobility metric described in Section \ref{sec4_2_3} and calculated the frequency of changes in research topics using JS divergence, ultimately determining the research topic mobility across five subfields. A higher mobility value indicates more frequent changes in research topics within a field, reflecting high efficiency in knowledge circulation and significant innovation potential. Figure \ref{fig:topic_mobility} illustrates the distribution of research topic mobility across the fields, with the gray dashed line representing the average of the median values for the five subfields. Overall, the research topic mobility of the five subfields is at a similar level, with medians concentrated in the 0.25–0.30 range, indicating that authors in each field generally maintain stable research topics across adjacent time periods. Notably, the box height for the Web\&IR subfield is slightly lower than that of the other subfields, suggesting a more concentrated data distribution in this subfield. We speculate that this is because researchers in the Web\&IR subfield tend to delve deeply into relatively fixed research directions, whereas research directions in other subfields undergo more diverse iterations or expansions.

\begin{figure}[H]
    \centering
    \includegraphics[width=0.8\textwidth]{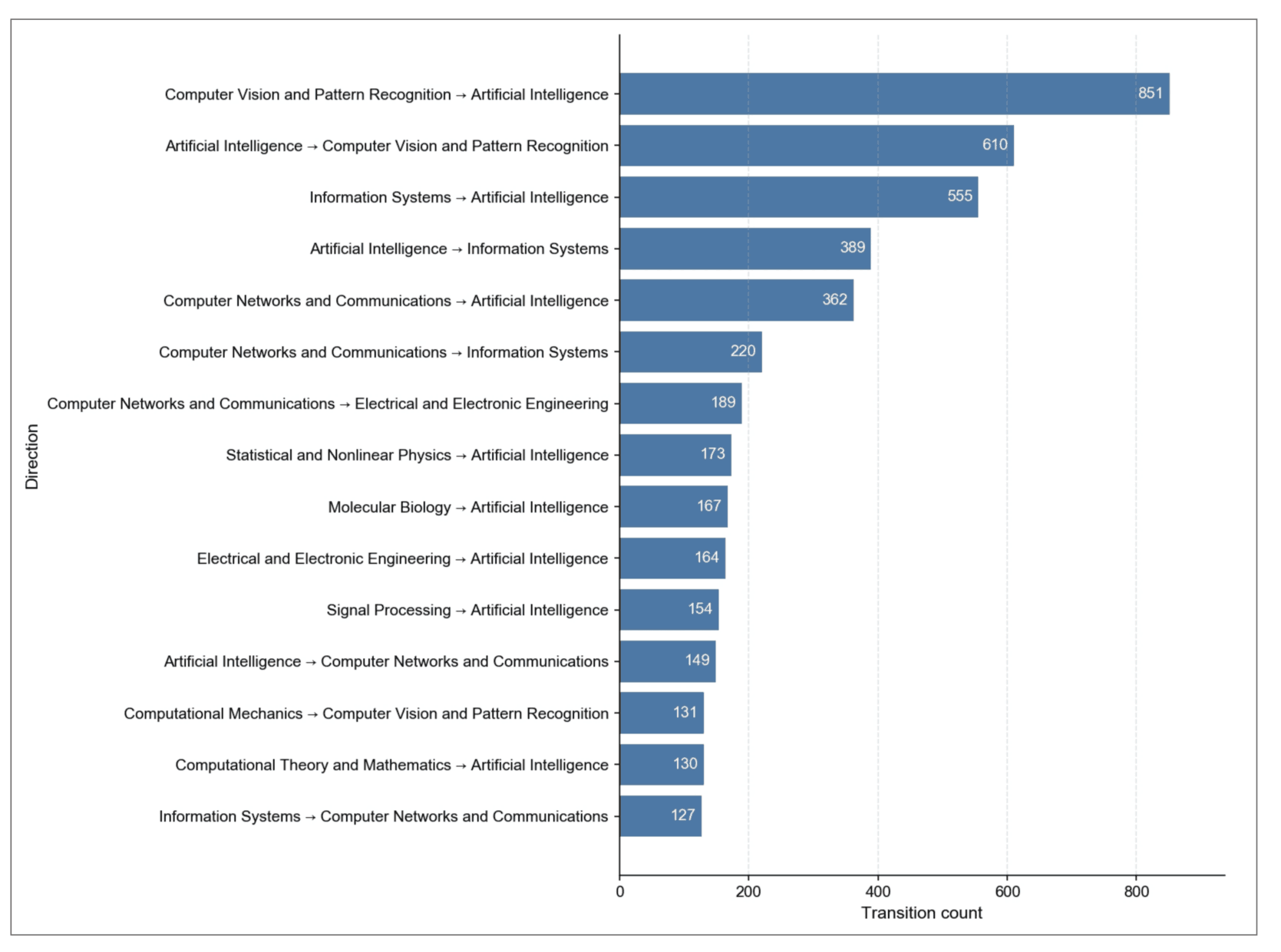}
    \caption{Top 15 Major Migration Directions of Research Topics}
    \label{fig:topic_migration_top15}
\end{figure}

To further examine the characteristics of research topic migration across the five subfields, Figure \ref{fig:topic_migration_top15} illustrates the 15 major migration directions of research topics. As shown in the figure, among the major migration directions, the majority of migrations have AI as their destination, while only three have AI as their origin. The remaining directions include cross-disciplinary pathways such as Computer Networks and Communications, Information Systems, Electrical and Electronic Engineering, Computational Mechanics, and Computer Vision and Pattern Recognition. In terms of bidirectional interaction, the frequency of two-way knowledge transfer between Computer Vision and Pattern Recognition and Artificial Intelligence, as well as between Information Systems and Artificial Intelligence, is notably high. Both fields exhibit a net inflow to AI (+241 and +116, respectively); Computer Networks and Communications also shows a net inflow to AI (+213). This indicates that AI serves as the primary convergence endpoint for interdisciplinary knowledge transfer, with the highest frequency of knowledge flow, while CVPR, IS, and CNC are the three high-frequency channels connecting to AI. At the same time, disciplines such as Statistical and Nonlinear Physics, Molecular Biology, Electrical and Electronic Engineering, and Signal Processing continue to provide a steady influx of research talent to the AI field.


\section{Discussion}\label{sec6}

\subsection{Shared Characteristics Across the Five Subfields}\label{sec6_1}

The five subfields of artificial intelligence show a broadly consistent increasing trend in impact and dissemination.The continuous growth in citation distribution, together with changes in citation velocity and the distribution of highly cited outputs, indicates that artificial intelligence research has entered a stage of high-intensity dissemination. The mode of knowledge dissemination has changed: the time for new findings to enter the core attention of the academic community is shortening, and key methods within subfields can more rapidly become shared reference points. Artificial Intelligence subfields no longer rely primarily on slow and long-term accumulation to build academic impact, but instead exhibit a knowledge evolution pattern characterized by rapid diffusion, absorption, and recombination. This change significantly accelerates impact and dissemination as well as technological application, enabling industry to participate more promptly in knowledge iteration, promoting efficient interaction among academia, industry, and research, and allowing emerging research to be more rapidly transformed into market-oriented technologies, thereby gaining advantages in global competition.

The five subfields of artificial intelligence have, to varying degrees, reduced their reliance on internal knowledge circulation while increasing their absorption of external disciplinary resources. Considering indicators such as self-citation intensity, cross-disciplinary linkages, and collaboration network structures, it can be observed that the advancement of artificial intelligence research is increasingly not the result of continuous accumulation within a closed discipline, but rather achieved through sustained interaction with multiple knowledge systems, including mathematics, statistics, control science, information science, as well as social and life sciences. This change indicates that although the subfields of artificial intelligence have distinct technical cores, their development increasingly depends on the introduction and integration of external knowledge. Artificial intelligence should no longer be understood as several parallel technical branches, but rather as an open knowledge system that continuously absorbs external theories, methods, and problem contexts. By actively engaging in interdisciplinary collaboration, enterprises can rapidly integrate external resources, enhance the adaptability and innovativeness of technologies, and thereby promote the diversification and commercialization of products and services. For example, the integration of artificial intelligence with social sciences can promote the rapid development of applications such as personalized recommendation and intelligent customer service, bringing stronger market competitiveness to enterprises.

The five subfields of artificial intelligence generally exhibit increasing complexity in collaboration networks and diversification of research actors. This change is reflected in increased inter-institutional collaboration, stronger inter-cluster connections, and rising industry participation, indicating that the organization of knowledge production in artificial intelligence is undergoing profound transformation. Traditionally, technological innovation mainly occurred within relatively stable academic communities, but the results of this study show that subfields of artificial intelligence increasingly rely on collaborative interactions among universities, enterprises, research institutions, and international cooperation. The basic unit of knowledge production is no longer an isolated team or single institution, but rather a networked collaborative structure composed of multiple actors. This change explains why certain subfields not only disseminate knowledge more rapidly but also more easily achieve technological spillover and application expansion.This is not because individual technological outputs are inherently stronger, but because the underlying collaboration networks provide more efficient channels for knowledge flow and resource integration, thereby accelerating the marketization of technologies.

Authors across the five subfields of artificial intelligence exhibit similar levels of author productivity, and the research directions in each subfield remain stable.At the same time, artificial intelligence serves as a central convergence node in the cross-disciplinary mobility of authors, exhibiting the highest frequency of knowledge flow and occupying a key position in interdisciplinary knowledge transfer. In particular, Computer Vision and Pattern Recognition (CVPR), Information Systems (IS), and Computer Networks and Communications (CNC) form high-frequency mobility channels closely associated with artificial intelligence, further highlighting its central role in promoting interdisciplinary collaboration. This change provides important implications for talent recruitment and resource allocation. As the influence of artificial intelligence technologies expands, the demand for interdisciplinary talent has increased sharply, especially in the integration of fields such as mathematics, statistics, life sciences, and social sciences.Enterprises and research institutions need to recognize that experts from a single discipline are no longer sufficient to support future development, and interdisciplinary talent will become key to competitiveness.Therefore, talent recruitment policies should place greater emphasis on experts with multidisciplinary backgrounds, especially innovative individuals who can promote collaboration across fields.

\subsection{Distinctive Features Across the Five Subfields}\label{sec6_2}

\subsubsection{AI: From Academic Breakthroughs to Rapid Interdisciplinary Integration}\label{sec6_2_1}

\textbf{1. Growth in Academic Impact and Tight Collaboration Network Structure}

The academic impact of AI has continued to grow, especially after breakthroughs in deep learning, showing a significant upward trend.According to the analysis in Section \ref{sec4_2_1}, although the H-index and citation velocity of artificial intelligence remain lower than those of fields such as CV, its academic impact is growing rapidly.From early symbolic reasoning to data-driven deep learning, the technological evolution of AI has enabled it to become a foundational pillar across multiple domains.Especially after the breakthrough of deep learning in 2012, AI has gradually formed a broad academic impact network.As discussed in Section \ref{sec4_2_1}, compared with other subfields, the technological breakthroughs of AI are not confined to computer science but also incorporate knowledge and techniques from fields such as sociology and the life sciences. This change indicates that AI is shifting from a research model based on isolated technological breakthroughs to one characterized by interdisciplinary integration, further expanding its application scope, especially in emerging areas such as smart healthcare and smart cities.

\textbf{2. Stability and Tightness of Collaboration Networks}

The stability of collaboration and the tightness of collaboration in AI have continuously increased in recent years (see Section \ref{sec4_2_1}). Compared with other fields, collaboration in AI shows not only a stable growth trend but also increasing tightness. Collaboration in AI is no longer characterized by loose academic exchanges, but has gradually formed a highly organized, interdisciplinary, multi-actor collaborative structure.This collaboration model not only accelerates technological innovation but also promotes the marketization of technologies. The rapid translation and widespread application of AI technologies are direct outcomes of this model.

\subsubsection{CV: Significant Academic Impact and Task-Oriented Interdisciplinarity}\label{sec6_2_2}

\textbf{1. Continuous Strengthening of Academic Impact and Its Path Continuity Mechanism}

CV exhibits the strongest academic impact among the five subfields, leading in H-index, citation distribution, and citation velocity. Over the past decade, CV has maintained consistently high citation accumulation and rapid dissemination, reflecting sustained attention and fast knowledge diffusion within the academic community. This advantage is closely related to its path-dependent technological paradigm, in which new methods are typically developed through incremental improvements on existing model architectures. Such an evolutionary pattern enhances comparability and reusability across studies, reduces the cost of understanding and applying new methods, and ultimately facilitates rapid knowledge dissemination and citation accumulation.

\textbf{2. Task-oriented and Uneven Structural Characteristics of Interdisciplinarity}

CV demonstrates a clear task-oriented pattern in its interdisciplinary connections. According to the Sankey diagram analysis in Figure~\ref{fig:sankey}, CV maintains multiple linkages with external disciplines; however, the strength of these connections is uneven, with stronger ties to vision-related and perception-oriented fields. This suggests that knowledge diffusion in CV is not uniformly distributed but instead concentrated around specific tasks and application scenarios. In these domains, the alignment between research problems and practical needs is relatively clear, enabling continuous methodological refinement and efficient application. Consequently, the development of CV relies not only on internal technical accumulation but also on sustained interaction with related disciplines.

\subsubsection{ML: Dispersed Industry Collaboration and Concentrated International Collaboration}\label{sec6_2_3}

\textbf{1. Dispersed Industry Collaboration and Its Constraints on Application Transformation}

According to Section \ref{sec4_2_2}, the proportion of collaboration between ML and industry has declined in recent years, indicating a slowdown in technology transfer. Meanwhile, Section \ref{sec4_2_2} shows that ML collaboration networks exhibit relatively low tightness, reflecting dispersed collaboration relationships and the absence of a highly coordinated structure. This pattern is partly associated with the resource-intensive nature of large-scale model development: companies such as Google and DeepMind possess substantial computational and data resources and tend to conduct intensive research internally, whereas universities face resource constraints, limiting their participation in such collaborations. In addition, as shown in Section \ref{sec4_2_1}, ML interdisciplinarity is primarily concentrated in foundational disciplines such as statistics, information theory, and control theory, with knowledge production focusing more on methodological and model development rather than direct application contexts.

\textbf{2. Concentration of International Collaboration and National Differences}

According to the chord diagram analysis in Section \ref{sec4_2_2}, ML international collaboration exhibits clear structural concentration. In recent years, the proportion of cross-national collaboration has declined, while collaboration ties have become increasingly concentrated among a limited number of countries, with the United States occupying a central position. Notably, although China demonstrates a high level of international collaboration in the broader artificial intelligence domain, its collaboration intensity with the United States in ML is comparatively lower than that observed between the United States and countries such as Germany or the United Kingdom. This pattern may be related to the high requirements of ML research in terms of computational resources and technical capacity, which tend to concentrate collaboration within countries possessing strong research infrastructures.

\subsubsection{NLP: Stable Structural Characteristics and Enterprise-Led Collaboration}\label{sec6_2_4}

\textbf{1. Stable Academic Impact and Interdisciplinarity}

Compared with other subfields, NLP does not exhibit particularly prominent academic impact, and its knowledge diffusion process remains relatively stable. Structurally, knowledge flows in NLP are primarily concentrated on language-related problems, without forming a broad pattern of interdisciplinary expansion. This suggests that NLP development is characterized more by the deepening of existing research frameworks than by extensive outward expansion. At the same time, NLP methods have been widely adopted in fields such as the social sciences for tasks including text analysis and knowledge discovery.

\textbf{2. Enterprise-led International and Industry Collaboration}

Compared with CV and ML, NLP international collaboration is more geographically dispersed but remains primarily concentrated in Europe and the United States. Industrial collaboration analysis shows that institutions such as Google Research dominate in both collaboration frequency and influence, reflecting their advantages in computational resources and large-scale data. Furthermore, compared with other subfields, technological iteration in NLP relies more heavily on large-scale pretraining practices led by industry.

\subsubsection{Web\&IR: Industry-Led Application Orientation and Concentrated, Stable Interdisciplinary 
Collaboration}\label{sec6_2_5}

\textbf{1. Industry-led Collaboration Concentration and Application Orientation}

Web\&IR exhibits the highest level of industry participation among the five subfields, indicating a strong linkage between research activities and application scenarios. As shown in Section \ref{sec4_2_2}, the field is largely dominated by major technology companies, with institutions such as Google Research leading in collaboration frequency, reflecting a high concentration of research resources and technical capacity. In addition, Web\&IR maintains close connections with CV and ML and continues to expand into large-scale application scenarios supported by deep learning technologies. In terms of research orientation, the field has gradually shifted from academic exploration toward application-driven domains, including search engine optimization, recommendation systems, and user behavior analysis, closely aligned with industry demands.

\textbf{2. Concentration and High Stability of Interdisciplinary Collaboration}

Interdisciplinary collaboration in Web\&IR is primarily based on the integration of computer science, data science, information systems, and user experience. Compared with CV and ML, its interdisciplinary connections are more concentrated within computer science and information systems, with strong linkages to statistics, artificial intelligence, and optimization. These interactions facilitate innovation in areas such as algorithm optimization, user interaction, and large-scale information processing. Moreover, as indicated in Section \ref{sec4_2_2}, Web\&IR exhibits high collaboration stability: once a research team establishes a strong position, it tends to maintain long-term influence, thereby providing a stable foundation for sustained technological advancement.

\section{Limitations and Future Work}\label{sec7}

\subsection{Data Sources}\label{sec7_1}

Although OpenAlex is a leading global academic platform widely used in bibliometric research, it still presents certain limitations in data processing and classification.

\textbf{1. Author Disambiguation}

OpenAlex assigns unique identifiers to authors using algorithmic approaches based on names, publication records, and citation relationships\footnote{https://help.openalex.org/hc/en-us/articles/24347048891543-Author-disambiguation?}. Previous studies have shown that automatic disambiguation may introduce errors, particularly for common names, name variants, and cases lacking ORCID identifiers, with this issue being especially pronounced for Chinese names. For instance, the author \textit{Xiaohong Chen} in OpenAlex is associated with an H-index of 254 and 19,030 publications, accompanied by heterogeneous aliases and institutional affiliations, which may indicate potential disambiguation inaccuracies.
It should be noted that similar issues also exist in widely used databases such as Web of Science and Scopus. For the purposes of this study, OpenAlex remains suitable as a unified data source. Potential biases from disambiguation were minimized during indicator selection, with only a limited impact on metrics such as H-index and topic mobility. In our sample of 111,329 authors, over 88\% have H-index $\leq 29$, whereas only 0.31\% and 0.01\% exceed thresholds of 100 and 200, respectively, indicating that extreme cases are rare. Therefore, the overall impact of disambiguation errors on the statistical results and comparative conclusions of this study is limited.

\textbf{2. Citation Counting Differences}

OpenAlex primarily counts citations from formally published literature and treats preprints, conference papers, and journal versions as separate records. In contrast, platforms such as Google Scholar tend to aggregate citations across multiple versions, leading to substantial differences in citation counts. For example, the paper \textit{"Attention Is All You Need"} has 206,819 citations in Google Scholar but only 6,466 in OpenAlex.
When multiple versions of the same work are recorded separately, citations may be distributed across different entries, potentially affecting the absolute citation count of individual papers. As this study relies on a consistent data source and counting scheme, such discrepancies mainly influence absolute values and may lead to underestimation. However, under a unified data standard, these biases do not significantly affect relative rankings or the overall comparative conclusions. Future research may integrate multi-source citation data and improve cross-version matching to enhance data completeness and accuracy.

\subsection{Analytical Dimensions}\label{sec7_2}

This study constructs an analytical framework based on Impact and Dissemination, Collaboration Characteristics, and Author Characteristics to characterize the development of artificial intelligence subfields. The dimension of research content is also essential for understanding the internal knowledge structure and evolutionary dynamics of a field. Compared with external indicators, content-based analysis can reveal topic diversity, overlap, innovation patterns, rates of change, and the diffusion of scientific concepts across subfields.However, there is currently a lack of mature and robust indicators suitable for large-scale cross-subfield comparison. Although recent advances in topic modeling and semantic embedding methods have improved analytical capabilities, these approaches often suffer from limited consistency and interpretability, particularly when applied across domains with substantial semantic differences (Cheng et al., 2023)\cite{Cheng2023SurveyTopicModels}. In addition, a unified framework for cross-domain semantic comparison has yet to be established, which limits reproducibility and standardization (Wang, 2024)\cite{Wang2024DualPerspectiveFramework}.Therefore, this study does not incorporate the research content dimension into the current analytical framework but instead treats it as an important direction for future work. Future studies may incorporate more robust content-based indicators to provide a more comprehensive understanding of the evolutionary characteristics of artificial intelligence subfields.

\subsection{Selection and Application of Indicators}\label{sec7_3}

In analyzing artificial intelligence subfields, this study prioritizes widely validated and comparable bibliometric indicators. For measuring interdisciplinarity, Shannon entropy $H(p)$ is employed to quantify the diversity and balance of disciplinary sources. This metric has been extensively applied in scientometric research and provides a stable representation of knowledge dispersion. Although distance-sensitive measures such as the Rao–Stirling index can capture cognitive differences between disciplines and improve resolution, Shannon entropy is sufficient for the large-scale comparative analysis conducted in this study. More complex indicators are therefore left for future exploration.For collaboration analysis, the weighted clustering coefficient is adopted to capture the frequency and closure of local collaboration structures. Edge weights are primarily defined based on co-authorship frequency, which effectively reflects fundamental collaboration patterns. Future research may incorporate additional dimensions—such as project types (e.g., large-scale programs versus basic research) and partner influence (e.g., institutional reputation or academic standing)—to refine the weighting of collaboration networks and provide a more nuanced characterization of scientific collaboration.


\section{Conclusion}\label{sec8}

This study examines five major subfields of artificial intelligence and analyzes their development from 2000 to 2024 across three dimensions: Impact and Dissemination, Collaboration Characteristics, and Author Characteristics. By integrating bibliometric indicators with visualization techniques---this study constructs a multidimensional analytical framework that enables systematic cross-subfield and longitudinal comparison. In doing so, it extends beyond the traditional scope of bibliometric research focused on single domains and provides a flexible and reproducible approach for analyzing interdisciplinary collaboration, global collaboration dynamics, and author-level characteristics.

The findings indicate that, over the past two decades, all five subfields have entered a stage of intensified knowledge diffusion, characterized by increased academic impact, accelerated dissemination, and a shortened diffusion cycle, alongside a shift from internally driven knowledge accumulation toward greater reliance on external disciplinary resources. Concurrently, knowledge production in each subfield has evolved from a closed, discipline-specific process into an open network structure involving interdisciplinary and multi-actor collaboration, resulting in increasingly complex collaboration networks and more diversified research participation. Despite these shared trends, substantial structural differentiation is observed across subfields: CV occupies a leading position in academic influence and shows significant advantages in indicators such as H-index, citation velocity, and citation distribution, while also presenting a concentrated development path around specific visual perception tasks. ML shows the characteristics of shrinking industrial collaboration and concentrated international collaboration, with a declining proportion of industry collaboration, a narrowing scope of cross-national collaboration, and collaboration relationships concentrated in a few countries, resulting in a relatively dispersed overall collaboration structure. Web\&IR shows the strongest industry-dominated feature, with its collaboration actors concentrated in large technology companies and forming a stable collaborative network structure. In contrast, AI as a whole is characterized by sustained growth, whereas NLP remains comparatively stable across multiple dimensions.

From a structural perspective, this study reveals that the evolution of artificial intelligence has transitioned from a pattern of unified diffusion to one of structural differentiation, in which subfields develop through distinct configurations of knowledge organization and collaboration. This finding highlights the intrinsic heterogeneity of artificial intelligence and suggests that rapidly evolving technological domains should be understood as complex systems shaped by multiple interacting structural mechanisms. By bridging bibliometric analysis and the computer science community, this work also fosters cross-disciplinary dialogue, laying a foundation for comparative and mechanism-oriented research on emerging technologies.
\begin{appendices}

\section{Supplementary tables}\label{sec:appendix_tables}

\begin{landscape}
\scriptsize
\setlength{\tabcolsep}{4pt}

\begin{longtable}{p{6cm}p{2.5cm}p{1cm}p{2.8cm}p{1.2cm}p{1.2cm}p{1.0cm}p{1.4cm}}
\caption{Top Five Fastest-Spreading Articles Within 5 Years (25 Citations)}\label{tab:fastest_spreading} \\
\toprule
\textbf{Title} & \textbf{Authors} & \textbf{Year} & \textbf{Inst-1} & \textbf{V25} & \textbf{Subfield} & \textbf{Days} & \textbf{Citations} \\
\midrule
\endfirsthead

\multicolumn{8}{c}{\tablename\ \thetable\ (continued)} \\
\toprule
\textbf{Title} & \textbf{Authors} & \textbf{Year} & \textbf{Inst-1} & \textbf{V25} & \textbf{Subfield} & \textbf{Days} & \textbf{Citations} \\
\midrule
\endhead

\bottomrule
\multicolumn{8}{p{18cm}}{\footnotesize \textit{Note}. First author only (et al.). Inst-1 = first author’s primary affiliation ("University" $\rightarrow$ Univ.).} \\
\endfoot

Random Erasing Data Augmentation & Zhun Zhong, et al. & 2020 & Xiamen Univ. & 154.767 & AI & 59 & 25 \\
Transformers in Time Series: A Survey & Qingsong Wen, et al.& 2023 & Alibaba Group & 140.481 & AI & 65 & 25 \\
Are Transformers Effective for Time Series Forecasting? & Ailing Zeng, et al. & 2023 & Chinese Univ.of Hong Kong & 123.395 & AI & 74 & 25 \\
Regularized Evolution for Image Classifier Architecture Search & Esteban Real, et al. & 2019 & Google & 120.148 & AI & 76 & 25 \\
T2I-Adapter: Learning Adapters to Dig Out More Controllable Ability for Text-to-Image Diffusion Models & Chong Mou, et al. & 2024 & Tencent & 108.705 & AI & 84 & 25 \\

CSWin Transformer: A General Vision Transformer Backbone with Cross-Shaped Windows & Xiaoyi Dong, et al. & 2022 & Univ.of Science and Technology of China & 9131.25 & CV & 1 & 25 \\
Delving Deep into Rectifiers: Surpassing Human-Level Performance on ImageNet Classification & Kaiming He, et al. & 2015 & Microsoft & 652.232 & CV & 14 & 25 \\
Pyramid Vision Transformer: A Versatile Backbone for Dense Prediction without Convolutions & Wenhai Wang, et al. & 2021 & Nanjing Univ. & 570.703 & CV & 16 & 25 \\
Photo-Realistic Single Image Super-Resolution Using a Generative Adversarial Network & Christian Ledig, et al. & 2017 & F{\i}rat Univ. & 537.132 & CV & 17 & 25 \\
Masked Autoencoders Are Scalable Vision Learners & Kaiming He, et al. & 2022 & Meta & 480.592 & CV & 19 & 25 \\

SNAS: Stochastic Neural Architecture Search & Sirui Xie, et al. & 2018 & N/A & 1141.406 & ML & 8 & 25 \\
GAN (Generative Adversarial Nets) & Ian Goodfellow, et al. & 2017 & Univ.de Montr{\'e}al & 537.132 & ML & 17 & 25 \\
ImageNet Classification with Deep Convolutional Neural Networks & Alex Krizhevsky, et al. & 2017 & Google & 456.562 & ML & 20 & 25 \\
Neural Word Embedding as Implicit Matrix Factorization & Omer Levy, et al. & 2014 & Bar-Ilan Univ. & 380.469 & ML & 24 & 25 \\
Learning Deep Features for Scene Recognition Using Places Database & Bolei Zhou, et al. & 2014 & Massachusetts Institute of Technology & 380.469 & ML & 24 & 25 \\

FNet: Mixing Tokens with Fourier Transforms & James Lee-Thorp, et al. & 2022 & Google & 1141.406 & NLP & 8 & 25 \\
The CoNLL 2007 Shared Task on Dependency Parsing & Joakim Nivre, et al. & 2007 & Empirical Methods in Natural Language Processing & 294.556 & NLP & 31 & 25 \\
Enhanced LSTM for Natural Language Inference & Qian Chen, et al. & 2017 & Univ. of Science and Technology of China & 130.446 & NLP & 70 & 25 \\
ERNIE: Enhanced Language Representation with Informative Entities & Zhengyan Zhang, et al. & 2019 & Beijing Academy of AI & 100.343 & NLP & 91 & 25 \\
Recurrent Continuous Translation Models & Nal Kalchbrenner, et al. & 2013 & Univ. of Oxford & 99.253 & NLP & 92 & 25 \\

Xirql & Norbert Fuhr, et al. & 2001 & TU Dortmund Univ. & 74.846 & Web\&IR & 122 & 25 \\
LightGCN & Xiangnan He, et al. & 2020 & Univ. of Science and Technology of China & 74.238 & Web\&IR & 123 & 25 \\
Neural Graph Collaborative Filtering & Xiang Wang, et al. & 2019 & National Univ. of Singapore & 67.142 & Web\&IR & 136 & 25 \\
Personalizing Search via Automated Analysis of Interests and Activities & Jaime Teevan, et al. & 2005 & Massachusetts Institute of Technology & 65.692 & Web\&IR & 139 & 25 \\
Proceedings of the 47th International ACM SIGIR Conference on Research and Development in Information Retrieval & N/A & 2024 & N/A & 64.305 & Web\&IR & 142 & 25 \\

\end{longtable}
\end{landscape}
\normalsize

\begin{landscape}
\scriptsize
\setlength{\tabcolsep}{4pt}

\begin{longtable}{p{6cm}p{2.5cm}p{1cm}p{3.5cm}p{1.5cm}p{1.0cm}p{1.5cm}}
\caption{Top Five Fastest-Spreading Articles Within 5 Years (100 Citations)}\label{tab:fastest_spreading_100} \\
\toprule
\textbf{Title} & \textbf{Authors} & \textbf{Year} & \textbf{Inst-1} & \textbf{Subfield} & \textbf{Days} & \textbf{Citations} \\
\midrule
\endfirsthead

\multicolumn{7}{c}{\tablename\ \thetable\ (continued)} \\
\toprule
\textbf{Title} & \textbf{Authors} & \textbf{Year} & \textbf{Inst-1} & \textbf{Subfield} & \textbf{Days} & \textbf{Citations} \\
\midrule
\endhead

\bottomrule
\multicolumn{7}{p{17cm}}{\footnotesize \textit{Note}. First author only (et al.). Inst-1 = first author’s primary affiliation ("University" $\rightarrow$ Univ.).} \\
\endfoot

T2I-Adapter: Learning Adapters to Dig Out More Controllable Ability for Text-to-Image Diffusion Models & Chong Mou, et al. & 2024 & Tencent & AI & 161 & 100 \\
Regularized Evolution for Image Classifier Architecture Search & Esteban Real, et al. & 2019 & Google & AI & 168 & 100 \\
Are Transformers Effective for Time Series Forecasting? & Ailing Zeng, et al. & 2023 & Chinese Univ. of Hong Kong & AI & 168 & 100 \\
Random Erasing Data Augmentation & Zhun Zhong, et al. & 2020 & Xiamen Univ. & AI & 212 & 100 \\
Inception-v4, Inception-ResNet and the Impact of Residual Connections on Learning & Christian Szegedy, et al. & 2017 & Google & AI & 231 & 100 \\

YOLOv7: Trainable Bag-of-Freebies Sets New State-of-the-Art for Real-Time Object Detectors & Chien-Yao Wang, et al. & 2023 & Institute of Information Science & CV & 15 & 100 \\
Delving Deep into Rectifiers: Surpassing Human-Level Performance on ImageNet Classification & Kaiming He, et al. & 2015 & Microsoft & CV & 31 & 100 \\
Mask R-CNN & Kaiming He, et al. & 2017 & Meta (Israel) & CV & 92 & 100 \\
Unpaired Image-to-Image Translation Using Cycle-Consistent Adversarial Networks & Jun-Yan Zhu, et al. & 2017 & Berkeley College & CV & 92 & 100 \\
Least Squares Generative Adversarial Networks & Xudong Mao, et al. & 2017 & City Univ. of Hong Kong & CV & 92 & 100 \\
Faster R-CNN: Towards Real-Time Object Detection with Region Proposal Networks & Shaoqing Ren, et al. & 2016 & Univ. of Science and Technology of China & ML & 30 & 100 \\
GAN (Generative Adversarial Nets) & Ian Goodfellow, et al. & 2017 & Univ. de Montr{\'e}al & ML & 74 & 100 \\
ImageNet Classification with Deep Convolutional Neural Networks & Alex Krizhevsky, et al. & 2017 & Google & ML & 74 & 100 \\
Faster R-CNN: Towards Real-Time Object Detection with Region Proposal Networks & Shaoqing Ren, et al. & 2015 & Microsoft Research & ML & 120 & 100 \\
Conditional Image Synthesis with Auxiliary Classifier GANs & Augustus Odena, et al. & 2017 & Google & ML & 148 & 100 \\

TinyBERT: Distilling BERT for Natural Language Understanding & Xiaoqi Jiao, et al. & 2020 & Wuhan National Laboratory for Optoelectronics & NLP & 87 & 100 \\
Recursive Deep Models for Semantic Compositionality Over a Sentiment Treebank & Richard Socher, et al. & 2013 & Stanford Univ. & NLP & 135 & 100 \\
fairseq: A Fast, Extensible Toolkit for Sequence Modeling & Myle Ott, et al. & 2019 & N/A & NLP & 136 & 100 \\
Is ChatGPT a General-Purpose Natural Language Processing Task Solver? & Chengwei Qin, et al. & 2023 & Nanyang Technological Univ. & NLP & 178 & 100 \\
Transformer-XL: Attentive Language Models beyond a Fixed-Length Context & Zihang Dai, et al. & 2019 & Carnegie Mellon Univ. & NLP & 209 & 100 \\

LightGCN & Xiangnan He, et al. & 2020 & Univ. of Science and Technology of China & Web\&IR & 160 & 100 \\
What is Twitter, a Social Network or a News Media? & Haewoon Kwak, et al. & 2010 & Korea Advanced Institute of Science and Technology & Web\&IR & 250 & 100 \\
Neural Graph Collaborative Filtering & Xiang Wang, et al. & 2019 & National Univ. of Singapore & Web\&IR & 258 & 100 \\
Heterogeneous Graph Attention Network & Xiao Wang, et al. & 2019 & Beijing Univ. of Posts and Telecommunications & Web\&IR & 268 & 100 \\
Neural Collaborative Filtering & Xiangnan He, et al. & 2017 & National Univ. of Singapore & Web\&IR & 273 & 100 \\

\end{longtable}
\end{landscape}
\normalsize

\begin{landscape}
\scriptsize
\setlength{\tabcolsep}{4pt}

\begin{longtable}{p{6cm}p{3cm}p{1cm}p{4cm}p{1.5cm}p{1.5cm}}
\caption{Top Five Papers by International Collaboration Proportion of Each Subfield}\label{tab:intl_collab_top5} \\
\toprule
\textbf{Title} & \textbf{Authors} & \textbf{Year} & \textbf{Inst-1} & \textbf{Subfield} & \textbf{Countries} \\
\midrule
\endfirsthead

\multicolumn{6}{c}{\tablename\ \thetable\ (continued)} \\
\toprule
\textbf{Title} & \textbf{Authors} & \textbf{Year} & \textbf{Inst-1} & \textbf{Subfield} & \textbf{Countries} \\
\midrule
\endhead

\bottomrule
\endfoot

The Price of Justified Representation & Edith Elkind, et al. & 2022 & University of Oxford & AI & 6 \\
The Parameterized Complexity of Global Constraints & Christian Bessi{\`e}re, et al. & 2008 & Laboratoire d'Informatique & AI & 6 \\
LCD: Learned Cross-Domain Descriptors for 2D--3D Matching & Quang-Hieu Pham, et al. & 2020 & Singapore University of Technology and Design & AI & 5 \\
Backdoors into Heterogeneous Classes of SAT and CSP & Serge Gaspers, et al. & 2014 & UNSW Sydney & AI & 5 \\
Egalitarian Committee Scoring Rules & Haris Aziz, et al. & 2018 & Data61 & AI & 5 \\

JSIS3D: Joint Semantic-Instance Segmentation of 3D Point Clouds With Multi-Task Pointwise Networks and Multi-Value Conditional Random Fields & Quang-Hieu Pham, et al. & 2019 & Singapore University of Technology and Design & CV & 5 \\
ERA: Enhanced Rational Activations & Martin Trimmel, et al. & 2022 & Lund University & CV & 4 \\
Geometric Visual Similarity Learning in 3D Medical Image Self-Supervised Pre-training & Yuting He, et al. & 2023 & Nanjing University of Aeronautics and Astronautics & CV & 4 \\
Learning High-Order Filters for Efficient Blind Deconvolution of Document Photographs & Lei Xiao, et al. & 2016 & King Abdullah University of Science and Technology & CV & 4 \\
The Light-Path Less Traveled & Srikumar Ramalingam, et al. & 2011 & Mitsubishi Electric (United States) & CV & 4 \\

Fast and Scalable Bayesian Deep Learning by Weight-Perturbation in Adam & Mohammad Emtiyaz Khan, et al. & 2018 & {\'E}cole Polytechnique F{\'e}d{\'e}rale de Lausanne & ML & 6 \\
Amortised MAP Inference for Image Super-resolution & Casper Kaae S{\o}nderby, et al.& 2016 & University of Copenhagen & ML & 5 \\
Self-supervised GAN: Analysis and Improvement with Multi-class Minimax Game & Ngoc-Trung Tran, et al. & 2019 & T{\'e}l{\'e}com Paris & ML & 4 \\
SDNA: Stochastic Dual Newton Ascent for Empirical Risk Minimization & Zheng Qu, et al. & 2015 & University of Hong Kong & ML & 4 \\

KONG: Kernels for Ordered-Neighborhood Graphs & Moez Draief, et al. & 2018 & Huawei Technologies (China) & ML & 4 \\

IndoNLI: A Natural Language Inference Dataset for Indonesian & Rahmad Mahendra, et al. & 2021 & University of Indonesia & NLP & 5 \\
The ROOTS Search Tool: Data Transparency for LLMs & Aleksandra Piktus, et al. & 2023 & Sapienza University of Rome & NLP & 5 \\
How Good Is NLP? A Sober Look at NLP Tasks through the Lens of Social Impact & Zhijing Jin, et al. & 2021 & Max Planck Institute for Intelligent Systems & NLP & 5 \\
Multilingual Multifaceted Understanding of Online News in Terms of Genre, Framing, and Persuasion Techniques & Jakub Piskorski, et al. & 2023 & Institute of Computer Science & NLP & 4 \\
JWSign: A Highly Multilingual Corpus of Bible Translations for More Diversity in Sign Language Processing & Shester Gueuwou, et al. & 2023 & Kwame Nkrumah University of Science and Technology & NLP & 4 \\

On Natural Language User Profiles for Transparent and Scrutable Recommendation & Filip Radlinski, et al. & 2022 & Google (United Kingdom) & Web\&IR & 5 \\
Leveraging Crowdsourcing Data for Deep Active Learning: An Application & Jie Yang, et al. & 2018 & Delft University of Technology & Web\&IR & 4 \\
Secure Centrality Computation Over Multiple Networks & Gilad Asharov, et al. & 2017 & Cornell University & Web\&IR & 4 \\
The Next Wave of the Web & Nigel Shadbolt, et al. & 2006 & University of Southampton & Web\&IR & 4 \\
Experiments on Generalizability of User-Oriented Fairness in Recommender Systems & Hossein A. Rahmani, et al. & 2022 & University College London & Web\&IR & 4 \\

\end{longtable}
\end{landscape}
\normalsize

\end{appendices}

\bibliography{references}

\end{document}